\documentclass[review,fleqn]{elsarticle}
\usepackage{booktabs}       
\usepackage{amsfonts}       
\usepackage{nicefrac}       

\usepackage{comment,amsfonts,mathtools,bm,booktabs,lmodern, microtype,xcolor,algorithm,graphicx, amsmath,amssymb, amsthm}
\usepackage{enumerate}
\graphicspath{ {./} {../code/output/} {./figures/} }
\usepackage[noend]{algorithmic}
\usepackage[colorlinks=false,linkbordercolor = {white},hidelinks]{hyperref}
\usepackage{boldline}
\usepackage{subcaption}

\usepackage{url}                  
\usepackage{nicefrac}         
\usepackage[section]{placeins}

\newtheorem{theorem}{Theorem}
\newtheorem{proposition}{Proposition}

\newtheorem{lemma}[theorem]{Lemma}
\theoremstyle{definition}

\newtheorem{assumption}{Assumption}
\theoremstyle{remark}

\newcommand{\paa}{p_{A}^{\alpha}}
\newcommand{\pab}{p_{A}^{\beta}}
\newcommand{\pbb}{p_{B}^{\beta}}
\newcommand{\pba}{p_{B}^{\alpha}}
\newcommand{\paao}{p_{A}^{\alpha*}}
\newcommand{\pabo}{p_{A}^{\beta*}}
\newcommand{\pbbo}{p_{B}^{\beta*}}
\newcommand{\pbao}{p_{B}^{\alpha*}}
\newcommand{\Fa}{F(\xi^{\alpha})}
\newcommand{\Fb}{F(\xi^{\beta})}
\newcommand{\fa}{f(\xi^{\alpha})}
\newcommand{\fb}{f(\xi^{\beta})}
\newcommand{\ha}{h_{\alpha}(\paa-\pba)}
\newcommand{\hb}{h_{\beta}(\pbb-\pab)}
\newcommand{\hap}{h_{\alpha}'(\paa-\pba)}
\newcommand{\hbp}{h_{\beta}'(\pbb-\pab)}
\newcommand{\ca}{c_{A}}
\newcommand{\cb}{c_{B}}
\newcommand{\camcb}{\ca-\cb}
\newcommand{\ta}{l_{\alpha}} 
\newcommand{\tb}{l_{\beta}} 
\newcommand{\xia}{\xi^{\alpha}}
\newcommand{\xib}{\xi^{\beta}}
\newcommand{\df}{\delta_{F}}
\newcommand{\da}{\delta_{A}}
\newcommand{\db}{\delta_{B}}
\newcommand{\vaa}{V_{A}^{\alpha}}
\newcommand{\vab}{V_{A}^{\beta}}
\newcommand{\vbb}{V_{B}^{\beta}}
\newcommand{\vba}{V_{B}^{\alpha}}
\newcommand{\vaao}{V_{A}^{\alpha*}}
\newcommand{\vabo}{V_{A}^{\beta*}}
\newcommand{\vbbo}{V_{B}^{\beta*}}
\newcommand{\vbao}{V_{B}^{\alpha*}}
\newcommand{\uaa}{U_{\alpha}^{A}} 
\newcommand{\uba}{U_{\alpha}^{B}}
\newcommand{\va}{\vaao-\vabo}
\newcommand{\vb}{\vbbo-\vbao}
\newcommand{\ga}{g_{\alpha}}
\newcommand{\gb}{g_{\beta}}
\newcommand{\la}{l_{\alpha}}
\newcommand{\lb}{l_{\beta}}
\newcommand{\oa}{s_{\alpha}}
\newcommand{\ob}{s_{\beta}}

\biboptions{authoryear}

\begin{document}

\begin{frontmatter}

\title{Price Discrimination in the Presence of Customer Loyalty and Differing Firm Costs}

\author[mymainaddress]{Theja Tulabandhula\corref{mycorrespondingauthor}}
\cortext[mycorrespondingauthor]{Corresponding author}
\ead{tt@theja.org}

\author[mymainaddress]{Aris Ouksel}
\ead{aris@uic.edu}

\author[mymainaddress]{Son Nguyen}
\ead{snguye65@uic.edu}

\address[mymainaddress]{601 S Morgan St, Information and Decision Sciences, University of Illinois at Chicago, Chicago 60607}

\begin{abstract}
We study how loyalty behavior of customers and differing costs to produce undifferentiated products by firms can influence market outcomes. In prior works that study such markets, firm costs have generally been assumed negligible or equal, and loyalty is modeled as an additive bias in customer valuations. We extend these previous treatments by explicitly considering cost asymmetry and richer customer loyalty behavior in a game-theoretic model. Thus, in the setting where firms incur different non-negligible product costs, and customers have firm-specific loyalty levels, we comprehensively characterize the effects of loyalty and product cost difference on market outcomes such as prices, market shares, and profits. Our analysis and numerical simulations provide new insights into how firms can price, how they can survive competition even with higher product costs, and how they can control these costs and/or increase customer loyalty to change their market position.
\end{abstract}

\begin{keyword}
Customer loyalty, product cost asymmetry, price discrimination, Markov equilibrium.
\end{keyword}

\end{frontmatter}

\section{Introduction}\label{sec:intro}

An established approach in competitive studies, especially ones that consider price discrimination, is to assume that product cost is either negligible or equal for all firms ~\citep{shaffer2000pay,chen2001individual,ouksel2011loyalty,MATSUBAYASHI2008571,WANG2017563,XIA2011194}. While this assumption may simplify mathematical derivations and shed light on certain aspects of such markets (e.g., how demand influences equilibrium prices), it is not reflective of reality. Firms routinely incur different product costs due to various reasons, including but not limited to differences in logistics, production, marketing, sales, distribution, service, technology, financial administration, information resources, and general administration.
This variation can potentially influence market outcomes through its interactions with other market properties. 

For example, in addition to product costs, customer loyalty is a key driver that determines much of the pricing and marketing decisions of a large variety of firms in today's world.  In industries such as internet services (e.g., AT\&T and Xfinity), telephone services (e.g., T-Mobile and Verizon), phones (e.g., Apple and Samsung), credit cards (e.g., Amex and Chase), airlines (e.g., Southwest and Delta), and subscription services (e.g., storage solutions Dropbox, Google Drive and Microsoft OneDrive, ride-hailing services Uber and Lyft, delivery services DoorDash and UberEats, media services YouTube TV, Hulu and traditional cable), firms sell largely similar good/service even if they have different product costs, and customers regularly exhibit loyalty and switching behavior. The impact of such customer behavior, which can be modeled by additive switching costs in certain cases, has received considerable attention in the recent literature~\citep{somaini2013model,rhodes2014re,villas2015short,cabral2016dynamic,MATSUBAYASHI2008571,WANG2017563}. As evidenced by these studies, the loyalty characteristics of consumers play a pivotal role in how firms make their decisions, for instance, how they price their products in the market. Nonetheless, most of these prior works only capture simple additive models of such behavior, artificially limiting the nature of market outcomes that are possible, especially with cost asymmetry thrown into the mix. 

Our work comprehensively shows that customer loyalty behavior can interact with product cost asymmetry and reveal new market characteristics that hitherto have been under-explored or completely missed in the literature. For firms, these results underscore the importance of controlling costs and managing customer loyalty to achieve desired market outcomes. For example, we show how market outcomes (such as prices and profits) vary as a function of cost asymmetry and loyalty behavior. Firms could use this information and decide to increase customer loyalty (e.g., by increasing marketing costs) and achieve a new desired profit level/market share. As another example, we show how forward looking firms can charge a premium to existing captive customers and initially provide discounts to rival firms' customers in order to potentially change a premium in later periods, and how these choices can be significantly affected by loyalty and cost considerations. Insights such as these (see our full contribution summary below) can only be derived by explicitly considering interactions between loyalty and costs.

We consider the setting of firms offering an undifferentiated good/service and competing in a market, which can be explicitly segmented into sub-markets. In particular, we consider a game theoretic model with two asymmetric firms, where firms incur different product costs and customers exhibit varying degrees of loyalty levels, and the latter is parametrized by a general parametric model that subsumes multiple prior works~\citep{somaini2013model,rhodes2014re,villas2015short,MATSUBAYASHI2008571}. Firms view the market as composed of two identifiable homogeneous sub-markets: a strong sub-market where loyal customers are ex ante willing to pay a premium; and a weak sub-market where customers ex ante prefer to purchase from the rival firm. One firm's strong sub-market is its rival's weak sub-market.  The loyalty levels are stochastic and not necessarily similar in the two sub-markets. 

The above setting is exceedingly commonplace. The two variables, namely costs and loyalty behavior, can be said to be prominent in influencing market structures in areas such as internet services, phone, digital storage, ride-hailing, food delivery, and media products, to name a few. Firms in these areas (such as AT\&T, Xfinity, T-Mobile, Verizon, Apple, Samsung, Dropbox, Google Drive, Microsoft OneDrive, Uber, Lyft, DoorDash, UberEats, YouTube, Hulu, traditional cable companies, etc.) routinely decide how to increase their profits and market share, while competing with rival firms in the presence of varied product costs and firm-specific customer behaviors. When competing in undifferentiated segments, this decision making can be quite challenging. Apart from the usual strategy of aiming to differentiate oneself (which oftentimes is not possible), the prominent levers they aim to control are minimizing costs and expanding marketing efforts/promotions to enhance customer loyalty. The economic insights provided by our work can help guide the stakeholders to make these decisions quantitatively, if and when they use these levers. 

The first key characteristic of our setup is that unlike ~\citep{shaffer2000pay}, who consider both pay-to-switch (offering discounts to new customers) and pay-to-stay (offering discounts to existing customers) phenomena, we primarily focus on the former. Charging a premium to existing customers is a popular strategy in certain industries (such as the above) and relies on customer stickiness and loyalty to boost profits. Multiple firms have successfully employed it to various degrees~\citep{howard_2021}. Not only are we able to capture this pricing strategy and related outcomes in our analysis, we also show how cost considerations can amplify or suppress its effectiveness, given our generalized loyalty model.

The second key characteristic is that there is no reduction between the results of our setting and a setting where product costs are assumed zero. In other words, one cannot recover the equilibria and market outcomes by reformulating our problem setting to one where cost differences are fully eliminated. If one were to remove costs, there would have to be a corresponding change in demand beyond what is captured by customer loyalty, which necessitates product differentiation and a change in how customers value the product. These changes are incompatible with the motivating applications discussed above, and keeping costs different and transparently tracing their influence on market outcomes makes our treatment more realistic and actionable.

The third key characteristic is that firms in our setting can identify and distinguish customers as being loyal to them or their rival firm. This is not an assumption needed for analytical tractability. Instead, in many real-life situations, this is quite naturally observed. For example, in a duopolistic setting, if a customer is not purchasing from a firm, the firm can reasonably assume that she is loyal to the rival firm. In practice, with the rise of pervasive tracking, customer profiling, and machine learning technologies, price discrimination has become an essential tool that firms routinely use to increase their market share and profits. For instance, both internet service providers AT\&T and Xfinity (in the USA) target discounted prices for new customers while simultaneously charging higher premiums for existing customers \citep{howard_2021}. They also know (through their data collection processes or via external data sources) which customers are loyal to their rival firm. Moreover, the same is true in the case of media providers (e.g., YouTube TV and Hulu), phone services (e.g., T-Mobile and Verizon), etc.

Our analysis involves deriving equilibrium outcomes that explicitly account for cost differences and variation in loyalty behavior across the two firms. 
Our key technical contributions are as follows:

\begin{itemize}
\item We provide a comprehensive study of the impact of product cost asymmetry and loyalty levels on competition in single and multi-period (specifically, infinite horizon) settings. The case where product cost is considered to be negligible or equal across firms represents a special case.
Our results point to nonlinear dependence of market outcomes on cost differences and loyalty parameters.
\item In the single stage setting, customers are a priori loyal to one of the two firms according to a novel parametric loyalty model, which we call the \emph{linear loyalty model}. We identify six market structures, which depend on the relationship between costs and loyalty levels, and derive explicit expressions for prices in each case. These results are further extended to the special cases of \emph{multiplicative loyalty} (\ref{sec:ml})  
and the previously known \emph{additive loyalty} (\ref{sec:al}) 
\item In the infinite horizon setting, firms are forward looking, and customers purchase in each period. Our results here are based on the notion of a Markov equilibrium~\citep{maskin2001markov}, and extend and complement the results in ~\citep{somaini2013model,rhodes2014re} and ~\citep{villas2015short}. For instance, customers are short-lived (e.g., two periods) in these prior works, unlike our setting. Our analytical results are also complemented by numerical simulations to cover cases where closed form expressions for market outcomes are hard to derive.
\end{itemize}

These results can be extended to multiple firms and can take into account product differentiation (e.g., by situating the products on a Hotelling line~\citep{XIA2011194}).
More importantly, showing that optimal prices vary non-linearly with cost differences and loyalty parameters also provides unique actionable insights for decision makers, some of which are as follows. First, a high-costing firm cannot only survive but thrive in an undifferentiated market as long as it cultivates a high degree of customer loyalty. In order to do so, it may have to increase its cost (e.g., due to additional marketing), which is counter-intuitive. More generally, any firm can affect its market position and the overall market outcome by controlling costs or by influencing customer behavior or both, and the choice depends on its existing market position and the sensitivity of the equilibrium. These considerations can be simultaneous to or follow the natural tendency of firms to differentiate their offerings from others. Second, even if the products are priced the same and the loyalty levels are the same, one firm can make a significantly different profit from the other based on product costs. But the firm that has a lower cost has much more flexibility in pricing and consequently in controlling the market outcomes as a whole. For instance, a motivated low costing firm can potentially drive its competitor out of the market.

Third, while we show that the equilibrium depends on loyalty parameters, the question of how firms learn these loyalty parameters (and not just the membership of each customer in the market) is left open. Depending on the sensitivity of profits to such parameters, which can be derived from our analysis, decision makers can choose to expend resources for learning/data science. Fourth, our computational framework can also accommodate time varying costs (e.g., decreasing costs due to technology improvements) and loyalty levels (e.g., due to delayed marketing effects) while computing market metrics. Finally, our setup allows for price discrimination beyond the coarse sub-market level that is presented here and can be re-purposed for individual customer level analysis (for example, see the infinite horizon setting in Section~\ref{sec:gl-ih} where we already do this). This implies that decision makers can use the same computational framework and analyze business outcomes at various granularities, and make further inroads into personalized pricing.

To summarize, our main contributions in this paper are the introduction of product cost asymmetry in a game-theoretic price discrimination model that captures a fairly general consumer loyalty behavior, and the analysis of the impact of such cost differences and loyalty characteristics on competition, prices and market share. We discuss additional related works in~\ref{sec:related-work}, justify various assumptions in~\ref{sec:assume-justify}, analyze a simpler infinite horizon setting in~\ref{sec:myopic}, discuss future directions in~\ref{sec:futurework}, and provide proofs in~\ref{sec:proofs}.

\section{Problem Setting: Competition, Costs, and Loyalty}\label{sec:gl}

We first discuss the single stage setting, where firms and customers interact in a single period, followed by the infinite horizon setting, where the firms and customers interact with each other over multiple periods. In both settings, we capture: (a) loyalty, which is a customer's ex ante preference towards a firm/product that measures the extent to which she is impervious to pricing enticements by a rival firm, and (b) product costs, which are the costs incurred by firms to produce a unit of the (single) product that they are offering in the market. 

\subsection{Single Stage Setting}\label{sec:gl-ss}

\noindent\textit{Firms with Costs:} There are two firms, $A$ and $B$, that produce a similar product. Their product costs are $\ca \geq 0$ and $\cb \geq 0$ respectively. We denote the prices offered by $A$ to two sets of customers as $(\paa, \pab)$. Customers in set $\alpha$ ex ante prefer $A$ (and thus belong to its strong sub-market) and are offered $\paa$, and customers in set $\beta$ ex ante prefer $B$ (so they belong to $A$'s weak sub-market) and are offered $\pab$. Analogously, firm $B$ offers prices $(\pbb, \pba)$.

\noindent\textit{Customers with Loyalty:} Each customer purchases exactly one unit (see Assumption~\ref{assume:customer-type}(iv)) and belongs to exactly one of the two sets: $\alpha$ and $\beta$, which expresses their ex ante preference for one firm over the other. If she belongs to set $\alpha$, she could buy from the non-preferred firm depending on her idiosyncratic loyalty level towards firm $A$, and the prices charged by the two firms (e.g., if firm $A$ charges a hefty premium). In other words, a customer may purchase from her non-preferred firm if the preferred firm decides to charge a premium higher than her loyalty level can tolerate. Because of such loyalty effects, her inclination to purchase from the non-preferred firm depends on how much discount she can obtain by \emph{switching}. 

Consider a generic customer belonging to set $\alpha$. Her idiosyncratic utilities (which are influenced by her loyalty to each of the firms) from purchasing the product from firms $A$ and $B$ are $\uaa$ and $\uba$ respectively. We will model the difference in utilities $\uaa-\uba$ using a random variable $\xi$ supported on the real line $\mathbb{R}$. That is, let $\uaa-\uba = \ga(\xi)$, where $\ga()$ is a scalar (potentially non-linear) invertible function that parameterizes the loyalty level. This customer will purchase from firm $A$ if the realization of random variable $\xi$ is such that $g_{\alpha}(\xi) \geq \paa-\pba$. That is, if the price premium charged by her preferred firm is less than her loyalty level, then she tolerates the premium and buys from her loyal firm. Otherwise, she switches to her non-preferred firm. 

Thus, given that she is in set $\alpha$, her ex ante probability of purchasing from $A$ is given by $1 - F(h_{\alpha}(\paa-\pba))$, where $F$ is the distribution function associated with $\xi$ and the scalar function $h_{\alpha}()$ is the inverse of $\ga()$. For notational convenience, let $\xia = \ha$ be a fixed scalar threshold, given the prices and $h_{\alpha}()$. Then, the probability of purchase can be re-expressed as $1 - \Fa$ (we hide the dependence of $\xia$ on $\paa$ on $\pba$ when the context is clear). Analogously, the ex ante probability that a customer from set $\beta$ purchases the product from firm $B$ is given by $1 - \Fb$, where the fixed scalar threshold $\xib = \hb$.

\noindent\textit{Dynamics:} All customers in the market ex ante belong to set $\alpha$ or $\beta$. In a single period of interaction, firms take into account the competition while maximizing their overall profit and simultaneously decide on a pair of prices to offer to their strong sub-market and weak sub-market. These prices are endogenously determined, in contrast to the product costs and loyalty levels, which are exogenous and are viewed as parameterizing the game instance. These latter quantities can be controlled (directly or indirectly) by the firms if they want to change their market position and profits but are fixed in the context of the game (See Section~\ref{sec:ll} for more details). Here is an initial list of key assumptions made:

\begin{assumption}\label{assume:customer-type} 
(i) Firms have the ability to offer different prices to different customers (e.g., $\paa$ and $\pab$ to their strong and weak sub-markets, respectively).
(ii) In addition to the customers knowing their membership in sets $\alpha$ and $\beta$, the firms have the ability to classify all customers in the market as being in set $\alpha$ or $\beta$ perfectly.
(iii) The firms have prior access to or are able to learn/estimate the loyalty model functions $\ga$ and $\gb$. 
(iv) The market is covered, and customers are ex ante members of either the strong sub-market of firm $A$ or the strong sub-market of firm $B$.
(v) The product costs are such that $\ca \geq \cb \geq 0$. By symmetry, the case where $\ca < \cb$ will not be examined. 
(vi) The functions $h_{\alpha}()$ and $h_{\beta}()$ are differentiable.
\end{assumption}

We again note that our analysis is restricted to two firms because we seek a minimal setting that captures the most salient features of the interaction between costs and loyalty. However, it can be extended to multiple firms without loss of generality. And while we characterize market outcomes, we will primarily be interested in pure equilibria (e.g., pure Nash equilibria), where firms don't randomize over prices. As we will show in Section~\ref{sec:ll}, in many cases, it is the only type of equilibrium in the game.

\noindent\textit{Demand:} Let the normalized initial market share of $A$ be $\theta$ (i.e., $\theta = |\alpha|/(|\alpha| + |\beta|)$, where $|\cdot|$ denotes the size of a set). Then, the demands from the strong and weak sub-markets for firm $A$'s product are $\theta(1-F(\xia))$ and $(1-\theta)F(\xib)$ respectively. Similarly, the demands from the strong and weak sub-markets for firm $B$ are $(1-\theta)(1-F(\xib))$ and $\theta F(\xia)$. 

\noindent\textit{Firm's Objective:} The objective of the firms is to post prices that maximize their profits in the presence of competition. Each firm has a single product (with infinite inventory) to sell and offers two prices, one to their loyal following (namely, their strong sub-market), another one to their rival's loyal customers (namely, their weak sub-market). Given the above demand functions, firm $A$'s profit maximization problem can be written as:
\begin{gather}
\max_{\paa \geq \ca,\pab \geq \ca} (\paa - \ca)\theta(1-F(\xia)) + (\pab - \ca)(1-\theta)F(\xib).
\label{eq:gl-ss-obj}
\end{gather}

It is easy to see that the problem above is separable across the two price variables, yielding two $1$-dimensional problems. An analogous optimization problem can be written for firm $B$.

Consider the case where loyalty level function $\ga(\xi) \in \mathbb{R}_+$. In this case, if a firm charges a premium to its loyal following, some of its least loyal customers (those with relatively low idiosyncratic loyalty levels) end up making purchases from its competitor. A higher premium improves profit margin; however, the firm's new market share for its loyal following also shrinks. Therefore, a firm should be mindful of the trade-off  between market share and profit margin. Further, the firm cannot charge a premium for its weak sub-market (its rival's loyal customers) as those customers already do not prefer its product. Therefore, the firm has to undercut its rival for its weak sub-market. If the firm offers a substantially lower price than its rival, its market share for its weak sub-market improves. However, its profit margin declines. 

\noindent\textit{Market Outcomes:} In our setting, firms independently and simultaneously determine their pricing strategies in a non-cooperative game. In this game, an equilibrium strategy profile (here, the four prices above constitute a strategy profile) is such that neither firm can improve its profits by unilaterally changing its own set of prices. 

To be precise, the equilibrium that we are concerned with in this sub-section is the Pure Nash Equilibrium (PNE). It is a strategy profile where neither firm $A$ or $B$ would benefit by changing their prices unilaterally. For instance, in the $\alpha$ sub-market, the price pair $(\paao$, $\pbao)$ is a PNE if and only if $\textrm{profit}_{A}^{\alpha}(\paao, \pbao) \geq \textrm{profit}_{A}^{\alpha}(\paa, \pbao)$ and $\textrm{profit}_{B}^{\alpha}(\paao, \pbao) \geq \textrm{profit}_{B}^{\alpha}(\paao, \pba)$
where $\textrm{profit}_{A}^{\alpha}(\paa, \pba) := (\paa - \ca)\theta(1-F(\xia))$ and $\textrm{profit}_{B}^{\alpha}(\paa, \pba) := (\pba - \cb)(1-\theta)F(\xia)$. A similar set of conditions can be defined for $(\pbbo$, $\pabo)$ in the $\beta$ sub-market. 
The following proposition states that if an equilibrium strategy exists, it satisfies the following first order conditions.

\begin{proposition}\label{prop:gl-ss-eq} Under Assumptions~\ref{assume:customer-type}(i)-(vi), the following implicit equations must be satisfied by any unconstrained equilibrium solution of prices offered by the firms $A$ and $B$:

\begin{align}
\paa = \ca + \frac{1 - \Fa}{\fa \hap}, \quad
\pab = \ca + \frac{\Fb}{\fb\hbp},\label{eq:gl-ss-eq}\\
\pbb = \cb + \frac{1 - \Fb}{\fb\hbp}, \text{ and } \pba = \cb + \frac{\Fa}{\fa\hap},\nonumber
\end{align}

where $\xia = \ha$ and $\xib = \hb$ respectively.
\end{proposition}

The proposition above can be used to prove the existence of an unique equilibrium for specific choices of the loyalty function and the distribution of the underlying random variable $\xi$, as long as the prices are unconstrained, i.e., they are not limited by the costs (for example, see Section~\ref{sec:ll-ss}, \ref{sec:ml-ss} and~\ref{sec:al-ss}). In reality, the prices are constrained by costs when the probabilities of purchase approach/equal $0$ or $1$. Additionally, cost lower bounds can also become binding constraints if the first derivatives of inverses of loyalty functions, viz., $\hap$ and $\hbp$ become negative over their domains. Constraints on prices can also manifest from customer behavior. For example, for any realistic loyalty model, we prefer that if $\pba>\paa$, then the probability of a customer from set $\alpha$ purchasing from firm $B$ is zero. This happens naturally, for instance, when $F$ is supported between $[0,\infty)$ and $\ga,\gb$ are the identity functions. In this case, the value $\max(0,\paao - \pbao)$ is the discount being offered by firm $B$ to its weak sub-market (where ${}^*$ denotes equilibrium/optimal prices) and similarly, $\max(0,\pbbo - \pabo)$ is the discount being offered by firm $A$ to its weak sub-market. 

\emph{It is precisely due to these binding constraints, which can manifest due to the loyalty functions and their relationship with the product costs, that we obtain new market structures, with potentially different equilibria}. 
These new market structures would not be easily discernible if one directly works with net prices (e.g., $\paa-\ca$) in their analysis. Further, as we will see later, the relationship between the equilibrium prices and the product cost differences is  \emph{non-linear} and depends on their interactions with the loyalty model parameters.

Given an equilibrium strategy profile $(\paao,\pabo,\pbbo,\pbao)$, the ex post market shares and profits in the single stage setting can be easily recovered from Equation~\ref{eq:gl-ss-obj}. For example, when $F$ is uniform between $[0,1]$ and $\ga,\gb$ are the identity functions, the size of the ex post strong sub-market of firm $A$ decreases proportionally to the difference in prices $\paao$ and $\pbao$ that customers loyal to $A$ observe. Assuming $\paao>\pbao$, the higher the difference, the larger is the premium being charged by firm $A$, and the smaller is the ex post market share. We can also infer from Equation~\ref{eq:gl-ss-eq} that the equilibrium prices do not depend on the ex ante market shares, and this is simply due to the firms being able to offer different prices for the two sub-markets, decoupling the pricing problems as a result.

\subsection{Infinite Horizon Setting}\label{sec:gl-ih}

We aim to derive characterizations of the prices that firms $A$ and $B$ offer in steady state in each period in a multi-period setting described below, along with their resulting profits and market shares. 


\noindent\textit{Forward Looking Firms with Costs:} Firms $A$ and $B$ price their products and customers respond by deciding their purchases, in a repeated fashion over an infinite horizon. Further, these firms are forward looking, i.e., they can consider future profits when they make their immediate pricing decisions in each period. One of the key consequences of firms being forward looking is that pricing decisions can now account for future outcomes. Analogous results for the case when firms are myopic is discussed in~\ref{sec:myopic}.

\noindent\textit{Customers with Loyalty:} While in each period $t$, a customer can either belong to set $\alpha$ or to set $\beta$, their loyalty level for that period, captured using the underlying random variable $\xi$, is assumed to be i.i.d across time. This is not a limitation because the loyalty functions $\ga$ and $\gb$ (which take $\xi$ as an input) depend on the ex ante membership of the customer in sets $\alpha$ and $\beta$ respectively, thus exhibiting a \emph{Markov property}. For a generic customer, let her loyalty random variables across time be denoted by the sequence $\{\xi_{i,t}\}$, where $i \in \{\alpha,\beta\}$ depending on which set she was part of immediately before time period $t$. If she belongs to the set $\alpha$ initially, then $i$ at time $t=0$ is equal to $\alpha$. We again assume that the market is covered (see Assumption~\ref{assume:customer-type}(iv)), and the customers are long-lived and continue to purchase from either firm $A$ or firm $B$. While variations such as customers being short-lived (e.g., entering and then exiting after two periods) and/or being strategic have been studied in prior works, our setting allows for a cleaner treatment of how loyalty interacts with costs without adding additional complexity and special cases (see~\ref{sec:futurework} for additional comments). 


\noindent\textit{Dynamics:} In each period, firms offer a pair of prices simultaneously to the market (composed of their strong and weak sub-markets) while accounting for the future state of the market and their discounted expected future profits. Customers purchase myopically based on their ex ante memberships in sets $\alpha$ and $\beta$, which can potentially change each period. That is, based on their current period purchases, the customers may switch their memberships (e.g., if they purchase from the rival firm). With updated membership information, the firms play their next set of state dependent actions (here, state refers to the loyalty set membership of each customer) in the subsequent time period ad infinitum.

\noindent\textit{Assumptions and their Justification:} Similar to the previous section, we will assume that both firms $A$ and $B$ know whether a given customer is in their strong sub-market or weak sub-market (see Assumption~\ref{assume:customer-type}(ii)). Because of this, the firms can focus on competing for business with each given customer, independent of other customers. Their profit from every customer in the market can then be aggregated to get the overall trends in profits, market share, etc. As an added benefit, this point of view makes the analysis of market outcomes tractable.

In addition to Assumptions~\ref{assume:customer-type}(i)-(vi), we make the following assumptions. First, we assume that firms $A$ and $B$ discount the expected value of their future profits using scalar valued time invariant discount factors $\da,\db \in (0,1)$ respectively. It is natural to assume that the firms follow the principle of \emph{time value of money}, where they value the profit obtained today higher than the same level of profit in the future, which results in the discount factors defined above. Second, for interpretation and analytical tractability, we will make the following assumption while discussing special cases.

\begin{assumption} The cost constraints ($\paa \geq \pab \geq \ca$ and $\pbb \geq \pba \geq \cb$) are non-binding.\label{assume:gl-ih-unconstrained}
\end{assumption}

Note that the above need not hold in general when firms solve for value maximizing steady-state prices, especially for arbitrary product costs and loyalty functions. Nonetheless, in the special cases where we assume this is true, it allows for a partial understanding of the market outcomes resulting from the infinite horizon game between the two firms.

Finally, we make a key regularity assumption about one of the primitives, namely the distribution function $F$, for analytical tractability, as shown below.

\begin{assumption}\label{assume:cdf} We assume that the random variable $\xi$ is such that $\frac{F(\xi)}{f(\xi)}$ and $\frac{F(\xi)-1}{f(\xi)}$ are strictly increasing functions of $\xi$ in its domain.
\end{assumption}

This assumption is satisfied by many natural choices for the distribution function, including the Beta and the Normal distributions.

\noindent\textit{Demand:} The demand in period each $t$ remains similar to the single stage setting. As mentioned before, we focus our analysis on a single myopic customer, which can be aggregated later to get the market shares and overall profits.

\noindent\textit{Firm's Objective:} The expected long-term value (LTV) that firm $A$ obtains by making a potential sale to a customer in its strong and weak sub-markets is as follows:

\begin{enumerate}
\item Firm $A$ can make a sale to a customer currently in its strong sub-market at a price $\paa$ to obtain the following expected long-term value:
\begin{gather}
\vaa(\paa) = (1-\Fa)(\paa -\ca +\da\vaao) + \Fa\da\vabo, \label{eqn:vaa}
\end{gather}
where $\vaao$ is the optimal value (in the context of the infinite horizon game) obtained by firm $A$ in the next time-step if this customer remains in its strong sub-market and $\vabo$ is the optimal value obtained by firm $A$ from the next time-step if the customer moves to its weak sub-market.
\item Firm $A$ can make a sale to a customer currently in its weak sub-market at a price $\pab$ to obtain the following expected long-term value:
\begin{gather}
\vab(\pab) = \Fb(\pab -\ca +\da\vaao) + (1-\Fb)\da\vabo. \label{eqn:vab}
\end{gather}
\end{enumerate}
Similarly, we can write the two expected long-term value functions of firm $B$, one each for a customer in its strong and weak sub-markets as follows:
\begin{gather}
\vbb(\pbb) = (1-\Fb)(\pbb -\cb +\db\vbbo) + \Fb\db\vbao, \text{ and} \label{eqn:vbb}\\
\vba(\pab) = \Fa(\pba -\cb +\db\vbbo) + (1-\Fa)\db\vbao. \label{eqn:vba}
\end{gather}

\noindent\textit{Market Outcomes:} Using the above expression, we can characterize the necessary conditions that any equilibrium strategy profile should satisfy under Assumptions~\ref{assume:customer-type} and~\ref{assume:gl-ih-unconstrained}, as shown below.

\begin{proposition}\label{prop:gl-ih-eq}
Under Assumptions~\ref{assume:customer-type} and ~\ref{assume:gl-ih-unconstrained}, the optimal prices should satisfy the following implicit equations:
\begin{gather}
\paa = \ca + \frac{1-\Fa}{\fa\hap} - \da(\vaao-\vabo), \label{eqn:paa}\\
\pab = \ca +\frac{\Fb}{\fb\hbp} -\da(\vaao-\vabo),\label{eqn:pab}\\
\pbb = \cb + \frac{1-\Fb}{\fb\hbp} -\db(\vbbo-\vbao), \text{ and} \label{eqn:pbb}\\
\pba = \cb + \frac{\Fa}{\fa\hap} -\db(\vbbo-\vbao). \label{eqn:pba}
\end{gather}
\end{proposition}

When we compare the above result to the unconstrained optimal prices obtained in Proposition~\ref{prop:gl-ss-eq}, we observe that each candidate optimal price has an additional additive term that is the discounted difference of optimal expected long term value of customers in a firm's strong sub-market and its weak sub-market.

From these expressions, we can obtain implicit equations relating thresholds $\xia$ and $\xib$ (which are functions of equilibrium prices) to the optimal values:
\begin{align}
\resizebox{.8\textwidth}{!}
     {%
$\xia = h_{\alpha}\left(\ca -\cb + \frac{1-2\Fa}{\fa\hap} - \da(\va)+\db(\vb))\right),$
}
\label{eqn:xia}\\
\resizebox{.8\textwidth}{!}
     {%
$\xib = h_{\beta}\left(\cb -\ca + \frac{1-2\Fb}{\fb\hbp} - \db(\vb) +\da(\va))\right).$
}
\label{eqn:xib}
\end{align}

These non-linear equations will be central for us in proving the existence and uniqueness of Markov equilibria under the linear loyalty model and its special cases in Section~\ref{sec:ll-ih} (also~\ref{sec:ml-ih-fm} and~\ref{sec:al-ih-fm}). From these equations, we can already observe that the thresholds (and thus prices) depend on the product costs as well as on the parameterization of the loyalty functions. In other words, depending on the primitives $\ga(), \gb(), F, \ca$ and $\cb$, market outcomes such as prices and market shares can be highly non-linear and/or discontinuous functions of costs and loyalty parameters.

Under Assumptions~\ref{assume:customer-type}, ~\ref{assume:gl-ih-unconstrained} and~\ref{assume:cdf} these equations turn out to be necessary conditions that any optimal strategy profile has to satisfy. They can be then used to show that the candidate optimal prices obtained by maximizing the value functions in Equations~\ref{eqn:vaa}-\ref{eqn:vba} constitute a unique \emph{Markov equilibrium}~\citep{maskin2001markov} of the infinite horizon non-cooperative game between the two firms. We complement these analytical claims with numerical results (that get rid of Assumption~\ref{assume:gl-ih-unconstrained}) to comprehensively demonstrate the combined non-trivial effects of loyalties and costs on market outcomes.

One plausible pricing strategy that emerges from our analysis of market outcomes and is also observed in practice is the following:  Firms can initially sway the customers in their weak sub-market by posting low enough prices and then charge premium prices once they have become part of their strong sub-market. In this strategy, the firms are willing to price lower today to have the option of pricing higher tomorrow, thus being able to reap larger aggregated profits overall. Many real world market dynamics across industries exhibit such a pricing strategy. For example, both YouTube TV and Hulu (two media providers in the USA) post lower prices to customers loyal to their rival firm. \cite{youtube} shows that the offered prices $\pab \approx$ US \$ 55 and $\paa \approx $ US \$ 65 do conform to the aforementioned strategy. The same is true with AT\&T and Xfinity (two internet service providers in the USA) as well as Uber and Lyft (two ride-hailing service providers in the USA). Of course, in reality, not all customers will be myopic, and the pricing strategies get coupled with other promotions and incentives in order to drive the overall sales and profits.
\section{Linear Loyalty Model: Analysis and Insights}\label{sec:ll}

In this section, we specialize the loyalty level function to be linear/affine. That is, for the linear loyalty (LL) model, we assume that the loyalty function has the functional form: $\ga(\xi) = \ta\xi +\oa$, where $\ta,\oa \in \mathbb{R}_{+}$ (its inverse is given by $h_{\alpha}(y) = (y-\oa)/\ta$). When $\oa=0$, we obtain the multiplicative special case (ML) of the linear loyalty model discussed in~\ref{sec:ml}, and when $\ta = 1$, we get the additive special case (AL) discussed in~\ref{sec:al}. One of the reasons we have separately considered these special cases is to highlight and contrast the equilibria that arise in each of these cases, and relate them to existing results in the literature (e.g., the additive special case is equivalent to \emph{switching costs} defined in prior works).

Given prices $\paa$ and $\pba$, the probability of a customer belonging to the set $\alpha$ purchasing from firm $A$ is $1 - \Fa$, where $\xi^{\alpha} = (\paa - \pba -\oa)/\ta$. When the support of random variable $\xi$ is restricted to $[0,1]$, the loyalty model provides two insights. First, the parameter $\ta+\oa$ (as well as $\tb+\ob$) can be interpreted as the maximum loyalty level a customer can have. Second, it also suggests the following constraint on pricing: if a customer is offered a higher price by a non-preferred firm, then she only purchases from her loyal firm. On the other hand, if she is offered a very high price by her preferred firm relative to the non-preferred firm (e.g., over and above the rival's price and parameter $\oa$ or $\ob$), then she may not purchase from her loyal firm at all. Since customers have varying degrees of loyalty levels, very loyal customers will tolerate higher premiums. 
Next, we analyze the single stage setting followed by the infinite horizon setting. 

\subsection{Single Stage Setting}\label{sec:ll-ss}

In this one-shot competition between firms, the demand functions for the strong and weak sub-markets of firm A under are as follows:
\begin{align}
D_{A}^{ss}(\paa, \pba) &= \theta\left(1-F\left(\frac{\paa - \pba - \oa}{\ta}\right)\right), \text{ and}\\
D_{A}^{ws}(\pab, \pbb) &= (1-\theta)F\left(\frac{\pbb - \pab - \ob}{\tb}\right).
\end{align}

We make the following assumption while providing some of the analytical results below (we will explicitly mention when this assumption is used).

\begin{assumption}\label{assume:uniform-01}
 Distribution function $F$ is uniform on $[0,1]$.
\end{assumption}

The above assumption allows for the interpretability of the relationships we seek to explore. Extension to other invertible distribution functions can be done in a similar manner.

\begin{figure}[h]
\centering
\includegraphics[width=0.65\textwidth]{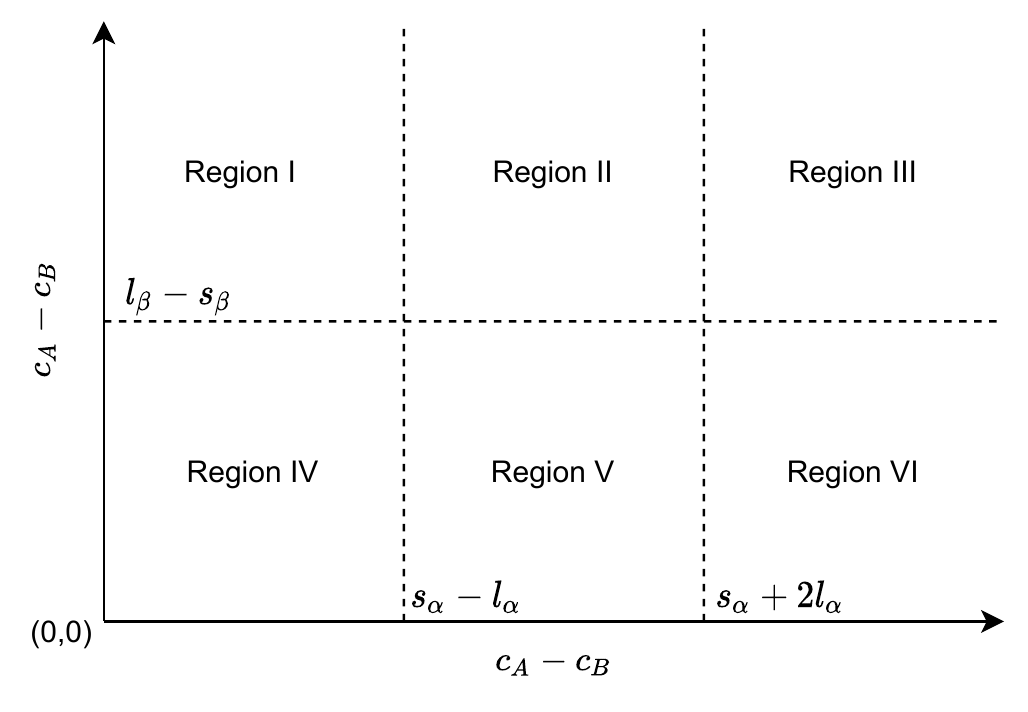}
\caption{Six regions that determine different equilibrium prices depending on the relationship between loyalty levels and product costs in the single stage linear loyalty setting.\label{fig:ll-ss-regions}}
\end{figure}

Under the choices made for $F$, $\ga$ and $\gb$ above, our analysis reveals six distinct price discrimination regions based on the interplay of loyalty parameters ($\ta$, $\tb$, $\oa$ and $\ob$) and the magnitude of product cost difference ($\ca-\cb$). Equilibrium conditions are determined for each of the 
following 
sub-cases in Propositions~\ref{prop:ll-ss-r1}-\ref{prop:ll-ss-r6} below, which are mutually exclusive and exhaustive (see Figure~\ref{fig:ll-ss-regions})
: 
\begin{itemize}
\item Region I: $\tb-\ob \leq \camcb \leq \oa-\ta$ (see Proposition~\ref{prop:ll-ss-r1}).
\item Region II: $\max(\tb-\ob,\oa-\ta) \leq \camcb \leq \oa+2\ta$ (see Proposition~\ref{prop:ll-ss-r2}).
\item Region III: $\max(\tb-\ob,\oa+2\ta) \leq \camcb$ (see Proposition~\ref{prop:ll-ss-r3}).
\item Region IV: $\camcb \leq \min(\oa-\ta,\tb-\ob)$ (see Proposition~\ref{prop:ll-ss-r4}).
\item Region V: $\oa-\ta \leq \camcb \leq \min(\oa+2\ta,\tb-\ob)$ (see Proposition~\ref{prop:ll-ss-r5}).
\item Region VI: $\oa+2\ta \leq \camcb \leq \tb-\ob$ (see Proposition~\ref{prop:ll-ss-r6}).
\end{itemize}

\begin{proposition}
\label{prop:ll-ss-r1}
Under Region I, with Assumptions~\ref{assume:customer-type} and~\ref{assume:uniform-01}, the unique pure Nash equilibrium prices for the strong and weak sub-markets for firms $A$ and $B$ for the LL model are as follows:
\begin{gather}
\paa = \cb+\oa,\;
\pab = \ca,\;
\pbb = \ca+\ob, \text{and }\;
\pba = \cb.
\end{gather}
\end{proposition}

\begin{proposition}
\label{prop:ll-ss-r2}
Under Region II, with Assumptions~\ref{assume:customer-type} and~\ref{assume:uniform-01}, the unique pure Nash equilibrium prices for the strong and weak sub-markets for firms $A$ and $B$ for the LL model are as follows:
\begin{gather}
\paa = \frac{1}{3}(2\ca+\cb+\oa+2\ta),\;
\pab = \ca,\;
\pbb = \ca+\ob, \text{and }\;\nonumber\\
\pba = \frac{1}{3}(\ca+2\cb-\oa+\ta).
\end{gather}
\end{proposition}

\begin{proposition}
\label{prop:ll-ss-r3}
Under Region III, with Assumptions~\ref{assume:customer-type} and~\ref{assume:uniform-01}, the unique pure Nash equilibrium prices for the strong and weak sub-markets for firms $A$ and $B$ for the LL model are as follows:
\begin{gather}
\paa = \ca,\;
\pab = \ca,\;
\pbb = \ca+\ob, \text{and }\;
\pba = \ca-\oa-\ta.
\end{gather}
\end{proposition}

\begin{proposition}
\label{prop:ll-ss-r4}
Under Region IV, with Assumptions~\ref{assume:customer-type} and~\ref{assume:uniform-01}, the unique pure Nash equilibrium prices for the strong and weak sub-markets for firms $A$ and $B$ for the LL model are as follows:
\begin{gather}
\paa = \cb+\oa,\;
\pab = \frac{1}{3}(\cb+2\ca-\ob+\tb),\;\nonumber\\
\pbb = \frac{1}{3}(2\cb+\ca+\ob+2\tb), \text{and }\;
\pba = \cb.
\end{gather}
\end{proposition}

\begin{proposition}
\label{prop:ll-ss-r5}
Under Region V, with Assumptions~\ref{assume:customer-type} and~\ref{assume:uniform-01}, the unique pure Nash equilibrium prices for the strong and weak sub-markets for firms $A$ and $B$ for the LL model are as follows:
\begin{gather}
\paa = \frac{1}{3}(2\ca+\cb+\oa+2\ta),\;
\pab = \frac{1}{3}(\cb+2\ca-\ob+\tb),\;\nonumber\\
\pbb = \frac{1}{3}(2\cb+\ca+\ob+2\tb), \text{and }\;
\pba = \frac{1}{3}(\ca+2\cb-\oa+\ta).
\end{gather}
\end{proposition}

\begin{proposition}
\label{prop:ll-ss-r6}
Under Region VI, with Assumptions~\ref{assume:customer-type} and~\ref{assume:uniform-01}, the unique pure Nash equilibrium prices for the strong and weak sub-markets for firms $A$ and $B$ for the LL model are as follows:
\begin{gather}
\paa = \ca,\;
\pab = \frac{1}{3}(\cb+2\ca-\ob+\tb),\;\nonumber\\
\pbb = \frac{1}{3}(2\cb+\ca+\ob+2\tb), \text{and }\;
\pba = \ca-\oa-\ta.
\end{gather}
\end{proposition}

The results above show that the game is in equilibrium regardless of the product cost difference and the loyalty model parameters. Figure~\ref{fig:ll-ss-regions} shows how the six regions related to each other. Both x- and y-axes represent $\camcb$, the horizontal dashed line corresponds to cases where $\tb-\ob = \camcb$, the left vertical dashed line corresponds to cases where $\oa-\ta = \camcb$, and the right vertical dashed line is for the cases where $\oa+2\ta = \camcb$. The figure captures the entire space of possible combinations of product cost differences divided into regions based on their relationship with the loyalty model parameters. The interplay of these two aspects (cost asymmetry and loyalty) together determines market equilibrium conditions, and we elaborate on them below.

\subsubsection{Discussion}

Propositions~\ref{prop:ll-ss-r1}-\ref{prop:ll-ss-r6} above illustrate that ignoring product cost differences in competitive price discrimination studies leads to disregarding many realistic competitive price discrimination market equilibria. Previous studies dealt only with the case represented by the origin in this graph, i.e., the case where product cost difference ($\camcb$) is zero (see Region IV). Region IV represents the cases where product cost difference is small compared to the maximum loyalty levels (recall that this interpretation is true when $F\sim U[0,1]$) to have any significant impact on the market structure. The two firms are able to sell to each other's strong and weak sub-markets. 
To the best of our knowledge, many of the previous competitive duopolistic price discrimination studies -- where product cost is either ignored or the difference in product costs is negligible – can be grouped within this class. 

\begin{figure}
\centering
	\begin{subfigure}{0.3\textwidth}
	\includegraphics[width=0.9\linewidth]{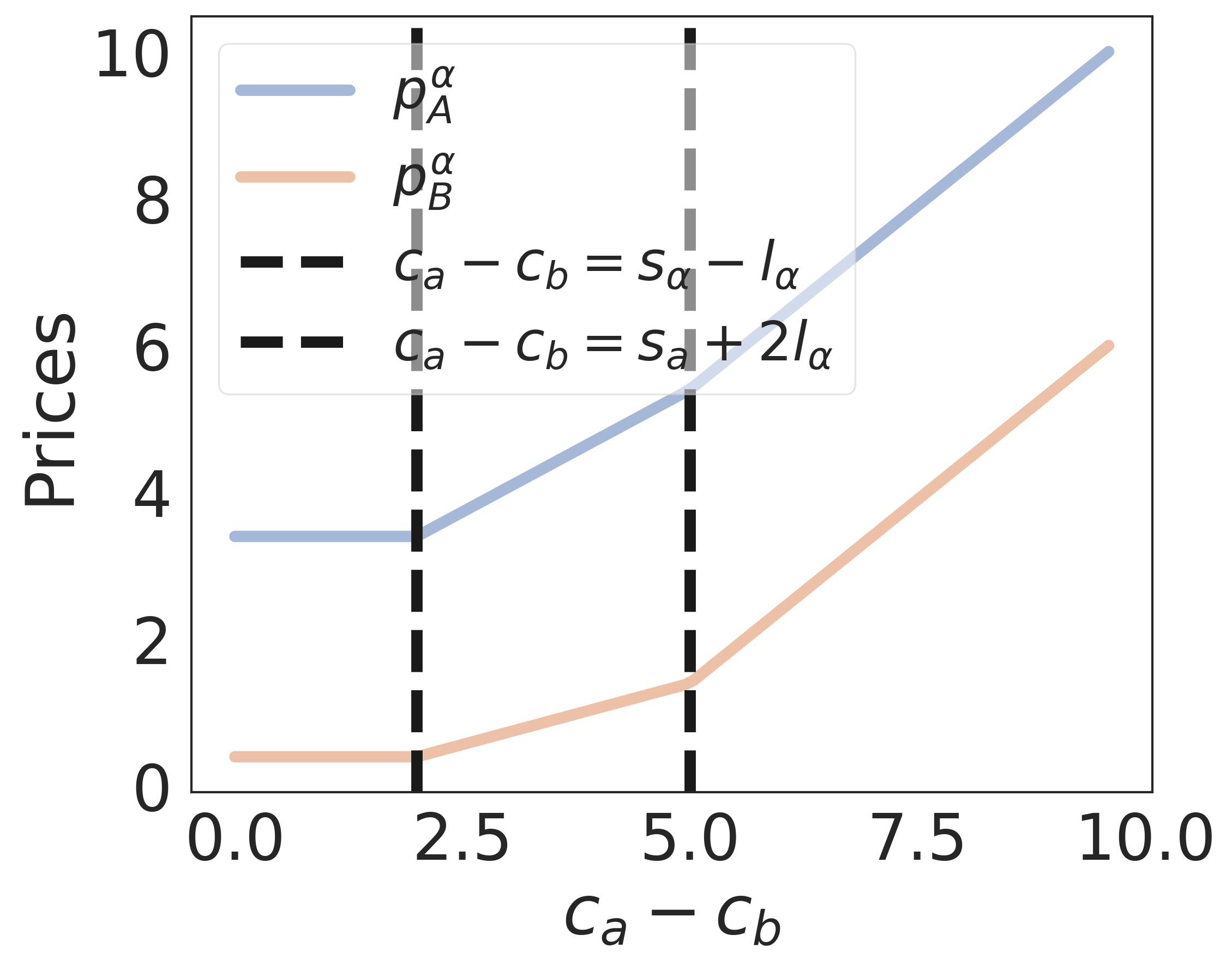} 
	\caption{Prices seen by firm $A$'s strong sub-market}
	\label{fig:ll_single_stage_paa_pba}
	\end{subfigure}
	\begin{subfigure}{0.3\textwidth}
	\includegraphics[width=0.9\linewidth]{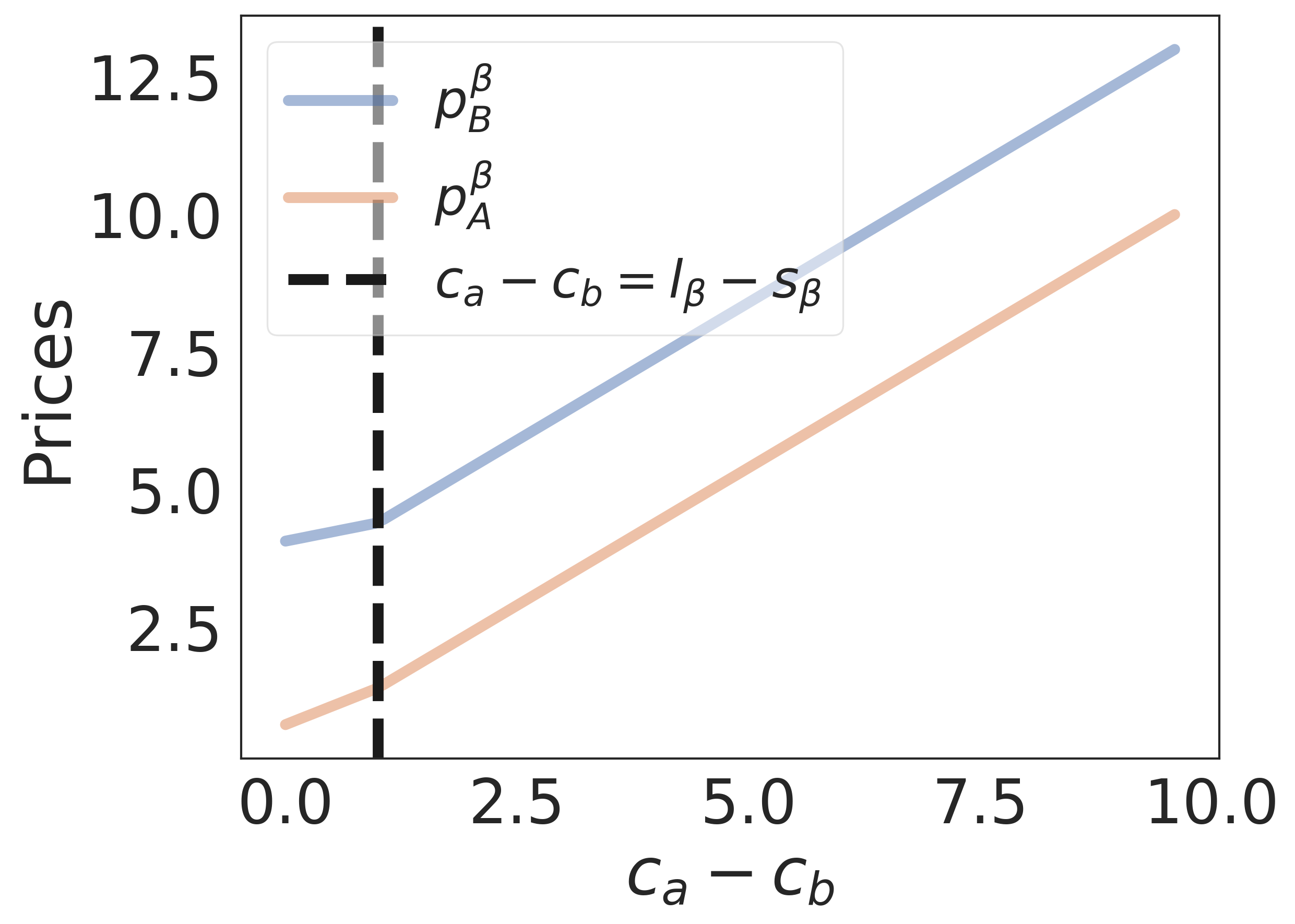}
	\caption{Prices seen by firm $A$'s weak sub-market}
	\label{fig:ll_single_stage_pbb_pab}
	\end{subfigure}
	\begin{subfigure}{0.3\textwidth}
	\includegraphics[width=0.9\linewidth]{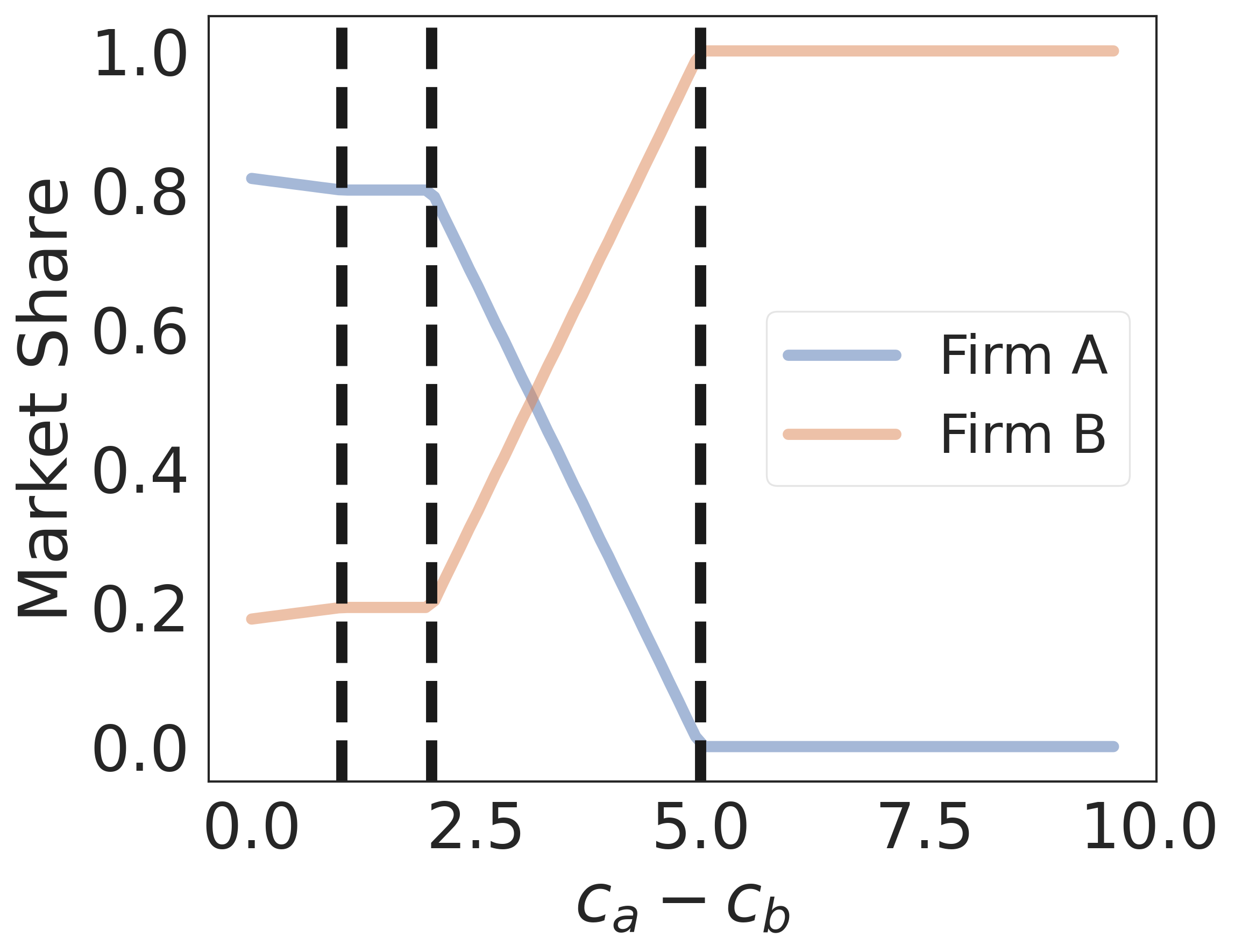} 
	\caption{Market share}
	\label{fig:ll_single_stage_market_share}
	\end{subfigure}
	\begin{subfigure}{0.3\textwidth}
	\includegraphics[width=0.9\linewidth]{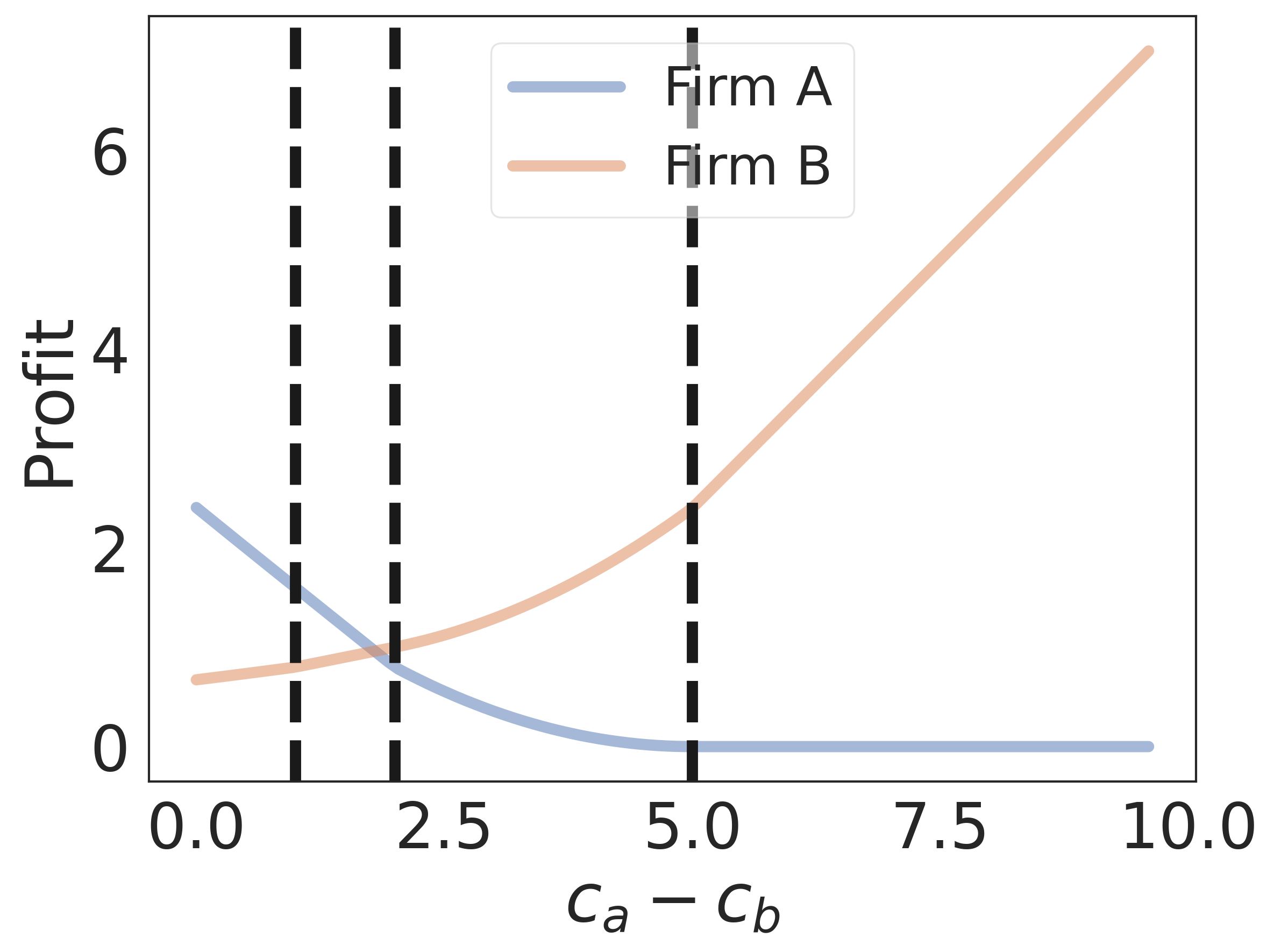}
	\caption{Profit}
	\label{fig:ll_single_stage_profits}
	\end{subfigure}
	\begin{subfigure}{0.3\textwidth}
	\includegraphics[width=0.9\linewidth]{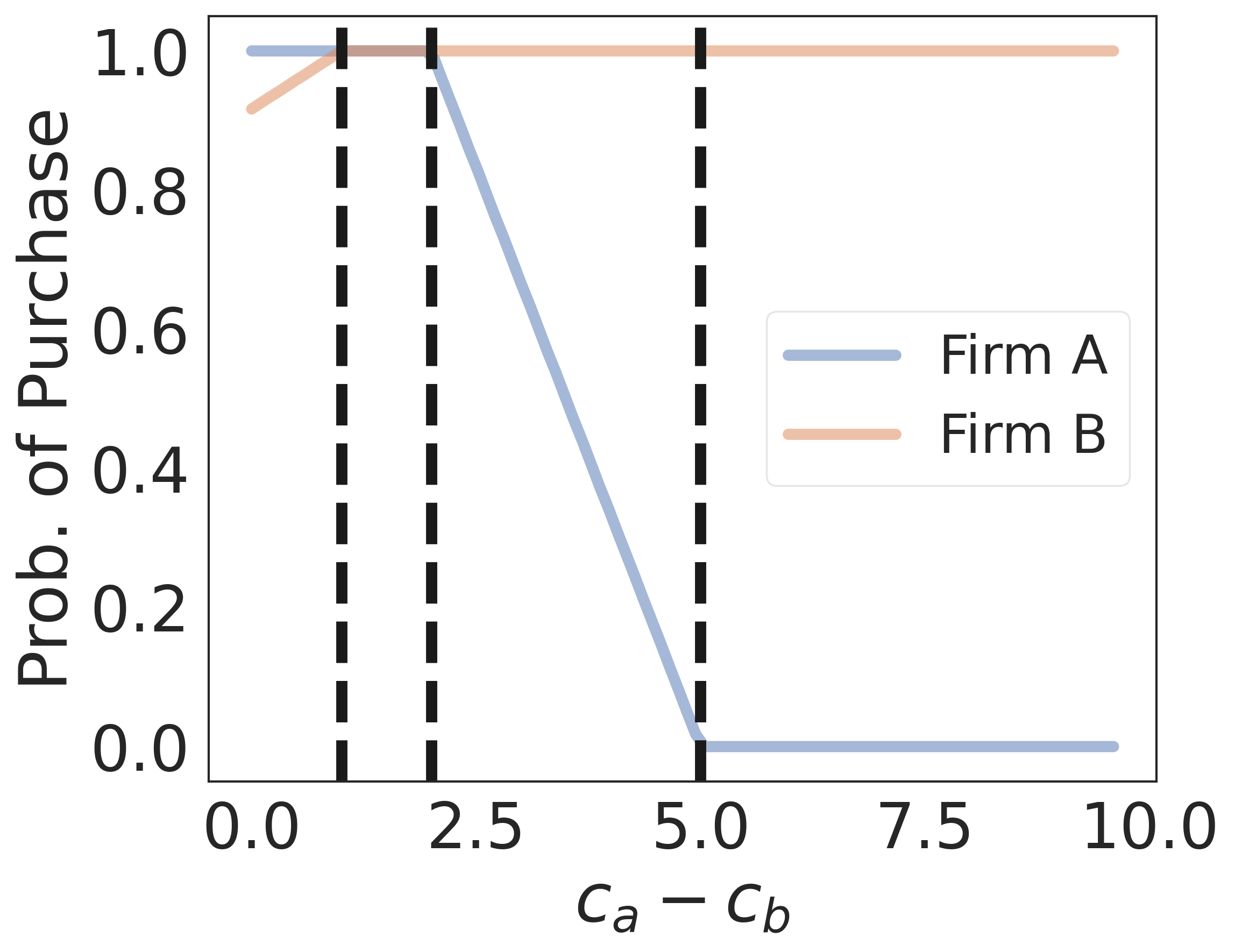}
	\caption{Probability of purchase}
	\label{fig:ll_single_stage_probabilities}
	\end{subfigure}
\caption{Single stage market outcomes under linear loyalty: Optimal prices, market shares, profits of firms, and probability of purchase of customers of types $\alpha$ and $\beta$ as a function of $\ca-\cb$. Here, $\cb=0.4, \oa=3, \ob=3, \ta=1, \tb=4$ and $\theta=0.8$.}
\label{fig:ll-ss-market-outcomes}
\end{figure}

We cannot expect the same outcome for the other regions. Nonlinear trends of market outcomes (prices, market shares, and profits) are evident from Figure~\ref{fig:ll-ss-market-outcomes}, which shows these for a specific choice of loyalty parameters and $\ob$. It reaffirms that different market outcomes can occur depending on the cost asymmetry and their relationship with loyalty parameters. For instance, as seen in Figure~\ref{fig:ll-ss-market-outcomes}, as the cost difference increases, high cost  firm $A$ has to increase its equilibrium price. Even with this increase, the high cost firm $A$ loses its market share and profit to its lower cost competitor firm $B$. It fails to sell in its own strong sub-market as well as its weak sub-market until it is driven out of business when the cost difference is equal to $\oa+2\ta$.

As mentioned earlier, the well-established approaches in competitive price discrimination literature, i.e., those assuming product cost as either negligible or equal for all firms, rule out the cases in Regions I, II, III, V, and VI, and yet, their market equilibrium conditions are significantly different in price, market share and profitability than in Region IV, as seen above. Thus, Regions I, II, III, V, and VI, which together represent a large class of competitive price discrimination market equilibrium cases, allow for a clearer understanding of the impact of costs and  loyalty on the market structure.

\subsection{Infinite Horizon Setting}\label{sec:ll-ih}

As in Section~\ref{sec:gl-ih}, we  characterize equilibrium prices when $\da=\db (=\df \textrm{, a common discount value}) >0$ (results for the setting where $\da=\db=0$ are in~\ref{sec:myopic}). 


It is harder to characterize the equilibrium conditions succinctly for the infinite horizon setting in general. Below, we ignore constraints on prices (as discussed in Section~\ref{sec:gl-ih}) and show that in this case, it is indeed possible to achieve a unique Markov equilibrium. Further, the result is obtained for any distribution function $F$ that satisfies Assumption~\ref{assume:cdf}. We complement this analysis with numerical evaluation of market trends when Assumption~\ref{assume:gl-ih-unconstrained} is relaxed.

\begin{proposition} For the linear loyalty model, under Assumptions~\ref{assume:customer-type},~\ref{assume:gl-ih-unconstrained} and ~\ref{assume:cdf}, there exists a unique Markov equilibrium when $\da=\db=\df > 0$, where firms price based on whether the customer bought their product in the immediate preceding time period. This equilibrium is characterized by the following fixed point equations for thresholds $\xia$ and $\xib$:
\begin{align*}
\left(\xia - \frac{\camcb-\oa}{\ta}\right)&\left(\frac{1-\df}{\df} +\Fb+1\right)\\
& + \frac{2\Fa-1}{\fa}\left(\frac{1-\df}{\df} +\Fa+\Fb\right) + \frac{\Fa}{\fa} \\
& = \frac{(1-\Fb)\tb}{\fb\ta} - \Fb\left(\frac{\tb}{\ta}\xib +\frac{\camcb}{\ta}+\frac{\ob}{\ta}\right),
\end{align*}
and
\begin{align*}
\left(\xib - \frac{\cb-\ca-\ob}{\tb}\right)&\left(\frac{1-\df}{\df} +\Fa+1\right)\\
& + \frac{2\Fb-1}{\fb}\left(\frac{1-\df}{\df} +\Fa+\Fb\right) + \frac{\Fb}{\fb} \\
& = \frac{(1-\Fa)\ta}{\fa\tb} - \Fa\left(\frac{\ta}{\tb}\xia +\frac{\cb-\ca}{\tb}+\frac{\oa}{\tb}\right).
\end{align*}

\label{prop:ll-ih-fm}
\end{proposition}

The above thresholds again can be used in conjunction with Equations~\ref{eqn:paa}-\ref{eqn:pba} to obtain the optimal prices, profits, and resulting market shares. The prices seen by the customer depend on the set they belong to (i.e., their state). Because the thresholds are implicitly defined, in the following, we numerically solve for them for a canonical market instance in order to obtain the dependence of key market metrics on cost asymmetry and loyalty parameters.

\subsubsection{Discussion} 

\begin{figure}
\centering
	\begin{subfigure}{0.3\textwidth}
	\includegraphics[width=0.9\linewidth]{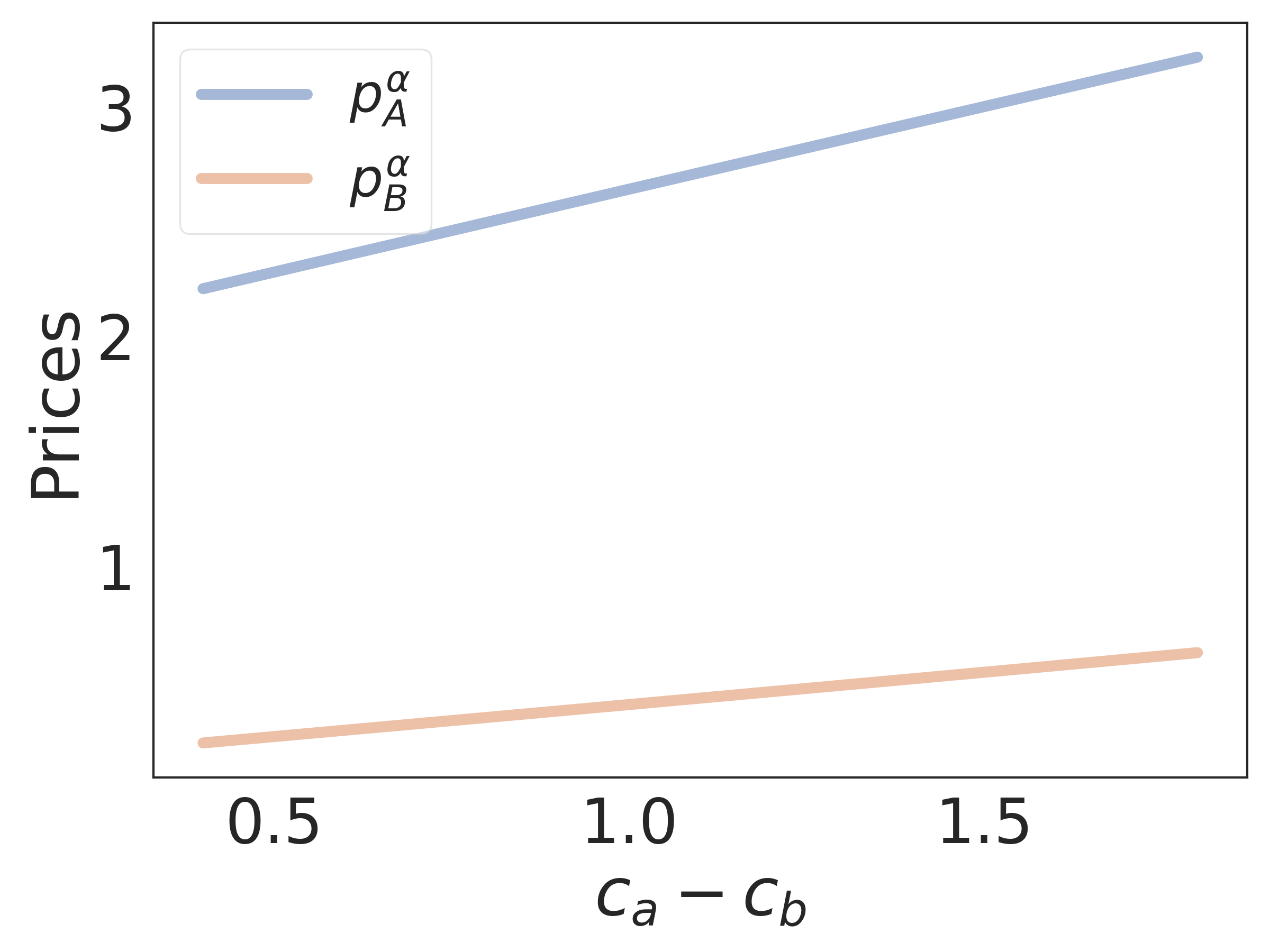} 
	\caption{Prices seen by firm $A$'s strong sub-market}
	\label{fig:ll_ih_fm_unc_paa_pba}
	\end{subfigure}
	\begin{subfigure}{0.3\textwidth}
	\includegraphics[width=0.9\linewidth]{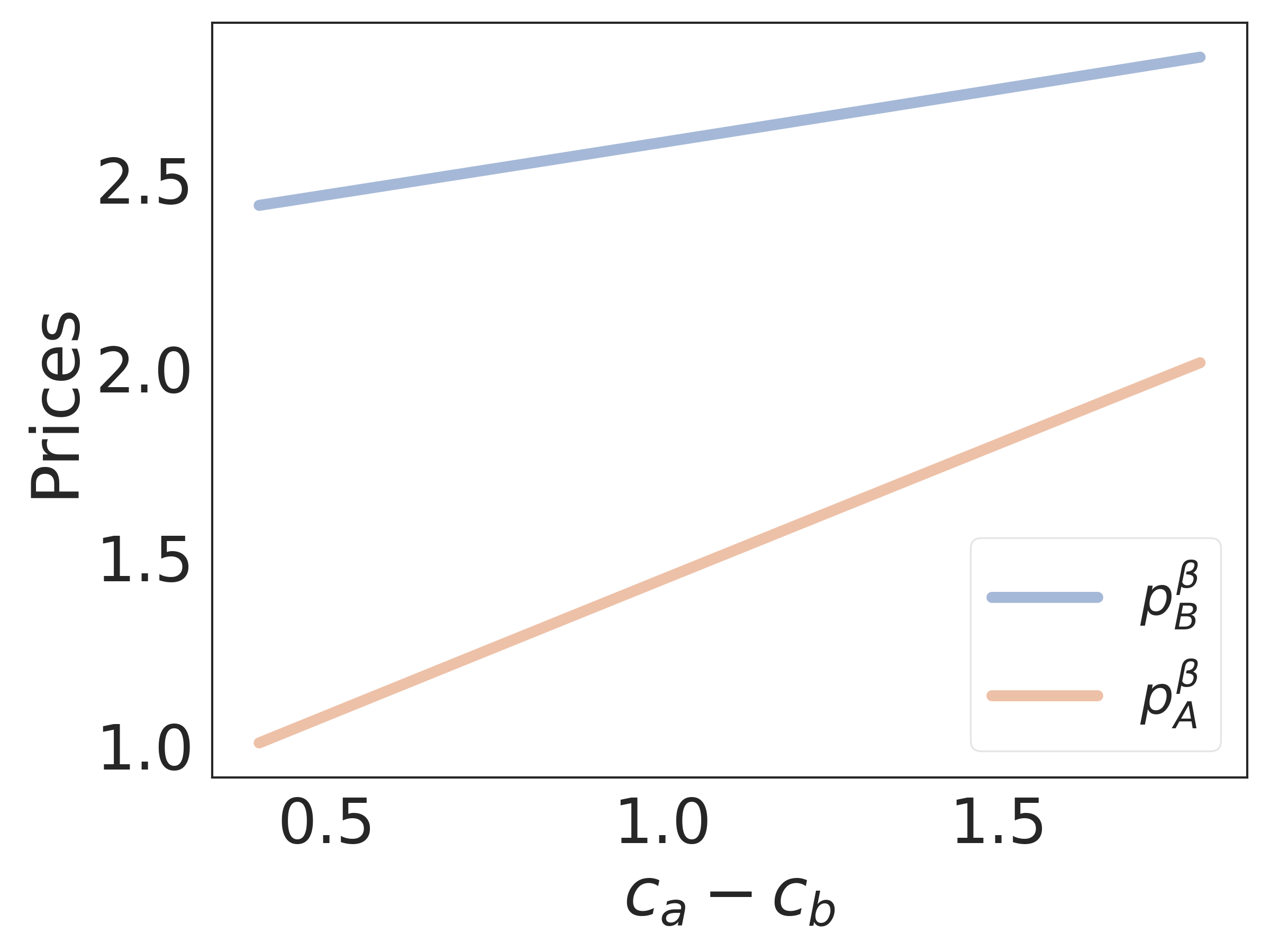}
	\caption{Prices seen by firm $A$'s weak sub-market}
	\label{fig:ll_ih_fm_unc_pbb_pab}
	\end{subfigure}
	\begin{subfigure}{0.3\textwidth}
	\includegraphics[width=0.9\linewidth]{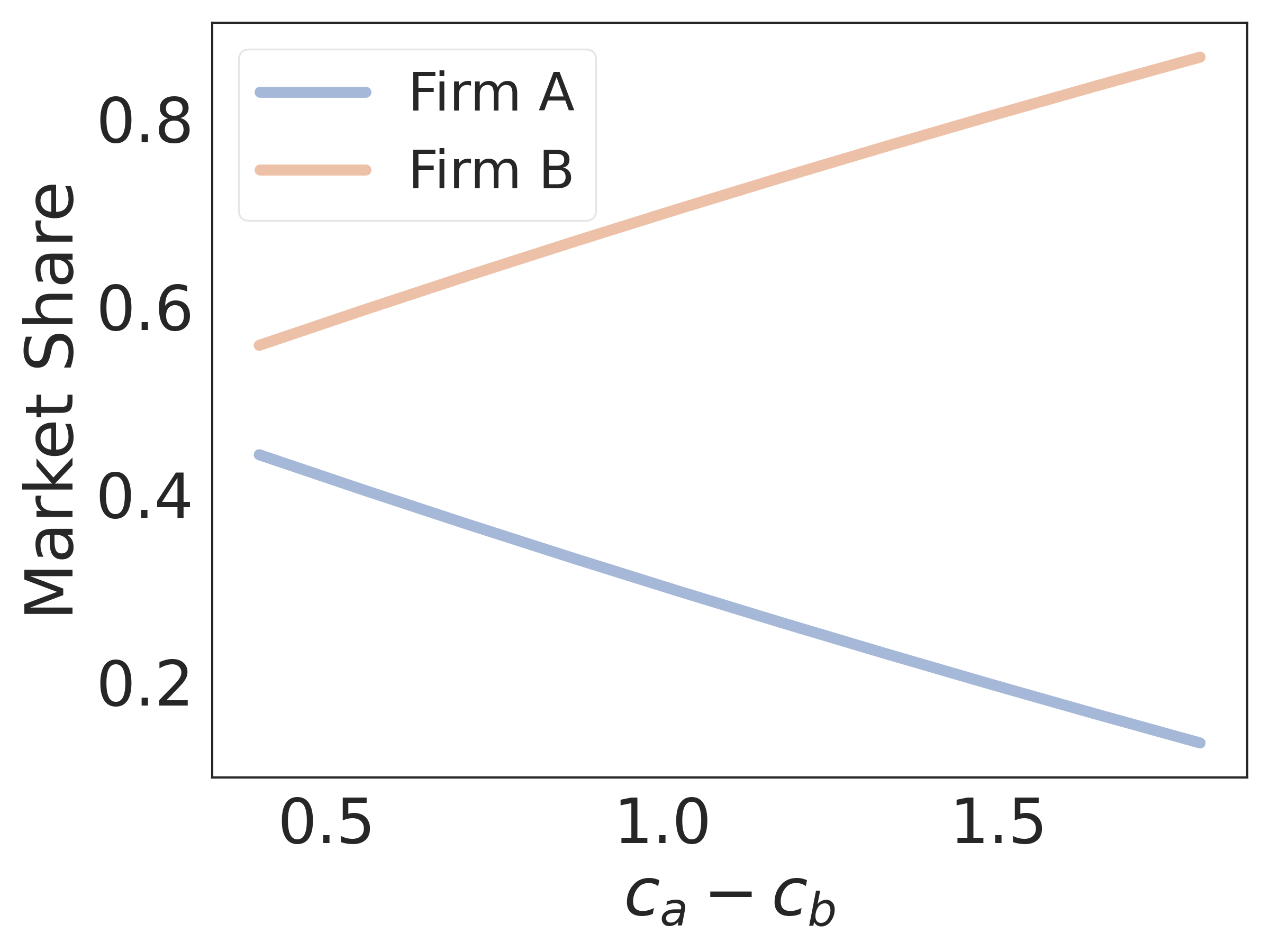} 
	\caption{Market share}
	\label{fig:ll_ih_fm_unc_mkt}
	\end{subfigure}
	\begin{subfigure}{0.3\textwidth}
	\includegraphics[width=0.9\linewidth]{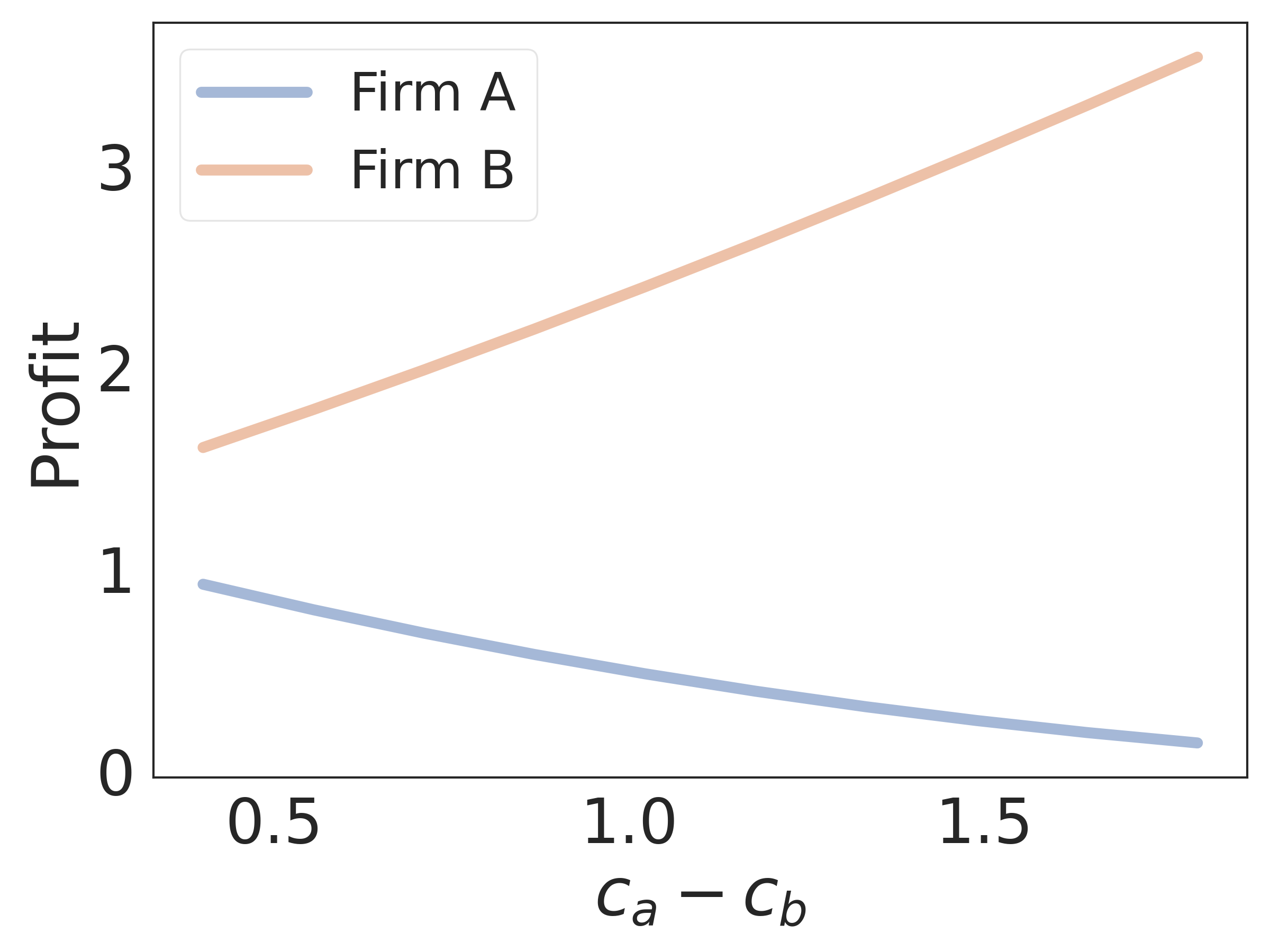}
	\caption{Profit}
	\label{fig:ll_ih_fm_unc_profit}
	\end{subfigure}
	\begin{subfigure}{0.3\textwidth}
	\includegraphics[width=0.9\linewidth]{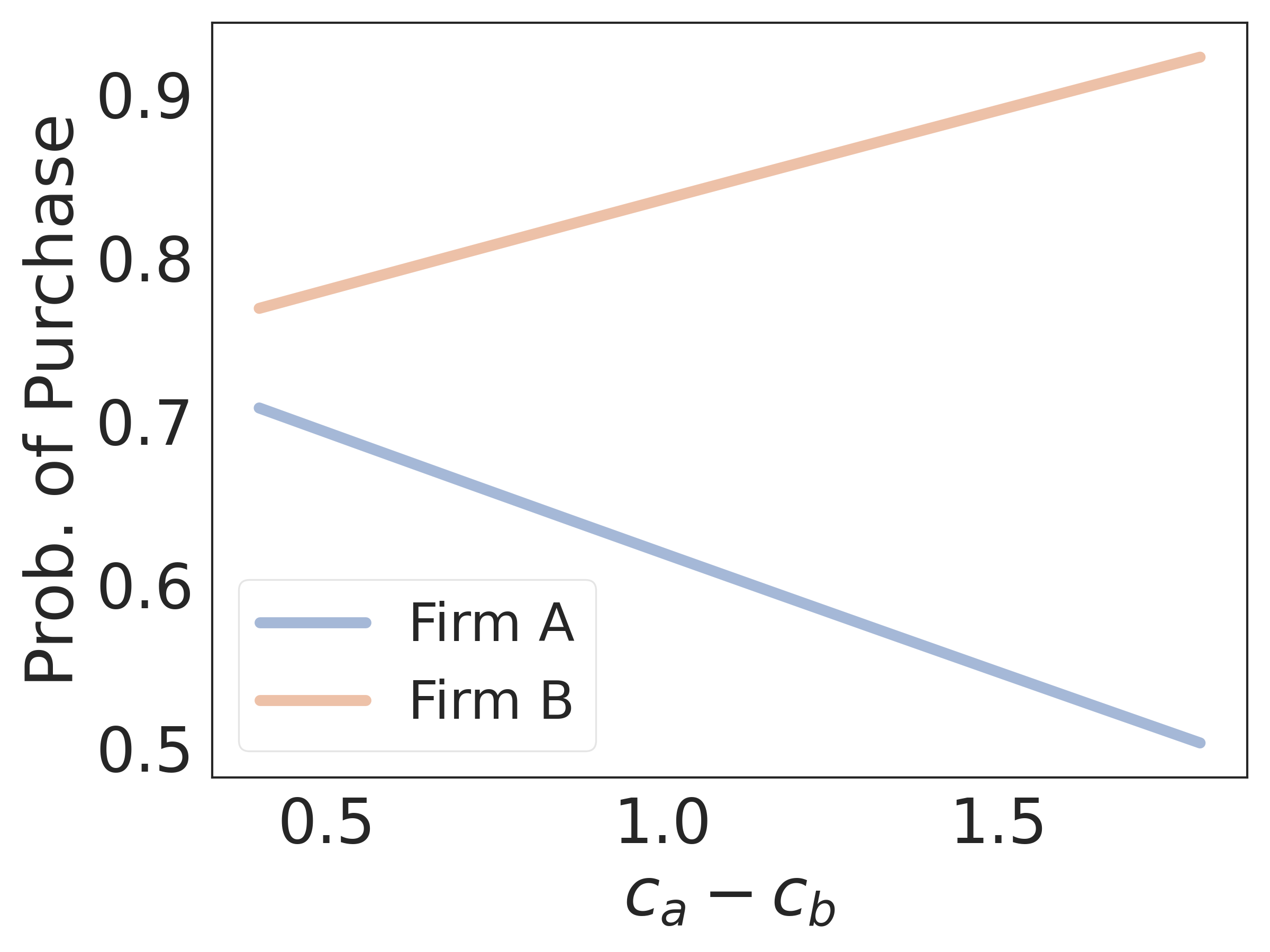}
	\caption{Probability of purchase}
	\label{fig:ll_ih_fm_unc_probabilities}
	\end{subfigure}
\caption{Infinite horizon setting market outcomes under linear loyalty where the constraints are non-binding: Optimal prices, market shares, profits of firms, and probability of purchase of customers of types $\alpha$ and $\beta$ as a function of $\ca-\cb$. Here, $\cb=.2, \ta=3, \tb=4, \oa = 1.1, \ob = .5, \df=.4$.}
\label{fig:ll_ih_fm_unc}
\end{figure}

In Figure~\ref{fig:ll_ih_fm_unc}, we plot the prices, market shares, and profits of firms for a specific set of loyalty parameter values and $\cb$ when $\da=\db=\df=0.4$. Here the prices are not constrained. That is, the values of $\ob$ and the loyalty parameters were selected in such a way that the constraints on prices are non-binding. From Figures~\ref{fig:ll_ih_fm_unc_paa_pba} and ~\ref{fig:ll_ih_fm_unc_pbb_pab}, we can infer that as the cost asymmetry increases, the prices change linearly with different slopes. In addition, as the cost difference between firm $A$ and firm $B$ increases, firm $A$ loses significant market share and profit because its loyal consumers increasingly prefer to buy from its competitor.

\begin{figure}
\centering
	\begin{subfigure}{0.3\textwidth}
	\includegraphics[width=0.9\linewidth]{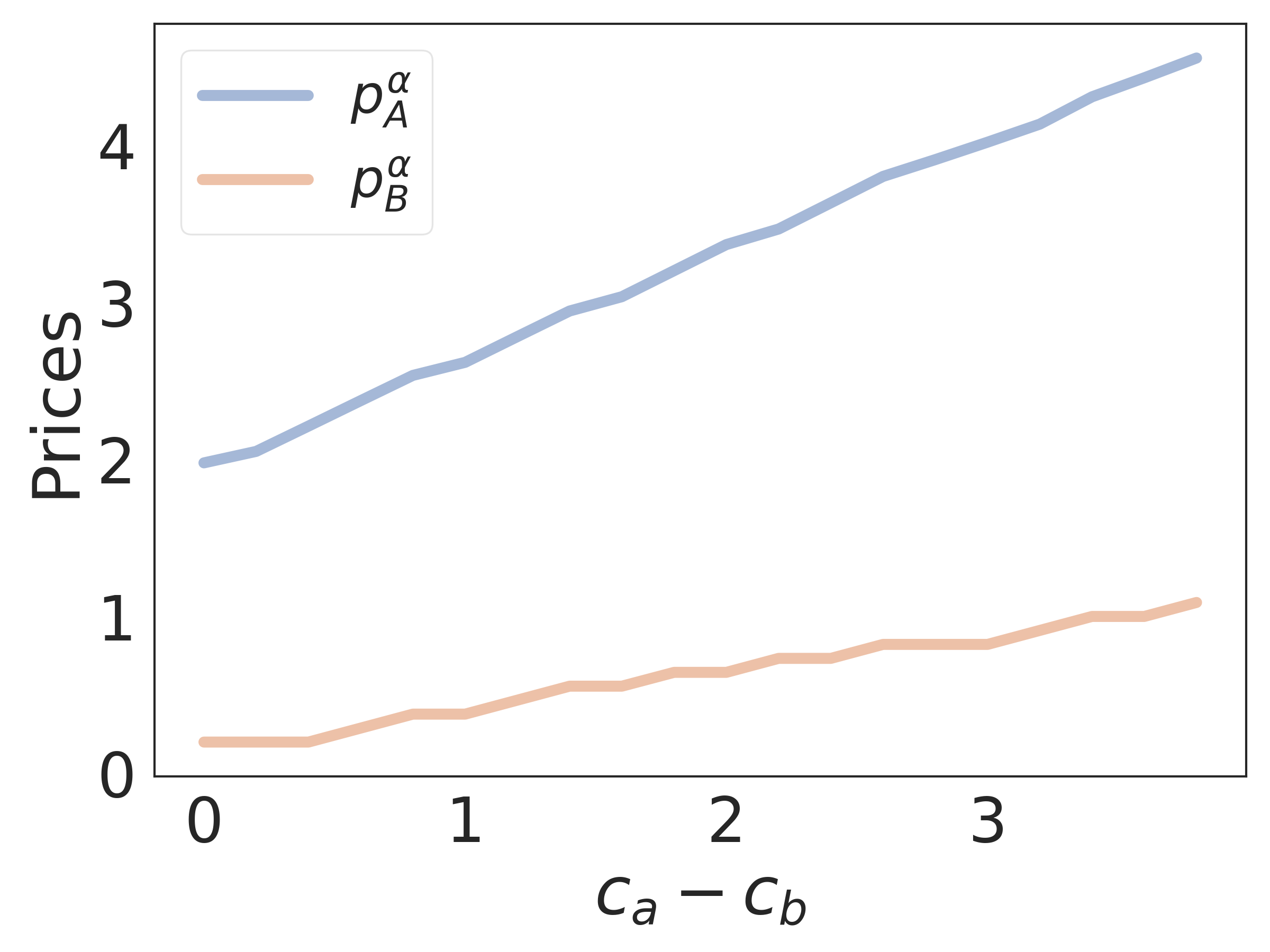} 
	\caption{Prices seen by firm $A$'s strong sub-market}
	\label{fig:ll_ih_fm_c_paa_pba}
	\end{subfigure}
	\begin{subfigure}{0.3\textwidth}
	\includegraphics[width=0.9\linewidth]{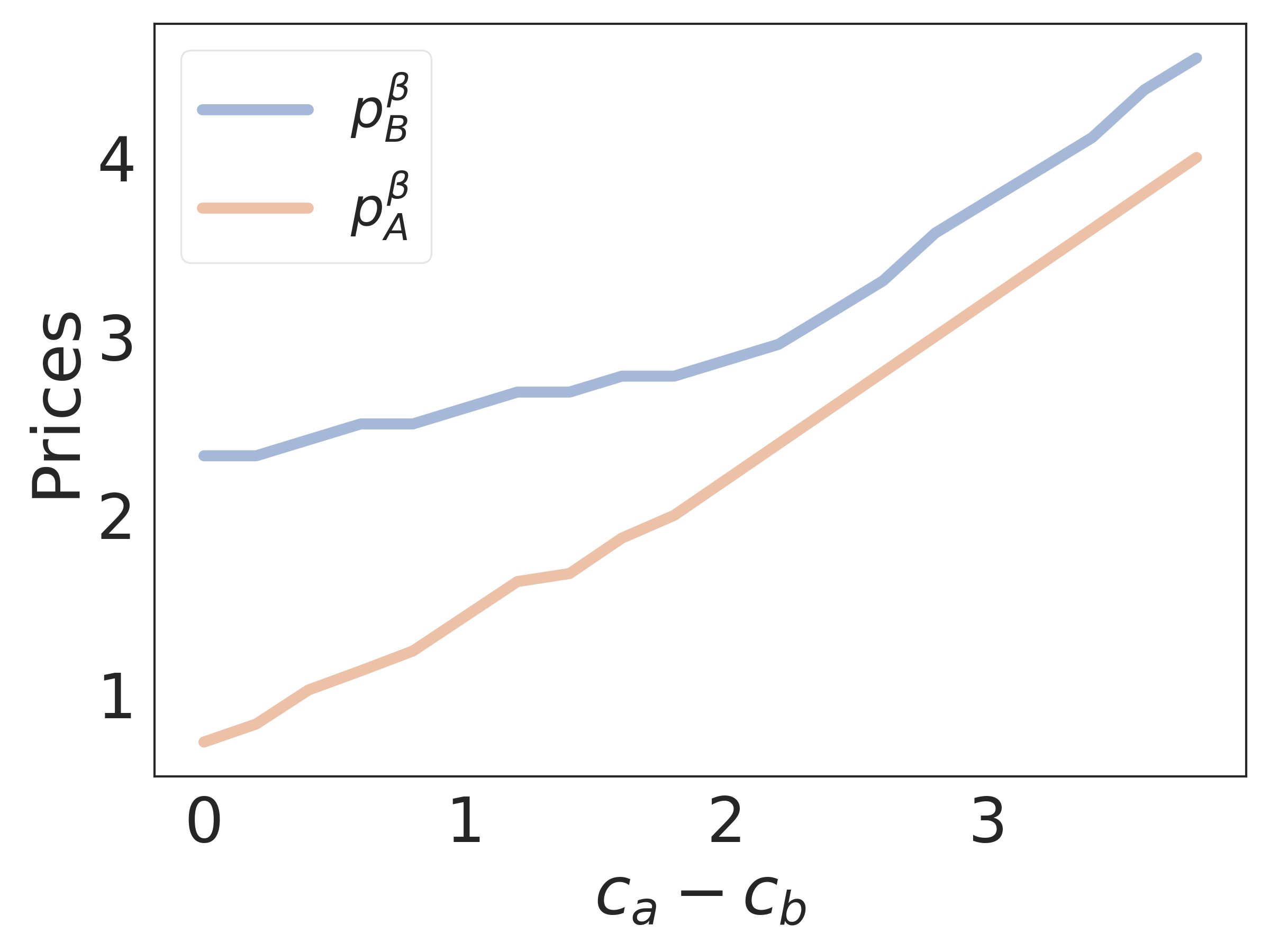}
	\caption{Prices seen by firm $A$'s weak sub-market}
	\label{fig:ll_ih_fm_c_pbb_pab}
	\end{subfigure}
	\begin{subfigure}{0.3\textwidth}
	\includegraphics[width=0.9\linewidth]{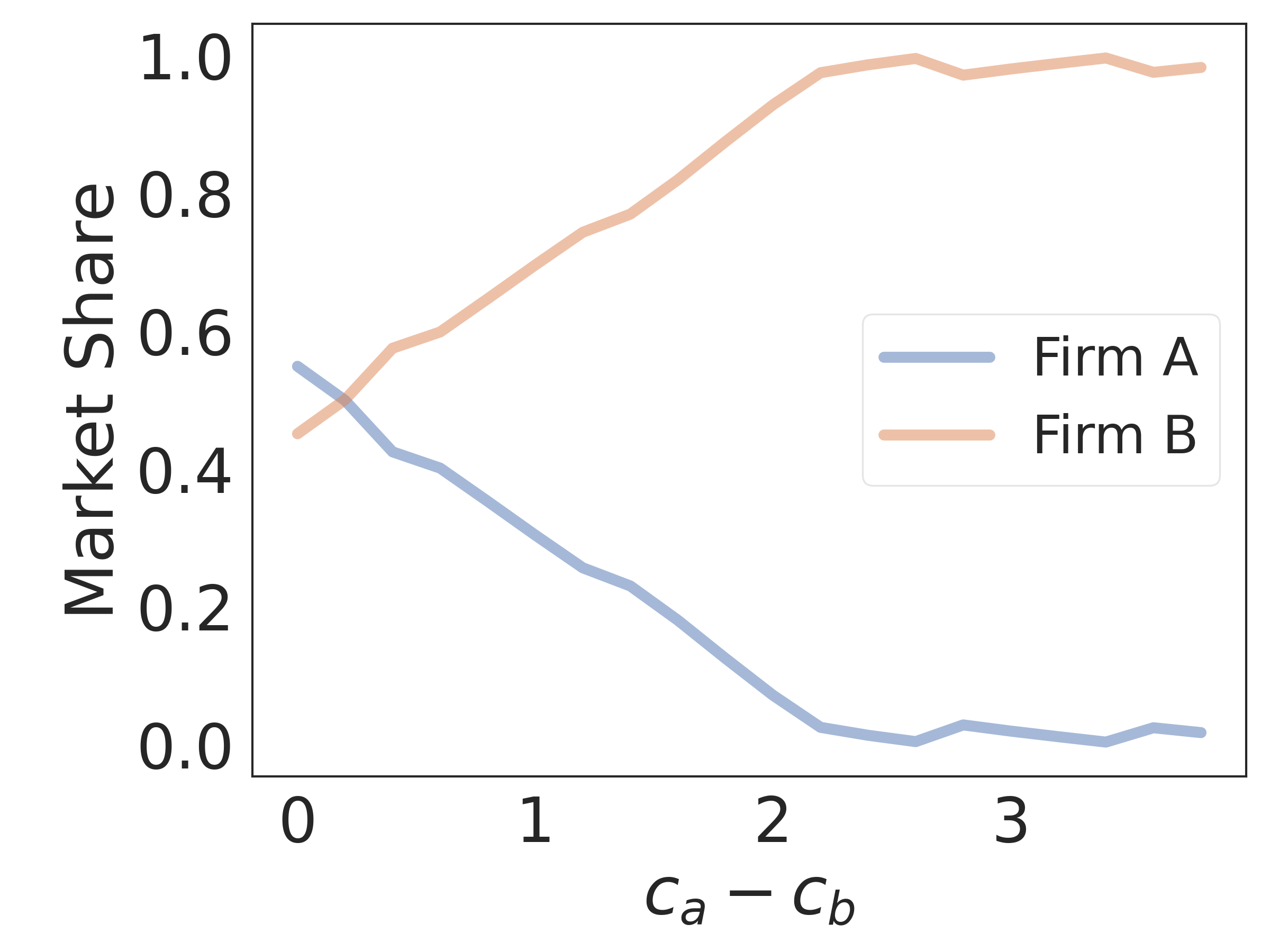} 
	\caption{Market share}
	\label{fig:ll_ih_fm_c_mkt}
	\end{subfigure}
	\begin{subfigure}{0.3\textwidth}
	\includegraphics[width=0.9\linewidth]{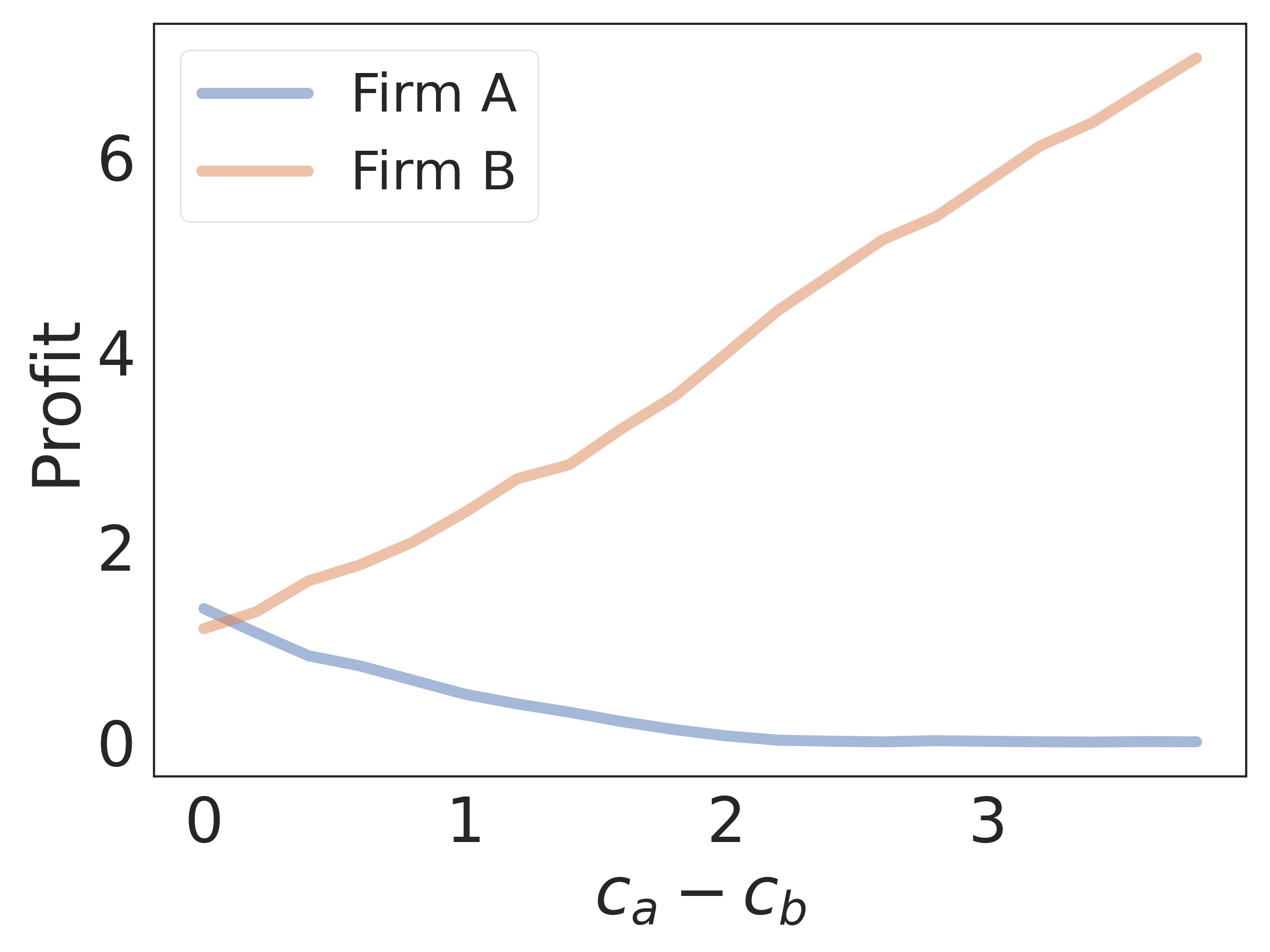}
	\caption{Profit}
	\label{fig:ll_ih_fm_c_profit}
	\end{subfigure}
	\begin{subfigure}{0.3\textwidth}
	\includegraphics[width=0.9\linewidth]{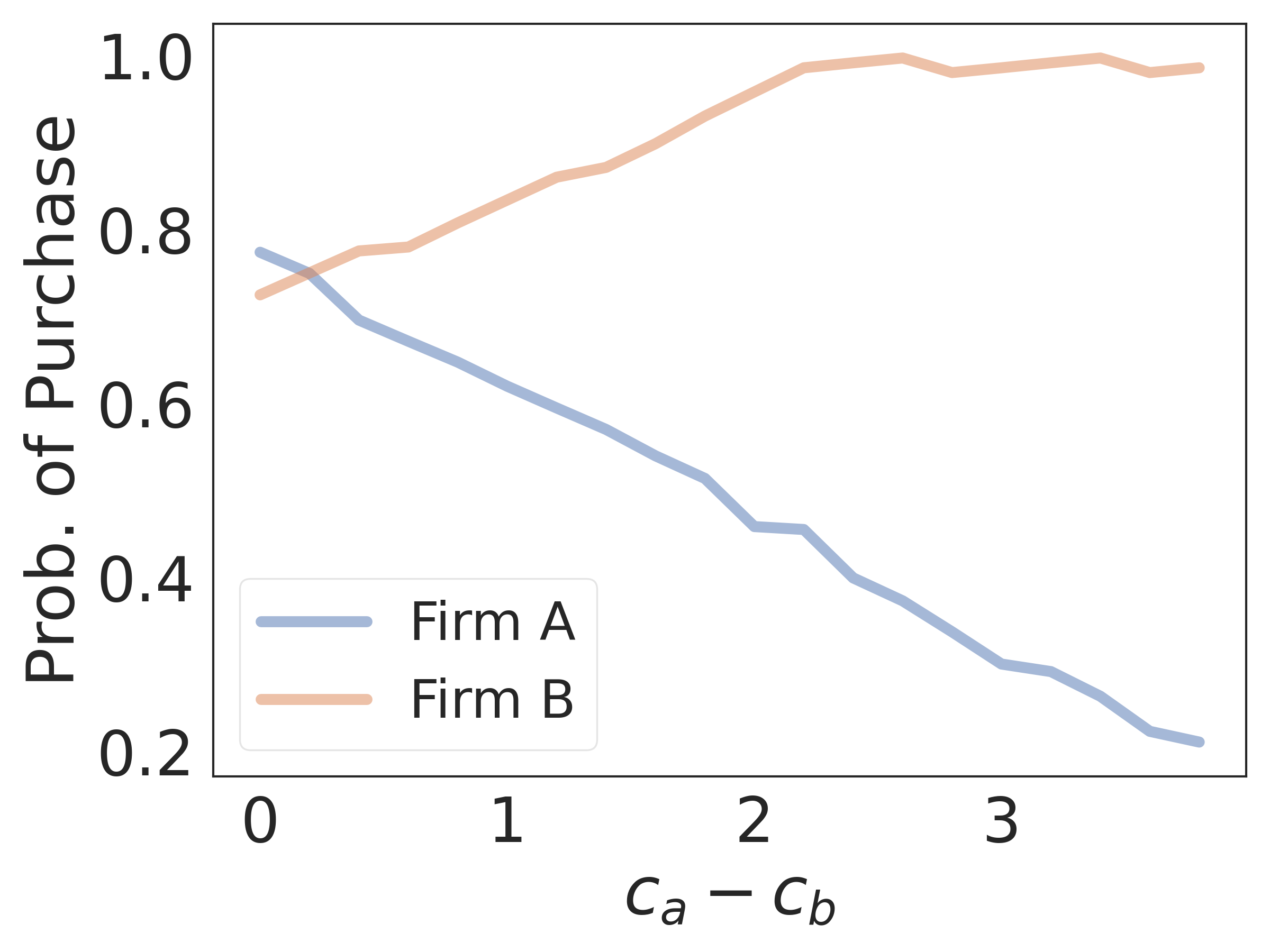}
	\caption{Probability of purchase}
	\label{fig:ll_ih_fm_c_probabilities}
	\end{subfigure}
\caption{Infinite horizon setting market outcomes using numerical simulations under linear loyalty where the constraints can be binding: Optimal prices, market shares, profits of firms, and probability of purchase of customers of types $\alpha$ and $\beta$ as a function of $\ca-\cb$. Here, $\cb=.2, \ta=3, \tb=4, \oa = 1.1, \ob = .5, \df=.4$ (same as Figure~\ref{fig:ll_ih_fm_unc}). In contrast to Figure~\ref{fig:ll_ih_fm_unc}, the equilibrium is computed for cost differences up to $4$ units.}
\label{fig:ll_ih_fm_c}
\end{figure}

When the prices are constrained, the situation changes quite a bit (see Figure~\ref{fig:ll_ih_fm_c}). For example, in Figure~\ref{fig:ll_ih_fm_c_pbb_pab}, we can observe that $\pbb$ converges to $\pab$ in a nonlinear way. Similarly, the rate of change of market share and profit as a function of cost asymmetry is also non-linear (see Figures~\ref{fig:ll_ih_fm_unc_mkt} and~\ref{fig:ll_ih_fm_unc_profit}). 

For both the constrained and the unconstrained instances, we omit the characterization of distinct regions (due to the non-availability of closed-form expressions describing the boundaries as seen in Section~\ref{sec:ll-ss} Figure~\ref{fig:ll-ss-regions}). Nonetheless, the non-linear trends of various market outcomes seen in Figure~\ref{fig:ll_ih_fm_c} provide convincing evidence of the non-trivial impact that cost asymmetry and loyalty can together have.

For the plots in Figure~\ref{fig:ll_ih_fm_c}, numerical computation of equilibria is performed using a dynamic stochastic game solver that uses the Homotopy method (from the numerical analysis literature) discussed in ~\cite{eibelshauser2019dsgamesolver}. The gist of the computational strategy is as follows. Instead of solving for the Markov equilibrium using, say, Kuhn-Tucker conditions, we simplify the underlying non-linear optimality equations using \emph{logit choice}. In particular, we discretize the action space (the price space) and assume that the action probabilities take a logit/softmax form. It turns out that the corresponding logit Markov quantal response equilibrium (QRE) is easier to solve computationally. We solve for QREs at various temperatures (similar to the simulation annealing procedure used for global optimization) that control the steepness of the logit functions. As the temperature parameter approaches infinity, the solution concept approximates the Markov equilibrium. Thus, the Homotopy method links the QRE solutions together and helps us compute the desired Markov equilibrium.

\subsection{Managerial Insights} 

Below, we list important managerial insights that can be gained from our analysis of non-linear market trends in Sections~\ref{sec:ll-ss} and~\ref{sec:ll-ih}.

\begin{enumerate}
\item A high costing firm can not only survive but thrive in an undifferentiated market as long as it cultivates a high degree of customer loyalty. In order to do so, it may have to increase its cost (e.g., due to additional marketing), which can be counter-intuitive at first. More generally, any firm can affect its market position and the overall market outcome by controlling costs or by influencing customer behavior or both, and the choice depends on its existing market position and the sensitivity of the equilibrium. These considerations can be simultaneous to (or follow) the natural tendency of firms to differentiate their offerings from others. 

\item Even if the products are priced the same, and the loyalty levels are the same, one firm can make a significantly different profit from the other based on product costs. But the firm that has a lower cost has much more flexibility in pricing and consequently in controlling the market outcomes as a whole. For instance, a motivated low costing firm can potentially drive its competitor out of the market.

\item While we show that the equilibrium depends on loyalty parameters, the question of how firms learn these loyalty parameters (and not just the membership of each customer in the market) is left open. Depending on the sensitivity of profits to such parameters, which can be derived from our analysis, decision makers can choose to expend resources for learning/data science.

\item Our analytical and computational framework can also accommodate time varying costs (e.g., decreasing costs due to technology improvements) and loyalty levels (e.g., due to delayed marketing effects) while computing market metrics. Being able to navigate the space of market outcomes under such realistic settings can help firms make better/robust short and long term  decisions.

\item Our setup allows for price discrimination beyond the coarse sub-market level analysis that is presented here, and can be re-purposed for individual customer level analysis as shown earlier in this section. This implies that decision makers can use the same computational framework and analyze business outcomes at various granularities, and make further inroads into personalized pricing.

\item The idea that firms can mark up prices for their loyal customers in the infinite horizon setting (see, for instance, Figures~\ref{fig:ll_ih_fm_unc_paa_pba} and~\ref{fig:ll_ih_fm_c_paa_pba}) seems to be at odds with pricing strategies that reward loyalty behavior by providing discounts. On a closer look, this is not the case. In our setting, a customer's loyalty does not increase or decrease due to previous prices (the parameters don't change) beyond determining which firm they will purchase from. Thus, discounting to encourage customers to be more loyal is ineffective (e.g., if for a given price the customer has decided to purchase from the firm, then further discounting does not help). In reality, firms employ both strategies to improve their market positions (e.g., ride-hailing providers Uber and Lyft) because customers exhibit richer behavioral patterns (e.g., they can be susceptible to reference price effects).

\item While in our model, as soon as loyal customers become part of the weak sub-market, the optimal action by the firm is to show a potentially discounted price; other exogenous factors may make this an inefficient strategy. For example, if repeated non-loyalty is strategic (e.g., customers are no longer myopic and are making purchasing decisions based on future pricing decisions of firms) or if it does not allow enough time for firms to re-target (i.e., learn the segment memberships), then our model's recommendations no longer apply. Nonetheless, in several practical applications, firms (e.g., delivery services UberEats and DoorDash and ride-hailing providers Uber and Lyft, among others) have routinely been able to detect repeated non-loyalty and are still able to price discriminate in order to improve their short and long term profits and market shares.
\end{enumerate}
\section{Conclusion}\label{sec:conclusion}

In this work, we show how firms can competitively price discriminate between their loyal customers and those of their rivals in a covered market by designing pricing strategies that explicitly take loyalty and product cost asymmetry into account. This approach vastly differs from extant literature, which has hitherto assumed virtually no product cost difference and a limited model of loyalty, viz., additive switching costs. We analyze market outcomes in both a single stage and an infinite horizon setting (the latter with forward looking firms), showing the non-linear relationship between prices, cost differences, and loyalty level parameters. When feasible, we support our analytical claims with numerical simulations. Our results show that the interplay of product cost difference and loyalty characteristics together can determine a variety of market outcomes at equilibrium. This rich characterization of market outcomes can help firms and decision makers work toward changing their positions for the better, either by cultivating a more loyal customer base or by controlling costs when compared to the competition.

\bibliography{references}
\newpage
\appendix
\section*{Appendix Outline}

The appendix/supplementary consists of the following sections:

\begin{enumerate}
    \item \ref{sec:related-work}: Describes additional related prior works.
    \item \ref{sec:assume-justify}: Provides justification for Assumptions~\ref{assume:customer-type}(i)-(vi).
    \item \ref{sec:myopic}: Analyzes the infinite horizon setting when firms are myopic.
    \item \ref{sec:futurework}: Presents several interesting directions that can be pursued based on the current work.
    \item \ref{sec:ml}: Analyzes market outcomes when the additive component of the linear loyalty model is not present.
    \item \ref{sec:al}: Analyzes market outcomes when the multiplicative component of the linear loyalty model is fixed to be one unit.
    \item \ref{sec:proofs}: Provides details of proofs of claims made in previous sections.
\end{enumerate}
\newpage
\section{Additional Literature Review}\label{sec:related-work}

\cite{dewan2003product} and ~\cite{liu2005imperfect} use Salop's Model~\citep{salop1979monopolistic} to study price discrimination.  In this model, points distributed around a circle represent both products/firms and customers. A customer incurs a transportation cost to acquire a product, which is proportional to the distance between the customer and the product/firm. Differences between transportation costs in Salop's model can be considered as capturing a type of loyalty, as the farther away firm has to match the difference in transportation costs to make a sale. Other works such as~\cite{somaini2013model,rhodes2014re,villas2015short,cabral2016dynamic, MATSUBAYASHI2008571, WANG2017563} define loyalty as switching costs which are additive in nature. Building on these previous foundational works, we work with a fairly general model of loyalty and describe various market structures that arise due to cost differences interacting with the loyalty model. Analogous to our work, the interaction between a type of loyalty and market entry difference (instead of cost difference) between two firms was the focus of~\cite{demirhan2007strategic}, although there the loyalty model considered is still a simplistic additive variant. Similarly, ~\cite{MATSUBAYASHI2008571} studied the interaction between loyalty and product quality and~\cite{WANG2017563} studied the interaction between loyalty and channel distribution structure in a manufacturing setting.

Unlike prior works such as~\citep{farrell2007coordination,villas2015short}, we consider an infinite horizon for both the firms and customers to ensure that there no end-of-game effects. \cite{somaini2013model} consider an infinite horizon oligopolistic dynamic price competition with switching costs (loyalty) and study customer retention and acquisition strategies, proving the existence and uniqueness of a specific Markov equilibrium. Their loyalty model is additive and subsumed by our more general analysis. While they do explicitly take product costs into account, the focus of the paper is not on exhibiting how these costs influence the equilibrium prices in the presence of loyalty. Moreover, the customers are short-lived, living for two time periods. Following this line of work,~\cite{rhodes2014re} also considers a two period setting with a very similar analysis for a duopoly. Unlike both these, we remove explicit product differentiation (Hotelling line) in order to better isolate the nature of loyalty effects on market outcomes. While~\cite{rhodes2014re} wants to answer the question of whether switching costs increase prices, our goals are focused on the impact of costs and loyalty on absolute prices and other market outcomes. Finally, we focus on customer segmentation, through the notions of strong and weak sub-markets, which are markedly absent from these two prior works. 

\cite{cabral2016dynamic} considers a setting similar to ours under an additive switching cost structure and focuses on the impact of switching costs on market outcomes ignoring product cost differences. We expand their analysis to  a more general loyalty setting, while at the same time characterizing additional equilibrium outcomes due to the interplay between product cost differences and loyalty, a phenomenon ignored in their analysis. In~\cite{choe2018pricing}, the authors consider a two-period duopoly setting where firms price in the first period without discrimination (i.e., there are no strong or weak sub-markets a priori unlike our setting) as a result of which market shares are established. In the second period, the firms post both a poaching price for their weak sub-markets as well as a personalized price for their strong sub-markets. Similar to our setting,~\cite{kehoe2018dynamic} consider an infinite horizon duopoly game with multiple products, where two firms interact with a single buyer. Firms price their varieties taking into account the ex-ante probability that the product is desired by the customer and update it using Bayes rule. Firms also choose which variety to offer in each period. Equilibrium conditions that dictate the prices and their effects on the efficiency of the market are considered, and while the decision space is richer, effects of cost asymmetry and loyalty are sidelined.

To summarize, our work uniquely fills the gap of understanding market outcomes in the undifferentiated competitive price discrimination setting, showing that cost asymmetry and loyalty lead to non-linear price, profit and market share trends. Along the way, we provide many motivating examples, analyses, and also draw insights that may be useful to decision makers interested in altering their market positions.
\section{Justifications for Assumptions~\ref{assume:customer-type}(i)-(vi)}\label{sec:assume-justify}

Below, we discuss the Assumptions~\ref{assume:customer-type}(i)-(vi) introduced in Section~\ref{sec:gl}. 

With respect to Assumption~\ref{assume:customer-type}(i), a natural question to ask is whether it is realistic for a firm to offer two different prices for the same product to two different segments of customers. As discussed in Section~\ref{sec:intro}, there are several industries where price discrimination is a prominent and widely employed strategy (both business-to-consumer/B2C and business-to-business/B2B are abound with specific examples of these). 

Further, we assume in Assumption~\ref{assume:customer-type}(ii) that these firms can identify customers in one segment (set $\alpha$) versus customers in the other segment (set $\beta$) in order to price discriminate. This assumption (of which customers belong to which sub-market) can be considered relatively mild because it is reflective of reality and is well supported by literature on price discrimination (see examples and references in Section~\ref{sec:intro}). For instance, it is known that firms routinely use customers' historical purchases and personal preferences to do market segmentation, such as by sending customized coupons and discounts to attract new customers
As a more concrete example, LL Bean (a popular retail company in the US) inserts into their catalogs \emph{special offers} that vary across households
Thus, it is not hard to imagine that firms are able to segment the market into their strong and weak sub-markets. 

Even though a firm may have less information about the loyal customers of its rival, in many cases, it can still obtain enough information from internal and external data sources to predict loyalty function parameters (i.e., parameters of $\ga$ and $\gb$), supporting Assumption~\ref{assume:customer-type}(iii). Note that we do not assume that the firms know the individual customer's idiosyncratic loyalty level, and our assumption is in line with much of the prior work in this area~\citep{shaffer2000pay,somaini2013model}. 

In Assumption~\ref{assume:customer-type}(iv), we assume that all consumers purchase a unit of the good and they exhibit loyalty behavior before the two-player game commences. The question of how firms build customer loyalty itself (e.g., through branding and marketing exercises) and estimate the parameters of the loyalty model is not a primary focus of this paper, and we assume that this capability exists. 

As mentioned in Assumption~\ref{assume:customer-type}(v), we assume that firm $A$ is the higher cost firm, and this is without loss of generality. The Assumption~\ref{assume:customer-type}(vi) that the loyalty functions $\ga$ and $\gb$ are invertible, and their inverse functions are first order differentiable is for notational and analytical convenience. This assumption is also natural, as evidenced by extant literature~\citep{somaini2013model,rhodes2014re,villas2015short,cabral2016dynamic}, and alleviates us from using a more mathematically heavy notation involving pre-images of functions and set based descriptions of stochastic events.

\section{Infinite Horizon and Myopic Firms}\label{sec:myopic}

We start with a generalized analysis of the infinite horizon setting where firms are myopic in ~\ref{subsec:mf}, and then specialize our results to the linear loyalty model in ~\ref{sec:ll-ih-mm}. The multiplicative and additive special cases are considered in their respective sections.

\subsubsection{General Loyalty}\label{subsec:mf}

When firms are myopic, their pricing decisions at time $t$ only depend on the immediate profits and not on the future profits and future state of the world (market share).

\noindent\textit{Dynamics:} In each period, firms offer a pair of prices simultaneously to the market (composed of their strong and weak sub-markets). Customers purchase based on their ex ante memberships in sets $\alpha$ and $\beta$. Based on their purchases, the customers may switch their memberships (e.g., if they purchase from the rival firm). With updated membership information, the game between the two firms is repeated in the subsequent time period ad infinitum.

\noindent\textit{Assumptions, Demand and Firm's Objective:} All the assumptions made in Section~\ref{sec:gl-ss}, namely Assumptions~\ref{assume:customer-type}(i)-(vi) are carried forward here. The same is true for the demand functions and the optimization problems that firms solve in each period. What changes from one period to another is the market share. For instance, assuming that initially, a customer was part of firm $A$'s strong sub-market, her probability of purchasing from $A$ at time period $t=1$ is given by $1-\Fa$ (see Section~\ref{sec:gl-ss}). This probability depends on the prices chosen by the two firms at $t=1$. Depending on the outcome (which is driven by the realization $\xi_{A,1}$), she can either remain in the set $\alpha$ or move to set $\beta$. If she does move to set $\beta$ at the end of $t=1$ by virtue of purchasing from firm $B$, then her probability of purchasing from set $A$ at $t=2$ now changes to $\Fb$, which depends on the prices set by the two firms at $t=2$. Her move to set $\beta$ at the end of $t=1$ implies that the ex ante market shares for $t=2$ have potentially changed.

\noindent\textit{Market Outcomes:} Recall that in the single stage setting (Section~\ref{sec:gl-ss}), the optimal prices of both firms did not depend on the initial market shares in their strong sub-markets ($\theta$ and $1-\theta$, respectively) because the problems in the two sub-markets were separable. This property is carried over to the infinite horizon setting with myopic firms, and hence the prices will remain the same in every period as long as other quantities (such as $\ca,\cb,\ga$ and $\gb$) remain the same. On the other hand, the market shares of both firms change over time, as we alluded to earlier. We will use $\theta_t$ to denote the time varying size/market share of firm $A$'s strong sub-market. Let $\paao, \pabo, \pbbo$, and $\pbao$ denote the time invariant equilibrium prices in any given period (they are invariant to time index $t$). Then the market shares and profit as a function of time $t$ can be characterized using the following lemma.

\begin{lemma}\label{lemma:gl-ih-mm}
Given $\paao,\pabo,\pbbo$ and $\pbao$, if $\Fa+\Fb \in (0,2)$, then the market share of firm $A$ at the end of time period $t$ is given as:
\begin{gather}
\theta_t = \theta(1-\Fa-\Fb)^t + \Fb\frac{1 - (1-\Fa-\Fb)^t}{\Fa + \Fb},
\end{gather}
where $\theta$ is the initial market share at $t=0$, $\xia = h_{\alpha}(\paao-\pbao)$ and $\xib = h_{\alpha}(\pbbo-\pabo)$. The market share of firm $B$ at the end of time period $t$ is simply $1-\theta_t$. Further,
\begin{gather}
\theta_{\infty} = \frac{\Fb}{\Fa + \Fb}.
\end{gather}
\end{lemma}

As shown above, the eventual market shares of firms $A$ and $B$ do not depend on the initial market shares. The size of the set of customers loyal to firm $A$ is proportional to the ratio $ \frac{\Fb}{\Fa + \Fb}$. This ratio is closer to $1$ if $\Fb$ is closer to $1$ and $\Fa$ is closer to $0$. In other words, if the prices $\paao,\pabo,\pbbo$, and $\pbao$ are such that the price $\pbbo$ charged by firm $B$ to its strong sub-market is much larger (i.e., a large premium) compared to the price offered by its rival, it leads to a large value for $\Fb$. At the same time, if the price $\paao$ charged by firm $A$ to its strong sub-market is comparable to the that offered by its competitor, it leads to a small value for $\Fa$. If the firms were symmetric (i.e., $\ca =\cb$) then $\paao = \pbbo$ and $\pabo = \pbao$ and $\theta_{\infty} = 1/2$, giving equal market shares to both firms. The expected profit of each firm at time $t$ is simply the previous period's market share multiplied by the current period's net price. Thus, the profit for firm $A$ at time $t$ is:
\begin{align*}
    \resizebox{.9\textwidth}{!}
     {%
$(\paao - \ca)\theta_{t-1}(1-F(h_{\alpha}(\paao-\pbao))) + (\pab - \ca)(1-\theta_{t-1})F(h_{\beta}(\pbbo-\pabo)),$
}
\end{align*}
where $\theta_t$ is given in Lemma~\ref{lemma:gl-ih-mm}. A similar profit expression can be written down for firm $B$. In summary, the market outcomes in the myopic case largely follow from the outcomes in the single state setting, in contrast to the forward looking setting discussed next.

\subsubsection{Linear Loyalty}\label{sec:ll-ih-mm}

Recall that the optimal prices as derived in Propositions~\ref{prop:ll-ss-r1}-\ref{prop:ll-ss-r6} remain valid in this setting. Using Lemma~\ref{lemma:gl-ih-mm}, and with $\theta$ as the initial market share of firm $A$ at time $t=0$, we can obtain the following expressions for the market shares (and profits can be derived analogously).

\begin{lemma} \label{lemma:ll-ih-mm} With Assumptions~\ref{assume:customer-type} and~\ref{assume:uniform-01}, the market share of firm $A$ at any time index $t$, namely $\theta_t$, under the linear loyalty model is as follows.

\begin{itemize}
\item Region I: $\theta_t = \theta$. Further, $\theta_{\infty} = \theta.$ 
\item Region II: $\theta_t = \theta\left(\frac{2\ta-\ca +\cb +\oa}{3\ta}\right)^t$. Further, $\theta_{\infty} = 0.$ In the ML special case when $\oa=\ob=0$, if $\ca=\cb$, $\theta_t = \theta(\frac{2}{3})^t$ and $\theta_{\infty} = 0$.
\item Region III: $\theta_t = 0$. Further, $\theta_{\infty} = 0.$ 
\item Region IV:
\begin{gather*}
\theta_t = 1-(1-\theta)\left(\frac{\camcb +\ob +2\tb}{3\tb}\right)^t. \textrm{ Further, } \theta_{\infty} = 1.
\end{gather*}
\item Region V:
\begin{align*}
\theta_t &= \theta\left(1/3 - \frac{\ca-\cb-\oa}{3\ta} + \frac{(\ca-\cb)+\ob}{3\tb}\right)^t \\
& + \frac{1-\frac{(\ca-\cb)+\ob}{\lb}}{\frac{\ca-\cb-\oa}{\la}-\frac{(\ca-\cb)+\ob}{\lb}+2}\\
&\quad \times \left(1-\left(1/3 - \frac{\ca-\cb-\oa}{3\ta} + \frac{(\ca-\cb)+\ob}{3\tb}\right)^t\right).
\end{align*}
Further, $\theta_{\infty} = \left(1-\frac{(\ca-\cb)+\ob}{\lb}\right)/\left(\frac{\ca-\cb-\oa}{\la}-\frac{(\ca-\cb)+\ob}{\lb}+2\right).$
There are two special cases: (a) When $\oa=\ob=0$, if $\ca=\cb$, then $\theta_t = \frac{\theta}{3^t} + \frac{1}{2}(1-\frac{1}{3^t})$ and $\theta_{\infty} = \frac{1}{2}$. (b) When $\oa=\ob=0$, if $\ta=\tb$, then $\theta_t = \frac{\theta}{3^t} + \frac{1}{2}(1-\frac{\ca-\cb}{\tb})(1-\frac{1}{3^t})$ and $\theta_{\infty} = \frac{1}{2}(1-\frac{\ca-\cb}{\tb})$.

\item Region VI:
\begin{align*}
\theta_t &= \theta\left(\frac{\camcb-\tb+\ob}{3\tb}\right)^t \\
& \quad+ \frac{\tb-\ob-(\ca-\cb)}{3\tb}\frac{1-\left(\frac{\camcb-\tb+\ob}{3\tb}\right)^t}{1+\frac{\tb-\ob-(\ca-\cb)}{3\tb}}.
\end{align*}
Further, $\theta_{\infty} = \left(\frac{\tb-\ob-(\ca-\cb)}{3\tb}\right)/\left(1+\frac{\tb-\ob-(\ca-\cb)}{3\tb}\right).$

\end{itemize}

The market share of firm $B$ at the end of time period $t$ is simply $1-\theta_t$ (similarly at steady state it is $1-\theta_{\infty}$). 
\end{lemma}

From the above, it is immediately clear that the steady state market shares of the high-cost firm are zero in Regions II and III. In Regions IV, V, and VI, the market share depends on the loyalty parameters and the cost asymmetry. For instance, in Region VI, which corresponds to $\oa+2\ta \leq \camcb \leq \tb-\ob$, the steady state market share of firm $A$ depends on how much smaller the product cost difference is when compared to the term $\tb-\ob$, where $\tb$ and $\ob$ parameterize its weak sub-market. Noticeably, its market share does not depend on the loyalty parameters of its loyal customers.

\section{Future Directions}\label{sec:futurework}

We have investigated the combined effects of cost asymmetry and loyalty for a fairly limited collection of market settings. First, we have assumed that both firms can identify customers with perfect accuracy. In practice, different firms have different insights about individual customer preferences due to the varying degree of customer data that is available to them. Despite continuous improvement in data collection and advances in information technology, firms mis-classify customers routinely. A firm with less customer information is more likely to classify customers erroneously. An incorrectly classified customer may not purchase the product as anticipated. One can incorporate classification errors while learning customer preferences in a suitable competitive market setup and study their implications. A significant direction of improvement along similar lines would be to consider an evolving learning process implemented by each firm and interleave it with firm decision making in each period.

Second, the analytical results presented here are restricted to affine loyalty functions. This can potentially be extended to include a non-loyal customer base (these always purchase the lowest priced product) and/or a collection of customers who can be segmented into weakly loyal, strongly loyal, and moderately loyal customers. This refined segmentation can capture markets that tend to be more volatile (e.g., certain consumer packaged goods), markets that are brand driven (e.g., airlines and products with firms such as Apple or Microsoft competing), as well as markets where there is almost no loyalty. It is interesting to analyze how firms \emph{harvest} their strongly loyal customers, \emph{pay to stay} the moderately loyal customers, and \emph{pay to switch} the weakly loyal customers of the rival firm, while still being able to profit from the non-loyal customer base. In practice, firms (such as Uber and Lyft for ride-hailing) use discounts to enroll new customers as well as retain existing customers (rewarding them for loyalty). Studying models that induce both these pricing outcomes would be interesting to pursue.

Lastly, a few further extensions of our analysis that are promising are as follows: (1) While we did not consider forward looking customers in this work, one can rely on prior works to build suitable extended games and study the interplay of costs and loyalty. (2) While we considered the parametric impact of a specific behavior (loyalty) on prices, one could attempt to combine multiple other behavioral factors that  have been considered in competitive pricing settings.
Combining these effects in a multidimensional parametric setting where prices depend on the entire customer's profile would improve the fidelity of the inferences drawn. (5) Capturing mixed-memberships of customer loyalties and their evolution over time as well as them changing due to network effects would be interesting. Similarly, costs can also evolve due to changes in the underlying technologies, marketing efforts, etc. (4) Relaxing the assumption that the market is covered and accounting for the fact that more than one firm (with the possibility of collusion) can have varied state dependent risk tolerances/discount factors can make the modeling more realistic and the conclusions more actionable.
\section{Multiplicative Special Case of Linear Loyalty: Analysis and Insights}\label{sec:ml}

In the multiplicative special case (ML) of the linear loyalty model, we assume that the loyalty function has the following form: $\ga(\xi) = \ta\xi$ (its inverse is given by $h_{\alpha}(y) = y/\ta$) where parameter $\ta \in \mathbb{R}_{+}$. Thus, given prices $\paa$ and $\pba$, the probability of a customer belonging to the set $\alpha$ purchasing from firm $A$ is $1 - \Fa$, where $\xi^{\alpha} = (\paa - \pba)/\ta$.  When the support of random variable $\xi$ is restricted to $[0,1]$, the loyalty model provides two insights. First, the parameter $\ta$ (as well as $\tb$) can be interpreted as the maximum loyalty level a customer can have. Second, it also suggests the following constraint on pricing: if a customer is offered a higher price by non-preferred firm, then she only purchases from its loyal firm. On the other hand, if she is offered a very high price by her preferred firm relative to the non-preferred firm (normalized by her maximum loyalty level), then she will definitely not purchase from her loyal firm. Since customers have varying degrees of loyalty levels, very loyal customers will tolerate higher premiums. 

\subsection{Single Stage Setting}\label{sec:ml-ss}

Recall that in this setting, the firms compete only once. The demand functions for the strong and weak sub-markets of firm $A$ under the ML special case are as follows:
\begin{gather*}
D_{A}^{ss}(\paa, \pba) = \theta\left(1-F\left(\frac{ \paa - \pba}{\ta}\right)\right), \text{ and}\\
D_{A}^{ws}(\pab, \pbb) = (1-\theta)F\left(\frac{ \pbb - \pab}{\tb}\right).
\end{gather*}

\begin{figure}[ht]
\centering
\includegraphics[width=0.45\textwidth]{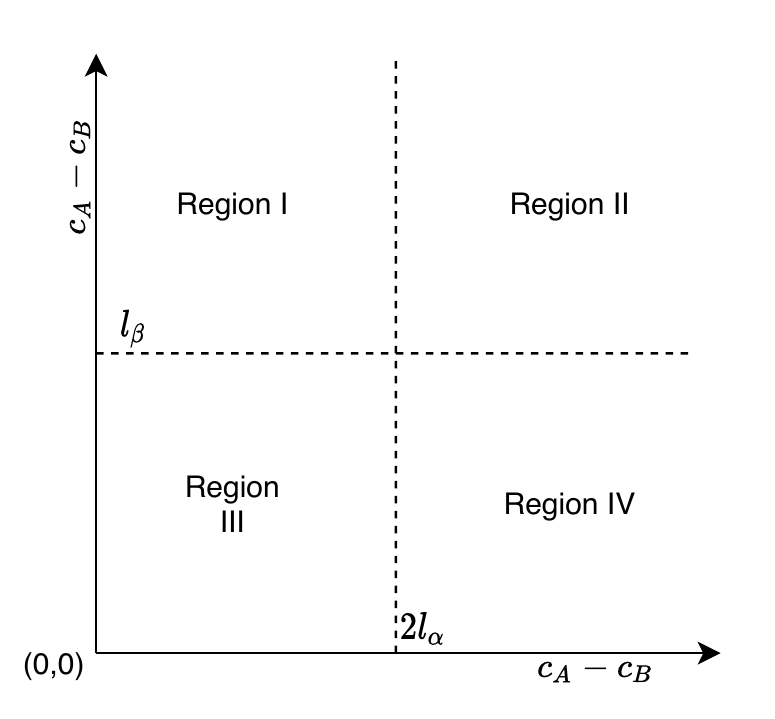}
\caption{Four regions that determine different equilibrium prices depending on the relationship between loyalty levels and product costs in the single stage ML setting.\label{fig:ml-ss-regions}}
\end{figure}

Under the choices made for $F$, $\ga$ and $\gb$ above, our analysis reveals four distinct price discrimination classes based on the interplay of maximum loyalty levels ($\ta$ and $\tb$) and the magnitude of product cost difference ($\ca-\cb$).  Equilibrium conditions are determined for each of the following sub-cases in Propositions~\ref{prop:ml-ss-r1}-\ref{prop:ml-ss-r4} below, which are mutually exclusive and exhaustive (see Figure~\ref{fig:ml-ss-regions}): 
\begin{itemize}
\item Region I: $\tb \leq \camcb \leq 2\ta$ (see Proposition~\ref{prop:ml-ss-r1}).
\item Region II: $\camcb \geq \max\{2\ta,\tb\}$ (see Proposition~\ref{prop:ml-ss-r2}).
\item Region III: $\camcb \leq \min\{2\ta,\tb\}$ (see Proposition~\ref{prop:ml-ss-r3}).
\item Region IV: $2\ta \leq \camcb \leq \tb$ (see Proposition~\ref{prop:ml-ss-r4}).
\end{itemize}

\begin{proposition}
\label{prop:ml-ss-r1}
Under Region I, with Assumptions~\ref{assume:customer-type} and~\ref{assume:uniform-01}, the unique pure Nash equilibrium prices for the strong and weak sub-markets for firms $A$ and $B$ for the ML special case are as follows:
\begin{gather}
\paa = \frac{1}{3}\left(2\ca  +\cb +2\ta \right),\;
\pab = \ca,\;\nonumber\\
\pbb = \ca, \text{and }\;
\pba = \frac{1}{3}\left(2\cb  +\ca +\ta \right).
\end{gather}
\end{proposition}

\begin{proposition}
\label{prop:ml-ss-r2}
Under Region II, with Assumptions~\ref{assume:customer-type} and~\ref{assume:uniform-01}, the unique pure Nash equilibrium prices for the strong and weak sub-markets for firms $A$ and $B$ for the ML special case are as follows:
\begin{gather}
\paa = \ca,\;
\pab = \ca,\;
\pbb = \ca, \text{and }
\pba = \ca-\ta.
\end{gather}
\end{proposition}

\begin{proposition}
\label{prop:ml-ss-r3}
Under Region III, with Assumptions~\ref{assume:customer-type} and~\ref{assume:uniform-01}, the unique pure Nash equilibrium prices for the strong and weak sub-markets for firms $A$ and $B$ for the ML special case are as follows:
\begin{gather}
\paa = \frac{1}{3}\left(2\ca  +\cb +2\ta \right),\;
\pab = \frac{1}{3}\left(2\ca  +\cb +\tb \right),\nonumber\\
\pbb = \frac{1}{3}\left(2\cb  +\ca +2\tb \right), \text{and }
\pba = \frac{1}{3}\left(2\cb  +\ca +\ta \right).
\end{gather}
\end{proposition}

\begin{proposition}
\label{prop:ml-ss-r4}
Under Region IV, with Assumptions~\ref{assume:customer-type} and~\ref{assume:uniform-01}, the unique pure Nash equilibrium prices for the strong and weak sub-markets for firms $A$ and $B$ for the ML special case are as follows:
\begin{gather}
\paa = \ca,\;
\pab = \frac{1}{3}\left(2\ca  +\cb +\tb \right),\;\nonumber\\
\pbb = \frac{1}{3}\left(2\cb  +\ca +2\tb \right),\; 
\pba = \ca - \ta.
\end{gather}
\end{proposition}

What is the impact of product costs and loyalty levels on market equilibrium?  The results above show that the game is in equilibrium regardless of the product cost difference and the loyalty model parameters. Figure~\ref{fig:ml-ss-regions} shows how the four regions related to each other. Both x- and y-axes represent $\camcb$, the horizontal dashed line corresponds to cases where $\tb = \camcb$, and the vertical dashed line is for the case $2\ta = \camcb$.  The figure captures the entire space of possible combinations of product cost differences divided into regions based on their relationship with the loyalty model parameters. The interplay of these two aspects (cost asymmetry and loyalty) together determines market equilibrium conditions, and we elaborate on them below.

\subsubsection{Insights}

Our results above illustrate that ignoring product cost in competitive price discrimination studies leads to disregarding a large number of realistic competitive price discrimination market equilibria.  Previous studies dealt only with the case represented by the origin in this graph, i.e., the case where product cost difference ($\camcb$) is zero (see Region III). Region III represents the cases where product cost difference is small compared to the maximum loyalty levels (recall that this interpretation is true when $F\sim U[0,1]$) to have any  significant impact on the market structure. The two firms are able to sell to each other's strong and weak sub-markets. 
To the best of our knowledge, many of the previous competitive duopolistic price discrimination studies -- where product cost is either ignored or the difference in product costs is negligible – can be grouped within this class. 

We cannot expect the same outcome for the other regions. In Region I, the high-cost firm is able to sell its product to some of its loyal following in its strong sub-market, but it is unable to make any inroads into its weak sub-market.The low-cost firm is able to prevent penetration of its rival into its strong sub-market.  In Region IV, the high-cost firm is able to sell into its weak sub-market, but, it cannot do the same for its own loyal customers in its strong sub-market. Region II represents the scenario where the low-cost firm drives its rival out of business.  

As mentioned earlier, the well-established approaches in competitive price discrimination literature, i.e., those assuming product cost as either negligible or equal for all firms, rule out the cases in Regions I, II or IV, and yet their market equilibrium conditions are significantly different in price, market share and profitability than in Region III, as seen above.  Thus, Regions I, II and IV, which together represent a large class of competitive price discrimination market equilibrium cases, allow for a clearer understanding of the impact of costs and the assumed loyalty model on the market structure.\\

\begin{figure}
\centering
	\begin{subfigure}{0.3\textwidth}
	\includegraphics[width=0.9\linewidth]{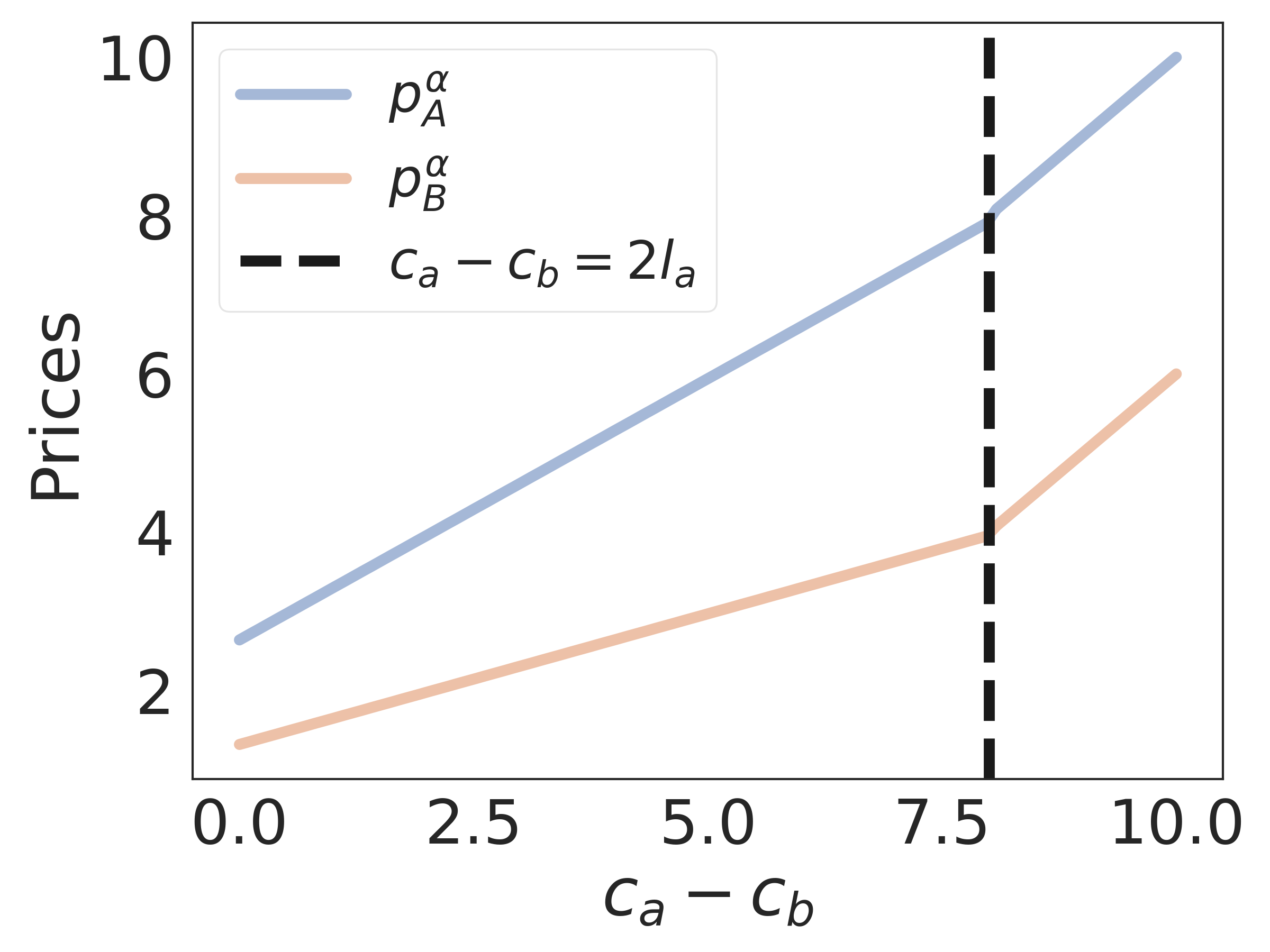} 
	\caption{Prices seen by firm $A$'s strong sub-market}
	\label{fig:ml_single_stage_paa_pba}
	\end{subfigure}
	\begin{subfigure}{0.3\textwidth}
	\includegraphics[width=0.9\linewidth]{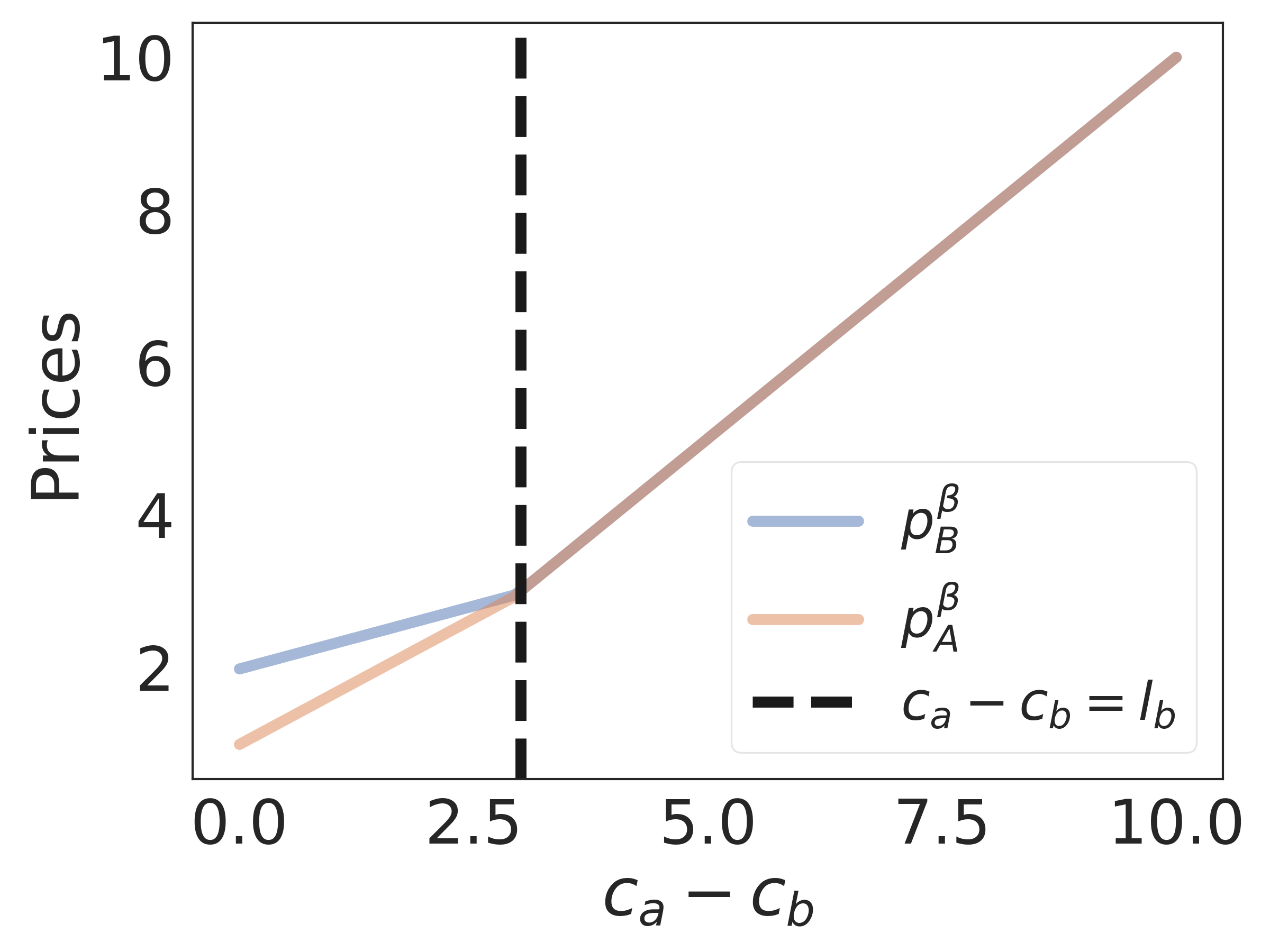}
	\caption{Prices seen by firm $A$'s weak sub-market}
	\label{fig:ml_single_stage_pbb_pab}
	\end{subfigure}
	\begin{subfigure}{0.3\textwidth}
	\includegraphics[width=0.9\linewidth]{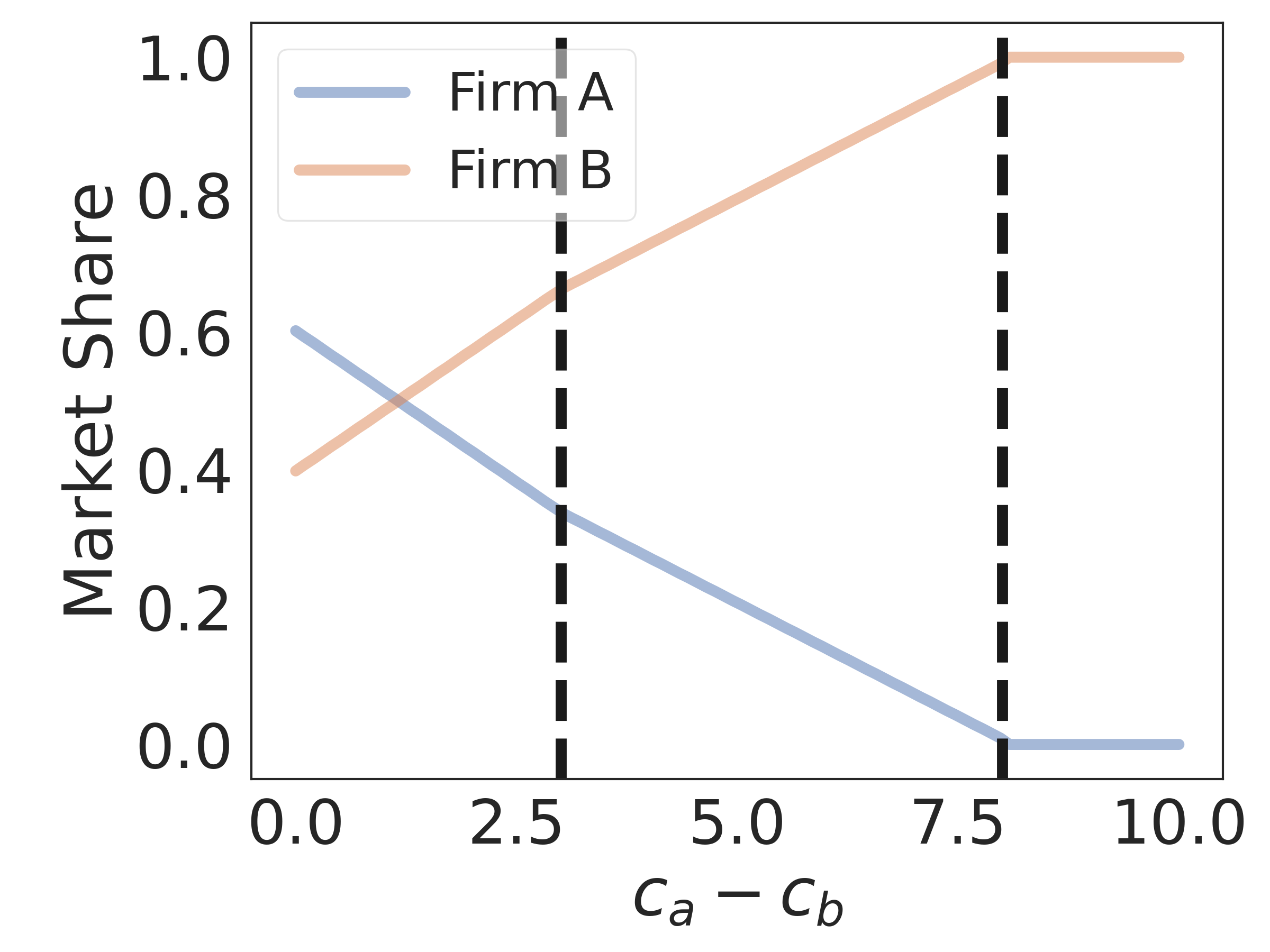} 
	\caption{Market share}
	\label{fig:ml_single_stage_market_share}
	\end{subfigure}
	\begin{subfigure}{0.3\textwidth}
	\includegraphics[width=0.9\linewidth]{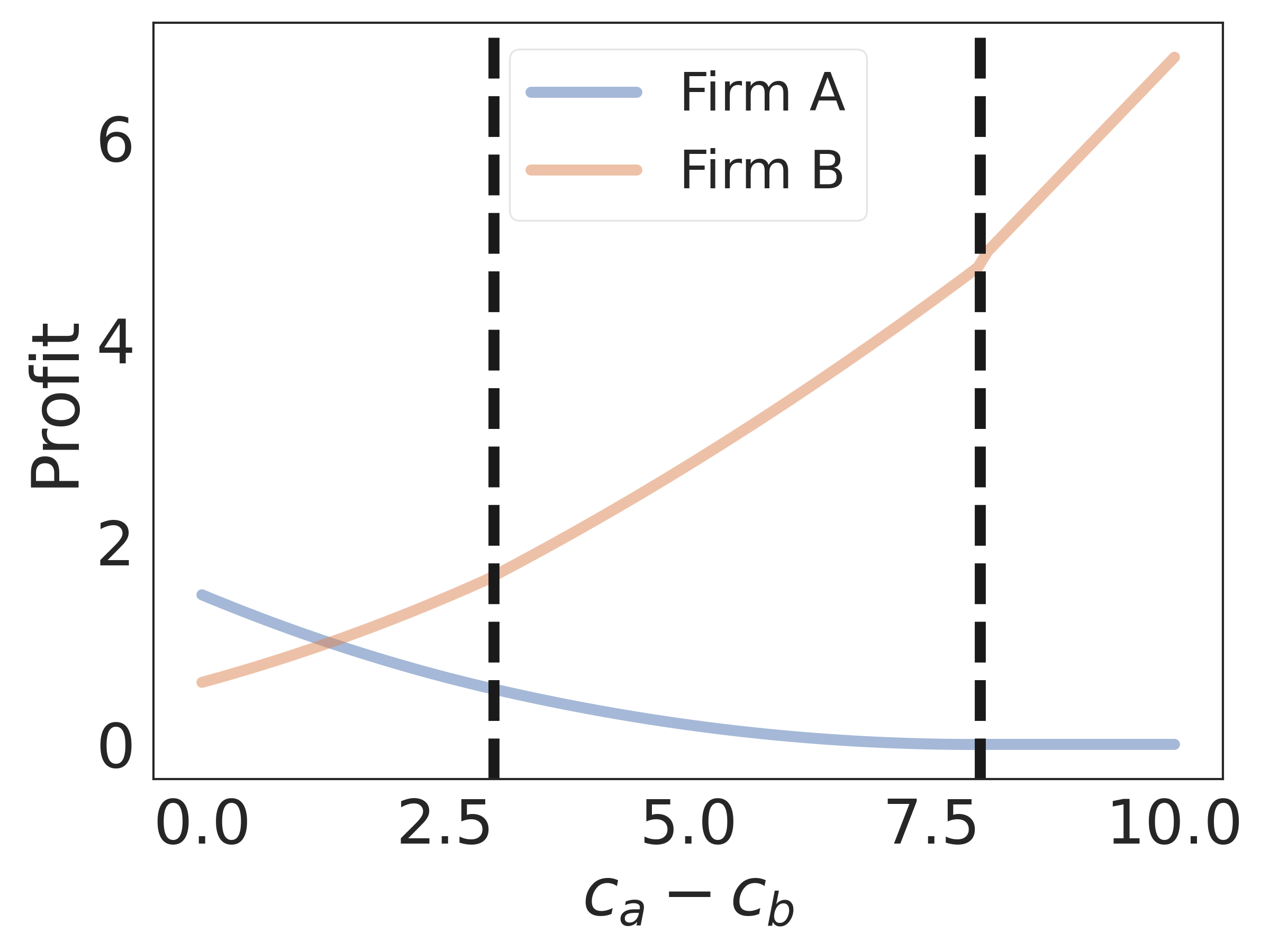}
	\caption{Profit}
	\label{fig:ml_single_stage_profits}
	\end{subfigure}
	\begin{subfigure}{0.3\textwidth}
	\includegraphics[width=0.9\linewidth]{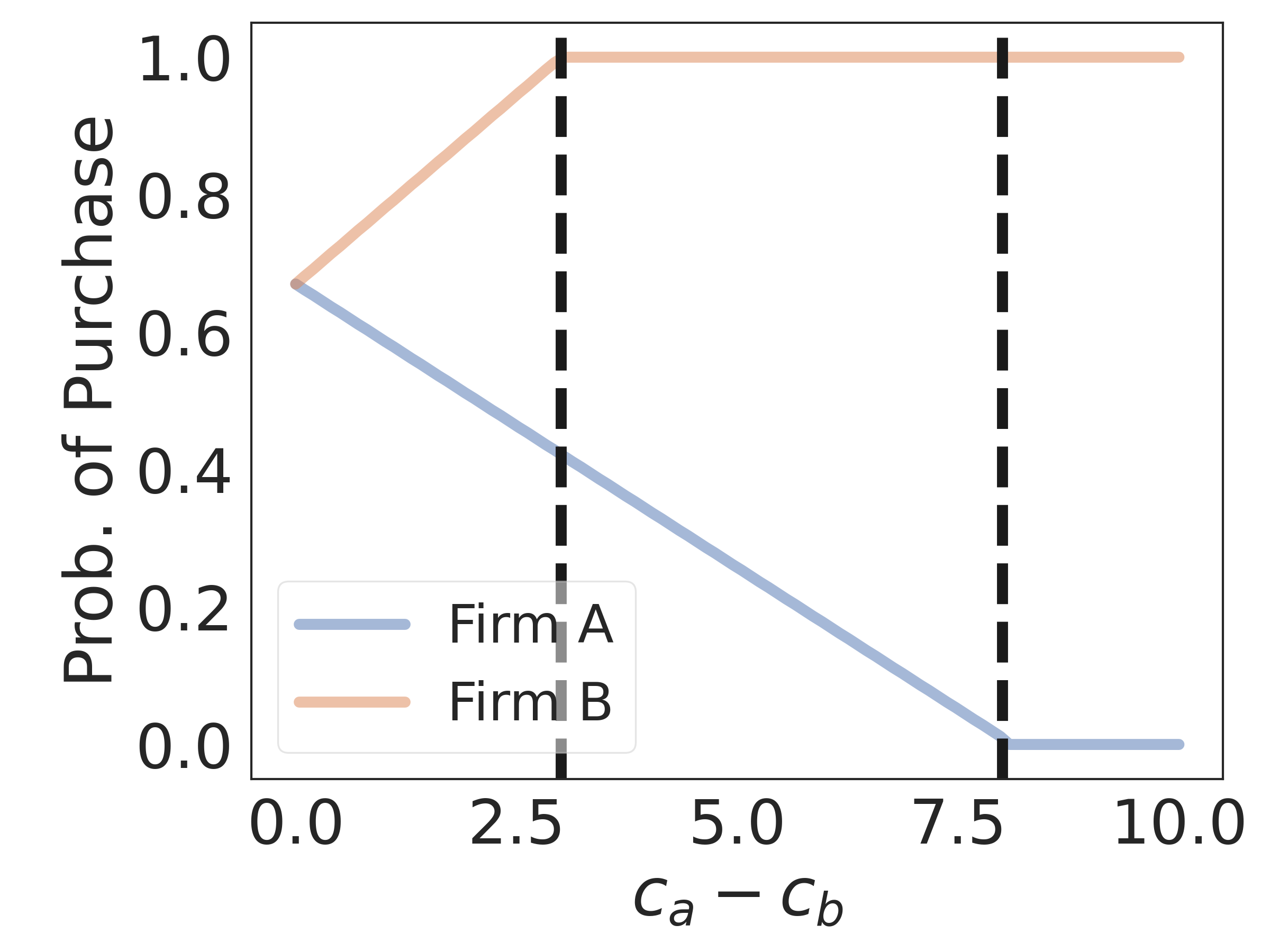}
	\caption{Probability of purchase}
	\label{fig:ml_single_stage_probabilities}
	\end{subfigure}
\caption{Single stage market outcomes under multiplicative loyalty: Optimal prices, market shares, profits of firms, and probability of purchase of customers of types $\alpha$ and $\beta$ as a function of $\ca-\cb$. Here, $\cb=0, \ta=4, \tb=3$ and $\theta=0.8$.}
\label{fig:ml-ss-market-outcomes}
\end{figure}

\noindent\textit{High-Cost Firm's Strong Sub-market}: We now discuss how firm prices and market shares change with product costs and loyalty, focusing on the high-cost firm's strong sub-market. As this is the strong market for the high-cost firm ($A$), it constitutes a loyal following that can pay a premium for its product. However, the high-cost firm is disadvantaged due to its high product cost. Even though the high-cost firm is the preferred firm for the customers in this set, it allows its rival to make inroads into its loyal customer base. This is not solely due to its disadvantage in product costs. Even for the case where both firms have exactly the same costs ($\ca=\cb=0$), firm $A$ cannot undercut firm $B$ to retain all of its loyal customers. Firm $A$ maximizes its profit by charging a higher premium to its loyal customers. Therefore, some of its least loyal customers cannot tolerate the premium and switch to the rival firm.  This scenario shows a natural trade-off between market share and prices: the premium charged by a firm may be increased only to the point where loss in market share starts having a negative impact on overall profitability. 

Figure~\ref{fig:ml_single_stage_paa_pba} illustrates firm prices with varying degrees of product cost difference.  While x-axis represent $\camcb$, y-axis is prices. The origin corresponds to studies assuming product cost as either negligible or equal for all firms. The solid lines show firm prices. The dashed vertical line represents the point where the cost difference is equal to  twice the loyalty parameter $\ta$. When the cost difference is less than $2\ta$, firms A and B charge $\frac{1}{3}\left(2\ca  +\cb +2\ta \right)$ and $\frac{1}{3}\left(2\cb  +\ca +\ta \right)$ respectively. Firm A, which is the high-cost firm, is also a high-priced firm in this sub-market. Regions I and III in Figure~\ref{fig:ml-ss-regions} correspond to the case where $\camcb < 2\ta$. Here, high-cost firm is able to charge a premium to some of its loyal customers. An increase in high-cost firm's product cost will cause both firms to raise their prices. For a  $\delta$ increase in firm $A$'s product cost, high-cost firm (firm $A$) increases its prices by $\frac{2}{3}\delta$. On the other hand, the low-cost firm (firm $B$) increments its price by $\frac{1}{3}\delta$. Thus, a change in product cost affects the prices differently, with high-cost firm increasing its price, lowering its profit margin and reducing market share. As firm $A$'s product cost keeps going up, its price will increase; however, the change in price does not match the increases in its costs. Therefore, the high-cost firm's profit margin slowly diminishes. At the same time, its market share decreases as its rival doesn't increase its prices as much as the high-cost firm does.  Therefore, more customers will switch to the low-cost firm. Thus, the low-cost firm improves its price, profit margin and market share. When product cost difference reaches $2\ta$, the high-cost firm is driven out of business in its strong sub-market. Even though, customers prefer its products, the high-cost firm cannot offer affordable prices anymore. When the cost difference is equal or greater than $2\ta$, firms $A$ and $B$ charge $\ca$  and $\ca-\ta$ respectively. Regions II and IV in Figure~\ref{fig:ml-ss-regions} correspond to the case where $\camcb \geq 2\ta$. The low-cost firm undercuts its rival and captures all of the customers in the market. As the product cost difference is considerably high, firm $A$ cannot protect its loyal customer base.  

There are three important implications for high cost firm's strong sub-market. First, the low-cost firm does not capture all of its rival's loyal customers when product cost difference reaches $\ta$. Why would the low-cost firm not undercut its rival and capture all of its rival's loyal customer when it has the ability to do it? After all, high-cost firm can no longer block its rival when product cost difference reaches $\ta$. And the answer is that both firms are mindful of the trade-off between price and market share. The low-cost firm can undercut its rival to increase its market share; however, the discounts offered are only to the point where the decrease in product price along with increase in market share starts having a negative impact on overall profitability.  Second, a change in product cost does not translate into a perfectly correlated adjustment of the prices. Recall that the high-cost firm increases its prices by $\frac{2}{3}\delta$ for a $\delta$ increase in its product cost. In other words, firm $A$ absorbs some of the increases in product cost.  Third, and most importantly, customer loyalty, which is captured by parameter $\ta$, is extremely vital for the survival of high-cost firm. As the loyalty level goes up, the vertical line where high-cost firm is driven out of business starts to move to the right. High-cost firm is able to tolerate greater difference in product costs. \\

\noindent\textit{High-Cost Firm's Weak Sub-market}:  We next look at the high-cost firm's weak sub-market. As this is the weak market for high-cost firm, customers prefer its rival's products. Moreover, high-cost firm is also disadvantaged due to its high product cost. 
Figure~\ref{fig:ml_single_stage_pbb_pab} illustrates firm prices with varying degrees of product cost difference.  While x-axis represents $\camcb$, y-axis is prices. Similar to Figure~\ref{fig:ml_single_stage_paa_pba}, the origin corresponds to previous studies where product cost as either negligible or equal for all firms. The solid lines again show firms' prices. The dashed vertical line is $\camcb = \tb$. The dashed vertical line in Figure~\ref{fig:ml_single_stage_pbb_pab} is the same dashed horizontal line in Figure~\ref{fig:ml-ss-regions}. Firm $B$, which is the low-cost firm, is the high-price firm in this sub-market.  When the cost difference is less than $\tb$, firms $A$ and $B$ charge $\frac{1}{3}\left(2\ca  +\cb +\tb \right)$ and $\frac{1}{3}\left(2\cb  +\ca +2\tb \right)$ respectively. 

Regions III and IV in Figure~\ref{fig:ml-ss-regions} correspond to the case where $\camcb < \tb$. The low-cost firm is able to charge a premium to some of its loyal customers. An increase in high-cost firm's product cost will cause both firms to raise their prices. Similar to the previous case, high-cost firm increases its prices by $\frac{2}{3}\delta$ for a $\delta$ increase in its product cost. On the other hand, the low-cost firm increases its price by $\frac{1}{3}\delta$. The change in high-cost firm's product cost also plays a similar role as it did in high-cost firm's strong sub-market. High-cost firm's price will go up; however, the change in price does not match the increase in its costs. Its profit margin slowly diminishes. At the same time, its market share decreases as its rival doesn't increase its prices as much as the high-cost firm does.  Therefore, more customers will switch to the low-cost firm, which improves its profit margin and market share at the same time. When product cost difference reaches $\tb$, high-cost firm is driven out of business in its weak sub-market. As the low-cost firm is customers' preferred brand in this segment and has a cost advantage, it can retain all of its loyal following by matching the high-cost firm's price. 

When the cost difference is equal or greater than $\tb$, firms $A$ and $B$ charge the same price $\ca$. Regions I and II in Figure~\ref{fig:ml-ss-regions} correspond to the case where $\camcb \geq \tb$.  The low-cost firm does not have to undercut its rival to retain its loyal customers in this case, as all of the customers in this segment already prefer the low-cost firm. As the product cost difference is considerably high, the low-cost firm is able to protect its loyal customer base. There is one important implication for the high-cost firm's weak sub-market. First, even though the low-cost firm is the preferred firm for the customers in this segment, it allows its rival to make inroads into its loyal customer base. Why would the low-cost firm allow its rival to make inroads into its loyal customer base? After all, as the preferred brand for the customers in this segment, the low-cost firm has the flexibility to offer a price that lets it retain all of its loyal customers. Yet, our results show that the low-cost firm is willing to lose some of its loyal customers. In other words, the low-cost firm can increase its profit simply by charging a higher premium to its loyal customers. As a consequence, some of its least loyal customers may find the premium high enough that they switch to the high-cost firm. This is again due to the trade-off between market share and prices.

\subsection{Infinite Horizon Setting}\label{sec:ml-ih}

As in Section~\ref{sec:gl-ih}, we first give results for the setting where $\da=\db=0$, and then characterize equilibrium prices in the unconstrained setting when  $\da=\db (=\df \textrm{, a common discount value}) >0$.

\subsubsection{Myopic Firms}

As before, let $\theta$ be the initial market share at $t=0$. Recall that the optimal prices as derived in Propositions~\ref{prop:ml-ss-r1}-\ref{prop:ml-ss-r4} remain valid in this setting. Thus, following Lemma~\ref{lemma:gl-ih-mm}, we can derive market shares and profits under the ML special case as shown below. 

\begin{lemma} \label{lemma:ml-ih-mm} With Assumptions~\ref{assume:customer-type} and~\ref{assume:uniform-01}, the market share of firm $A$ at any time index $t$, namely $\theta_t$, under the multiplicative loyalty special case is given as $ \theta_t = \theta\eta_1^t + \eta_2(1 - \eta_1^t)$, where $\eta_1$ and $\eta_2$ are defined as follows.

\begin{itemize}
\item Region I: $
\eta_1 = \frac{2}{3} -\frac{\camcb}{3\ta}, \textrm{ and }\; \eta_2 = 0. \textrm{ Further, } \theta_{\infty} = 0$. In the special case when $\ca=\cb$, $\theta_t = \theta(\frac{2}{3})^t$ and $\theta_{\infty} = 0$. 
\item Region II: $\eta_1 = 0$, and $\eta_2 = 0$. Further, $\theta_{\infty} = 0$.
\item Region III:
\begin{gather*}
\eta_1 = \frac{1}{3} + (\frac{1}{3\tb} - \frac{1}{3\ta})(\camcb), \textrm{ and }\; \eta_2 = \frac{(\tb-(\camcb))}{3\tb(1-\eta_1)}.
\end{gather*}
Further, $\theta_{\infty} = \eta_2$. There are two special cases: (a) when $\ca=\cb$, $\theta_t = \frac{\theta}{3^t} + \frac{1}{2}(1-\frac{1}{3^t})$ and $\theta_{\infty} = \frac{1}{2}$; and (b) when $\ta=\tb$, $\theta_t = \frac{\theta}{3^t} + \frac{1}{2}(1-\frac{\ca-\cb}{\tb})(1-\frac{1}{3^t})$ and $\theta_{\infty} = \frac{1}{2}(1-\frac{\ca-\cb}{\tb})$.

\item Region IV:
\begin{gather*}
\eta_1 = \frac{\camcb - \tb}{3\tb}, \textrm{ and }\; \eta_2 = \frac{\tb-(\camcb)}{4\tb-(\camcb)}. \textrm{ Further, } \theta_{\infty} = \eta_2.
\end{gather*}
In the special case when $\ca=\cb$, $\theta_t = \theta(\frac{-1}{3})^t + \frac{1}{4}(1-(\frac{-1}{3})^t) $ and $\theta_{\infty} = \frac{1}{4}$. 
\end{itemize}

The market share of firm $B$ at the end of time period $t$ is simply $1-\theta_t$ (similarly at steady state it is $1-\theta_{\infty}$). 
\end{lemma}

From the above, it is immediately clear that the steady state market shares of the high-cost firm are zero in Regions I and II. In Regions III and IV, the market share depends on the loyalty parameters and the cost asymmetry. For instance, in Region IV, which corresponds to $2\ta \leq \camcb \leq \tb$, the steady state market share of firm $A$ depends on how much smaller the product cost difference is when compared to the loyalty level of its weak sub-market. 
Noticeably, its market share does not depend on the maximum loyalty level of its loyal customers.

\subsubsection{Forward Looking Firms}\label{sec:ml-ih-fm}

It is harder to characterize succinctly the equilibrium conditions for the infinite horizon in general. Below, we ignore constraints on prices (as discussed in Section~\ref{sec:gl-ih}) and show that in this case, it is indeed possible to achieve a unique Markov equilibrium. Further, the result is obtained for any distribution function $F$ that satisfies Assumption~\ref{assume:cdf}.

\begin{proposition} For the multiplicative loyalty special case, under Assumptions~\ref{assume:customer-type},~\ref{assume:gl-ih-unconstrained} and ~\ref{assume:cdf}, there exists a unique Markov equilibrium when $\da=\db=\df > 0$, where firms price based on whether the customer bought their product in the immediate preceding time period. This equilibrium is characterized by the following fixed point equations for thresholds $\xia$ and $\xib$:
\begin{align*}
\left(\xia - \frac{\camcb}{\ta}\right)&\left(\frac{1-\df}{\df} +\Fb+1\right)\\
& + \frac{2\Fa-1}{\fa}\left(\frac{1-\df}{\df} +\Fa+\Fb\right) +  \frac{\Fa}{\fa} \\
& = \frac{(1-\Fb)\tb}{\fb\ta} - \Fb\left(\frac{\tb}{\ta}\xib +\frac{\camcb}{\ta}\right),
\end{align*}
and
\begin{align*}
\left(\xib - \frac{\cb-\ca}{\tb}\right)&\left(\frac{1-\df}{\df} +\Fa+1\right)\\
& +  \frac{2\Fb-1}{\fb}\left(\frac{1-\df}{\df} +\Fa+\Fb\right) +  \frac{\Fb}{\fb} \\
& = \frac{(1-\Fa)\ta}{\fa\tb} - \Fa\left(\frac{\ta}{\tb}\xia +\frac{\cb-\ca}{\tb}\right).
\end{align*}

\label{prop:ml-ih-fm}
\end{proposition}
The above thresholds can be used in conjunction with Equations~\ref{eqn:paa}-\ref{eqn:pba} to obtain the optimal prices, profits and resulting market shares. Because the thresholds are implicitly defined, in the following, we numerically solve for them for a canonical market instance in order to obtain the dependence of key market metrics on cost asymmetry and loyalty parameters.

\subsubsection{Insights}

\begin{figure}
\centering
	\begin{subfigure}{0.3\textwidth}
	\includegraphics[width=0.9\linewidth]{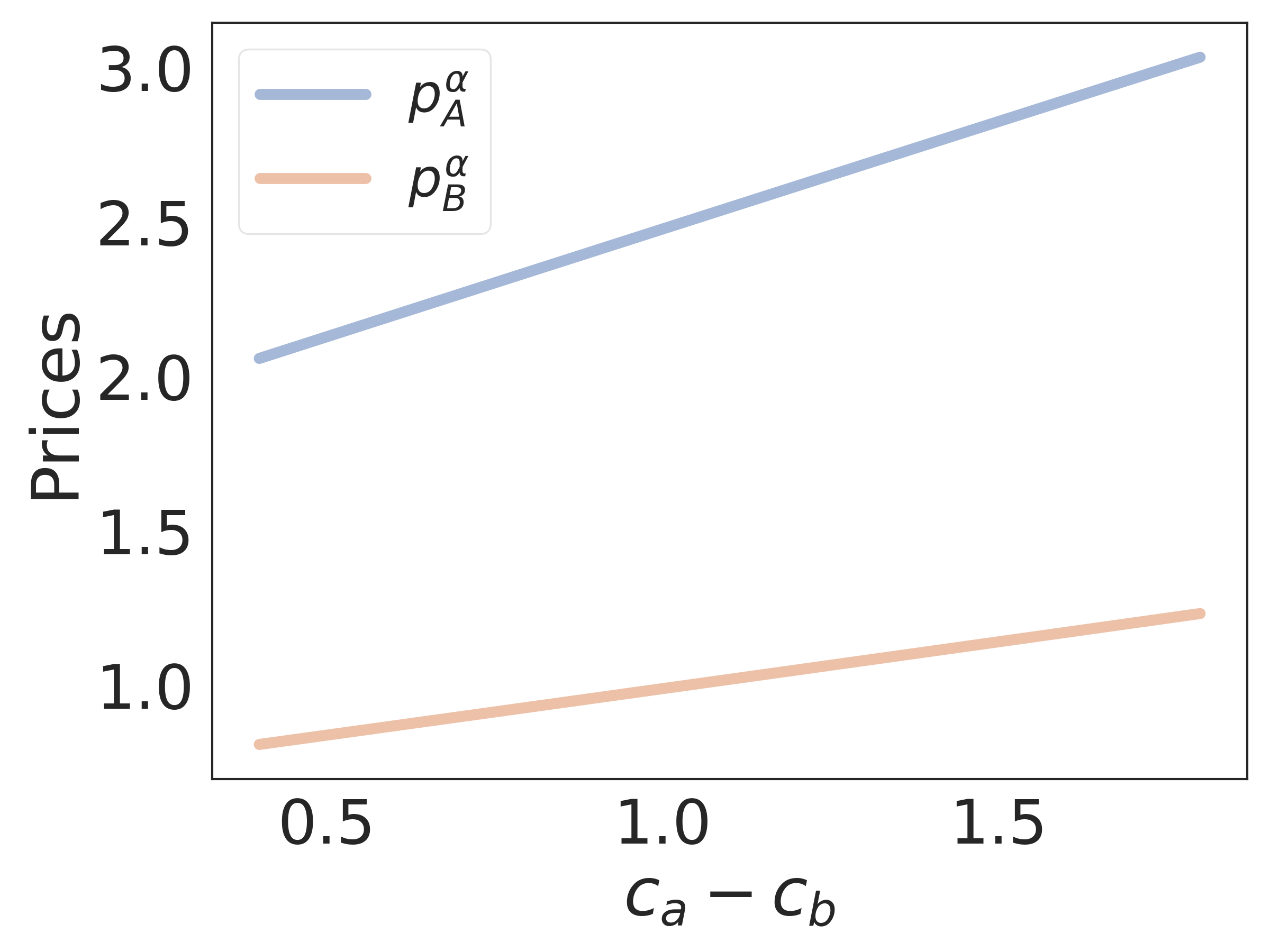} 
	\caption{Prices seen by firm $A$'s strong sub-market}
	\label{fig:ml_ih_fm_unc_paa_pba}
	\end{subfigure}
	\begin{subfigure}{0.3\textwidth}
	\includegraphics[width=0.9\linewidth]{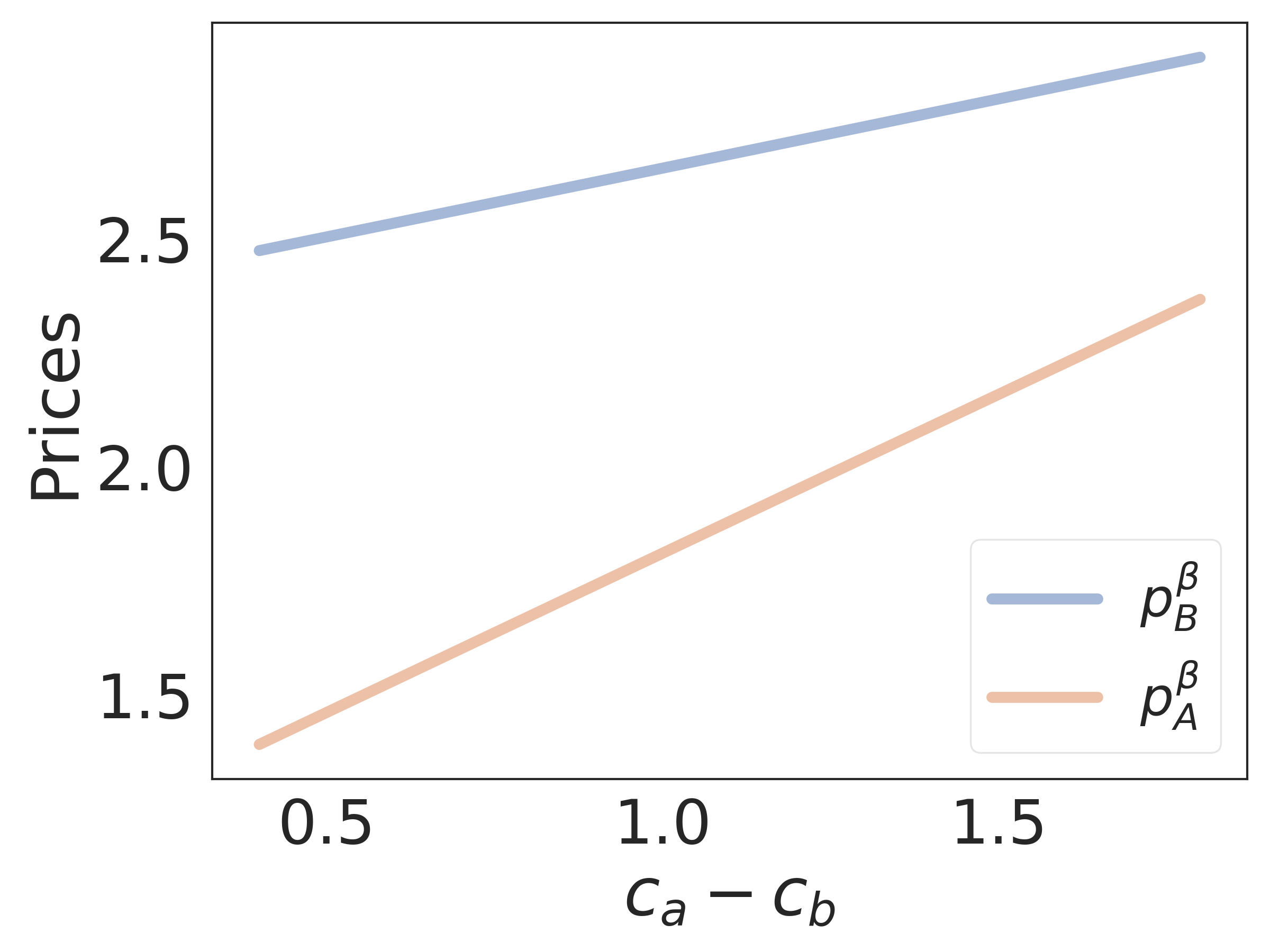}
	\caption{Prices seen by firm $A$'s weak sub-market}
	\label{fig:ml_ih_fm_unc_pbb_pab}
	\end{subfigure}
	\begin{subfigure}{0.3\textwidth}
	\includegraphics[width=0.9\linewidth]{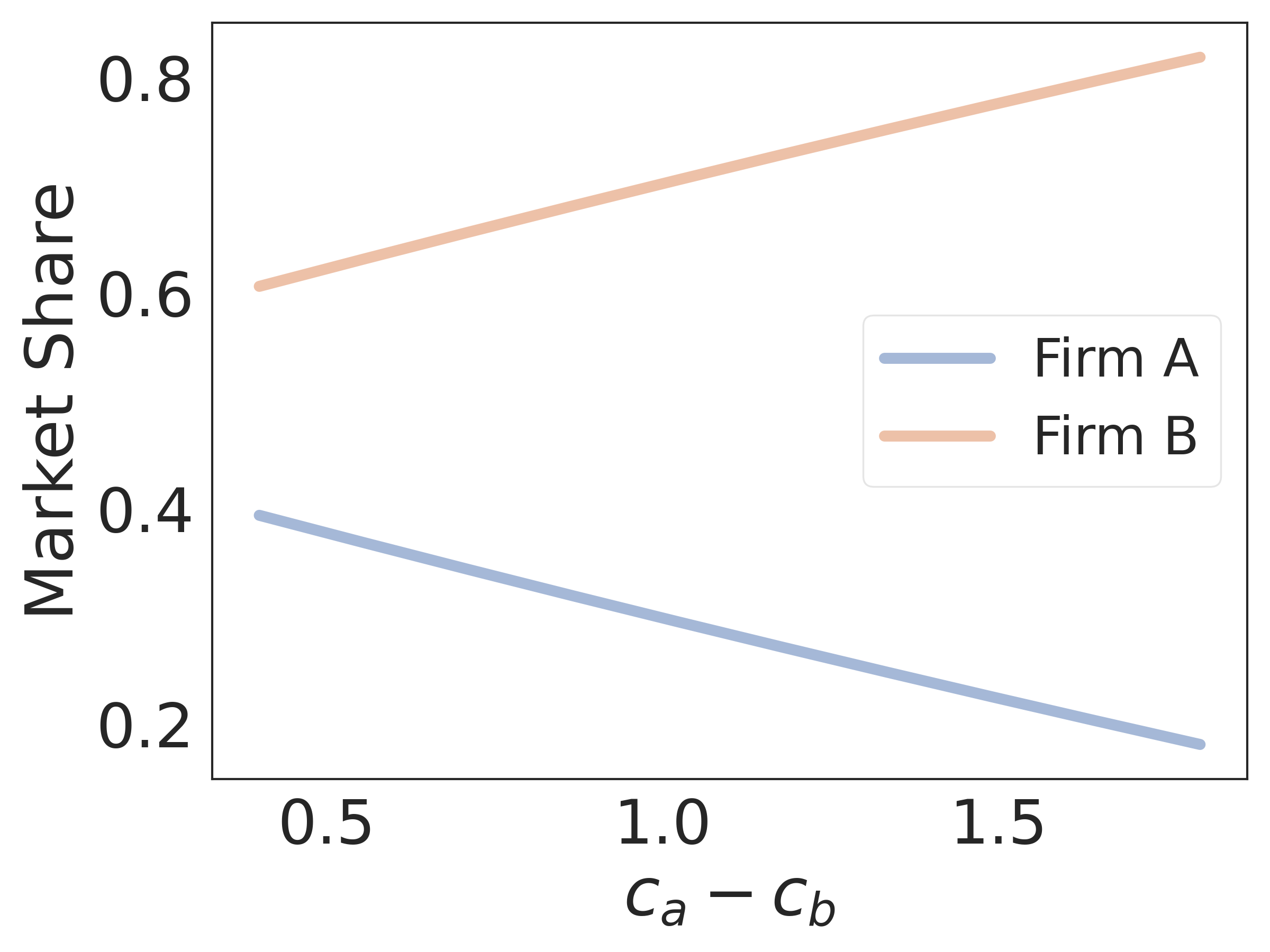} 
	\caption{Market share}
	\label{fig:ml_ih_fm_unc_mkt}
	\end{subfigure}
	\begin{subfigure}{0.3\textwidth}
	\includegraphics[width=0.9\linewidth]{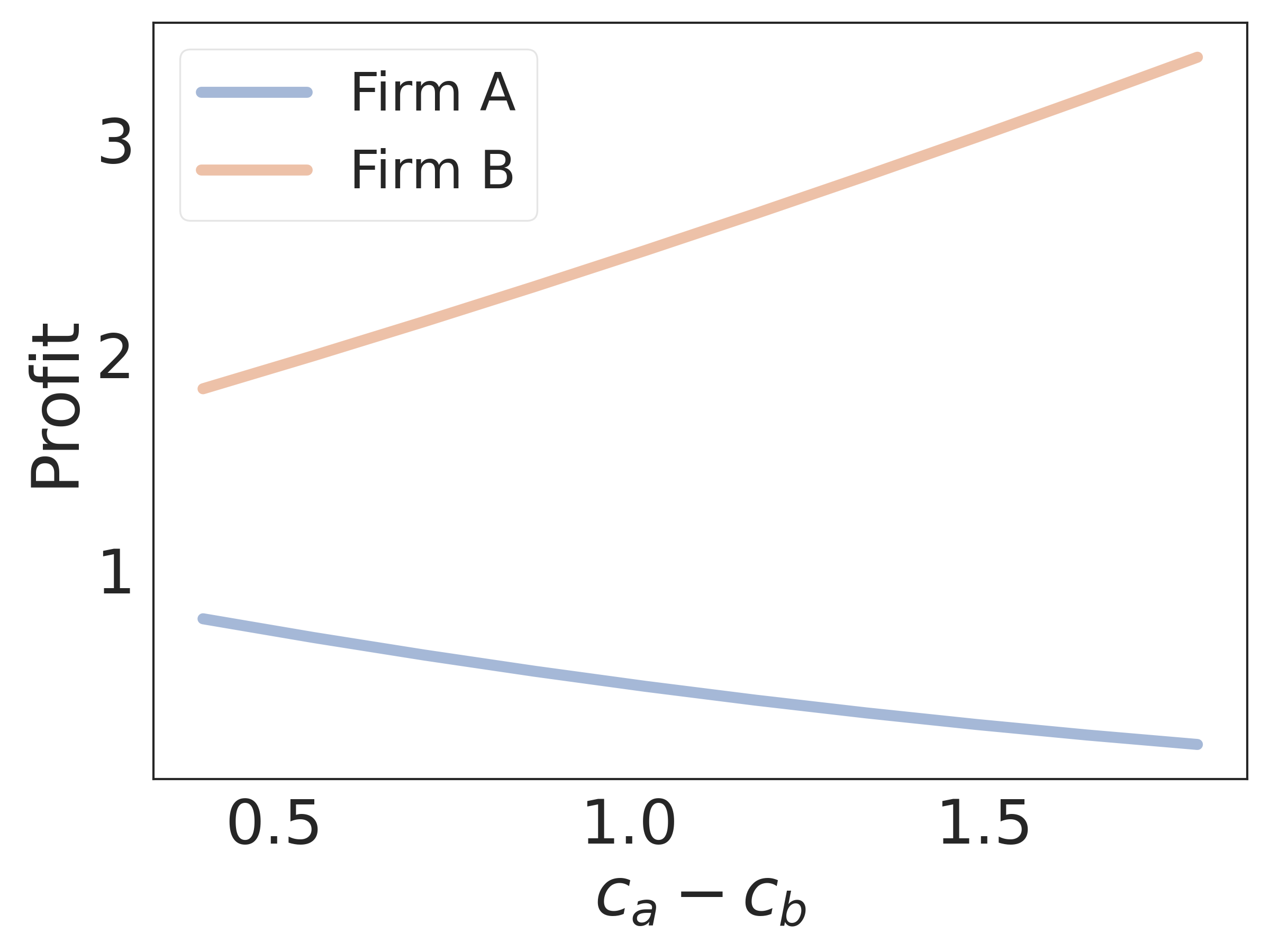}
	\caption{Profit}
	\label{fig:ml_ih_fm_unc_profit}
	\end{subfigure}
	\begin{subfigure}{0.3\textwidth}
	\includegraphics[width=0.9\linewidth]{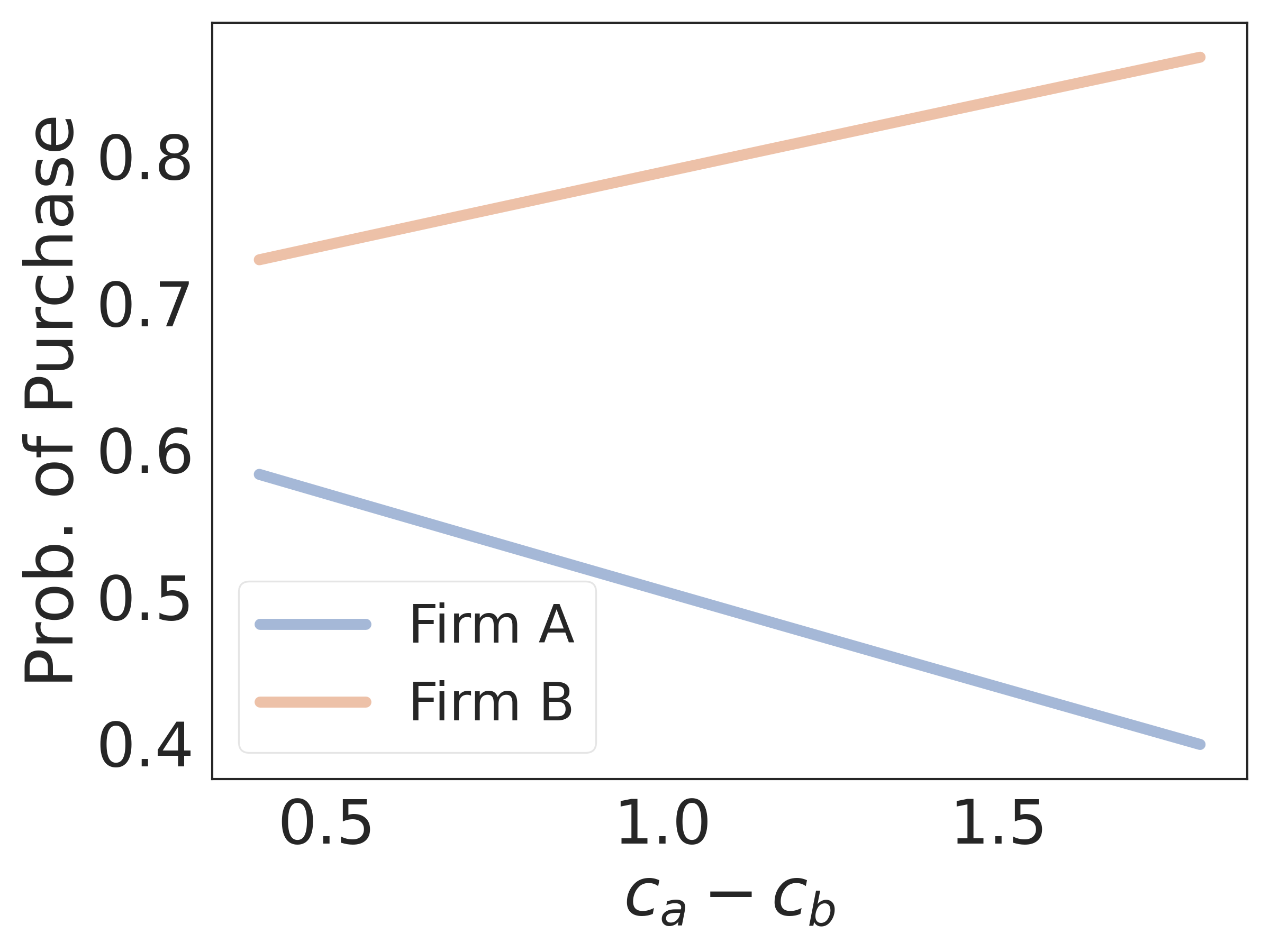}
	\caption{Probability of purchase}
	\label{fig:ml_ih_fm_unc_probabilities}
	\end{subfigure}
\caption{Infinite horizon setting market outcomes under multiplicative loyalty where the constraints are non-binding: Optimal prices, market shares, profits of firms, and probability of purchase of customers of types $\alpha$ and $\beta$ as a function of $\ca-\cb$. Here, $\cb=.2, \ta=3, \tb=4, \df=.4$.}
\label{fig:ml_ih_fm_unc}
\end{figure}

In Figure~\ref{fig:ml_ih_fm_unc}, we plot the prices, market shares and profits of firms when $\da=\db=\df=0.4$ and Assumption~\ref{assume:gl-ih-unconstrained} holds. That is, the values of costs and the loyalty parameters are chosen such that the constraints on prices are non-binding. From Figures~\ref{fig:ml_ih_fm_unc_paa_pba} and ~\ref{fig:ml_ih_fm_unc_pbb_pab}, we can infer that the prices vary linearly with different slopes as the cost asymmetry increases. Further, as the cost difference between firm $A$ and firm $B$ increases, firm $A$ loses significant market share and profit, as its myopic loyal consumers increasingly prefer to purchase from its rival instead. 

\begin{figure}
\centering
	\begin{subfigure}{0.3\textwidth}
	\includegraphics[width=0.9\linewidth]{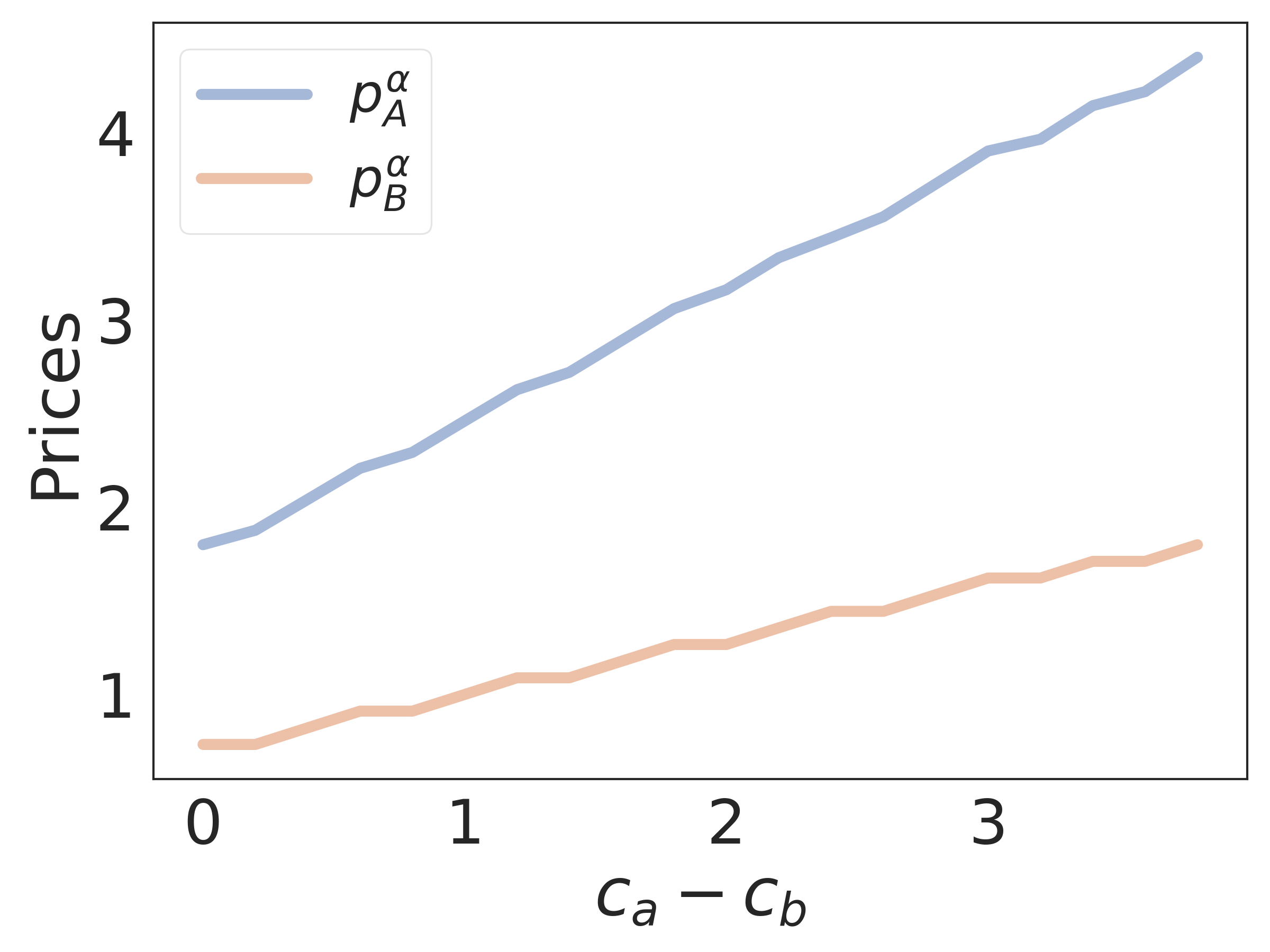} 
	\caption{Prices seen by firm $A$'s strong sub-market}
	\label{fig:ml_ih_fm_c_paa_pba}
	\end{subfigure}
	\begin{subfigure}{0.3\textwidth}
	\includegraphics[width=0.9\linewidth]{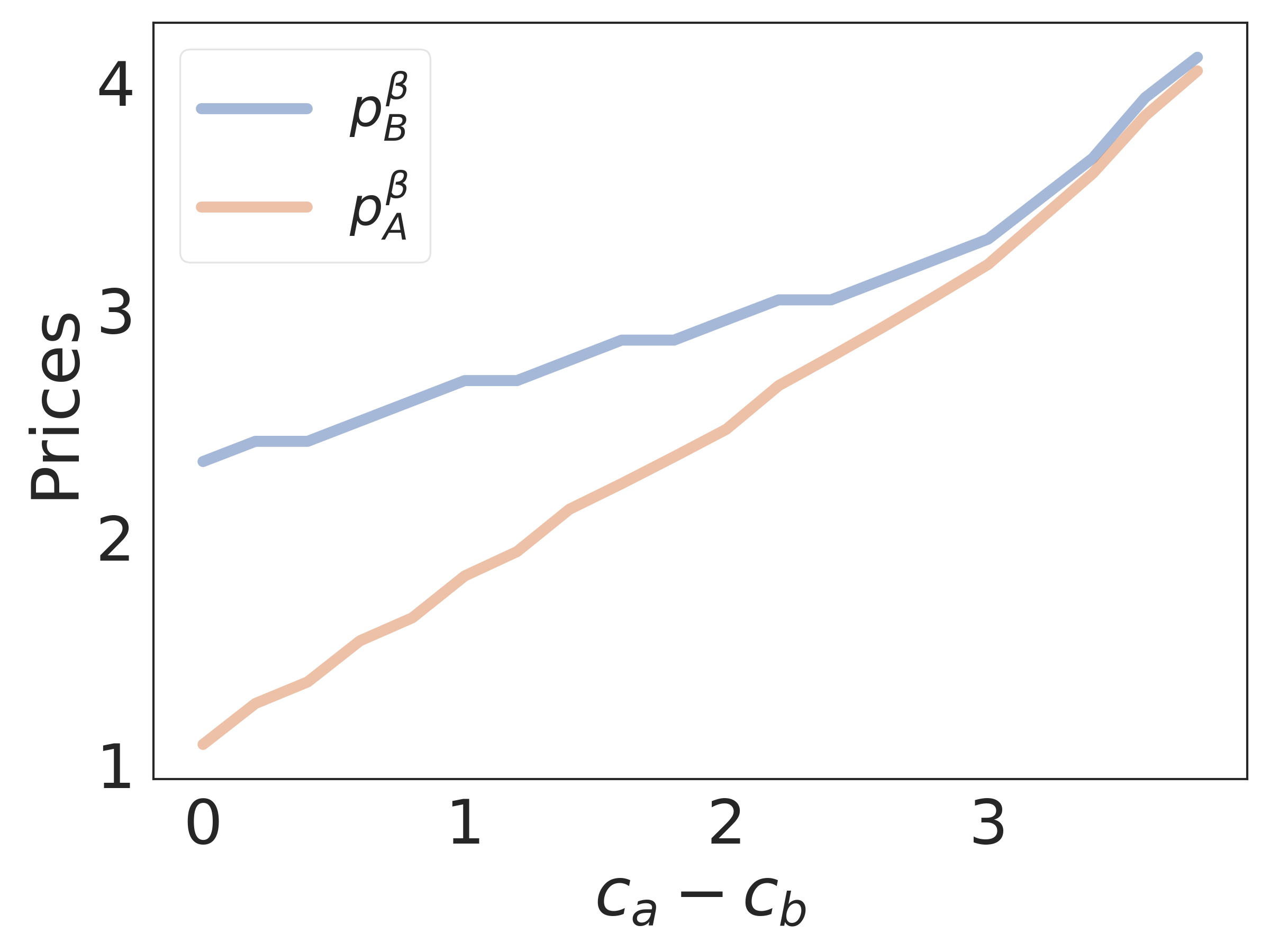}
	\caption{Prices seen by firm $A$'s weak sub-market}
	\label{fig:ml_ih_fm_c_pbb_pab}
	\end{subfigure}
	\begin{subfigure}{0.3\textwidth}
	\includegraphics[width=0.9\linewidth]{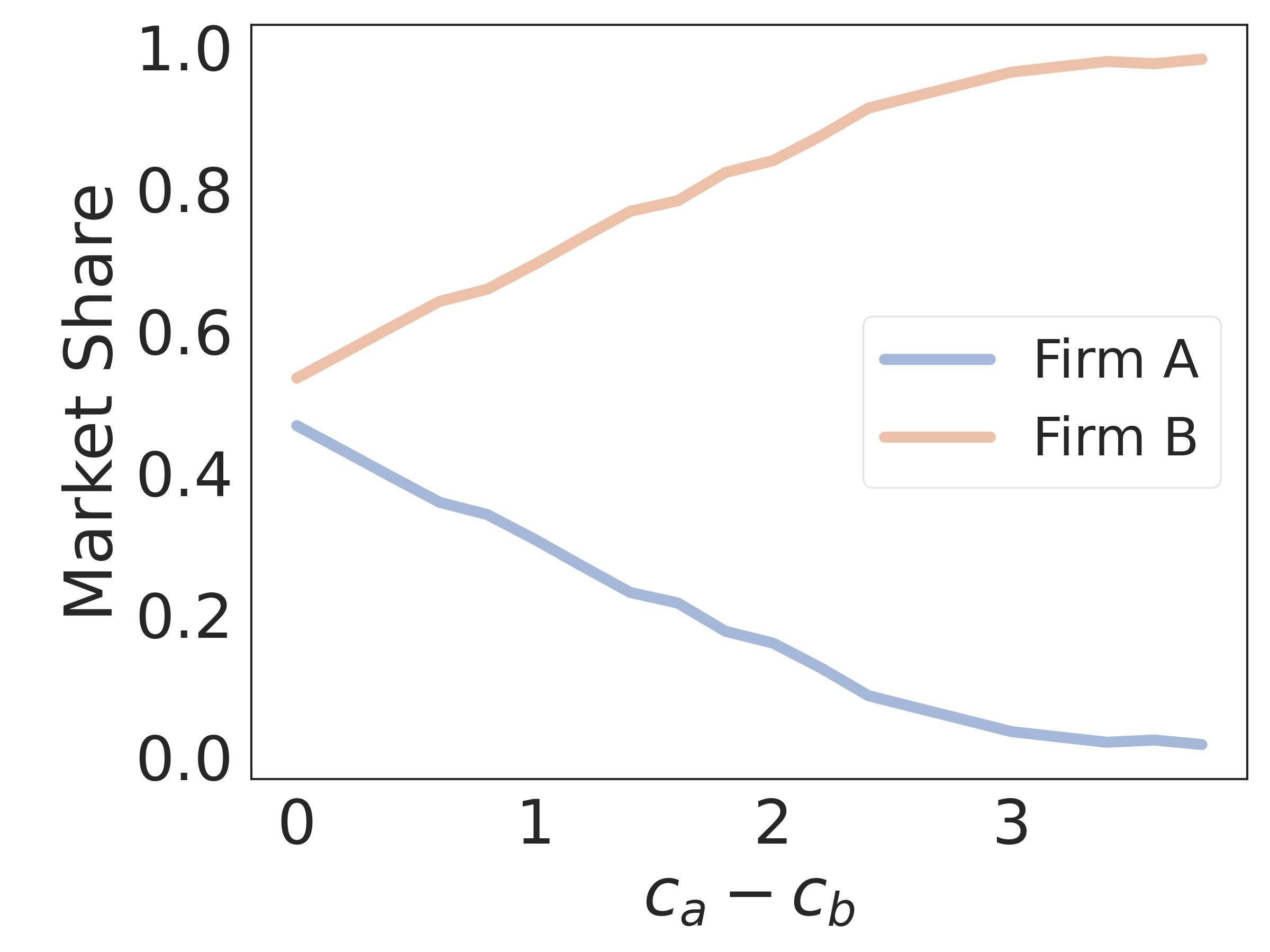} 
	\caption{Market share}
	\label{fig:ml_ih_fm_c_mkt}
	\end{subfigure}
	\begin{subfigure}{0.3\textwidth}
	\includegraphics[width=0.9\linewidth]{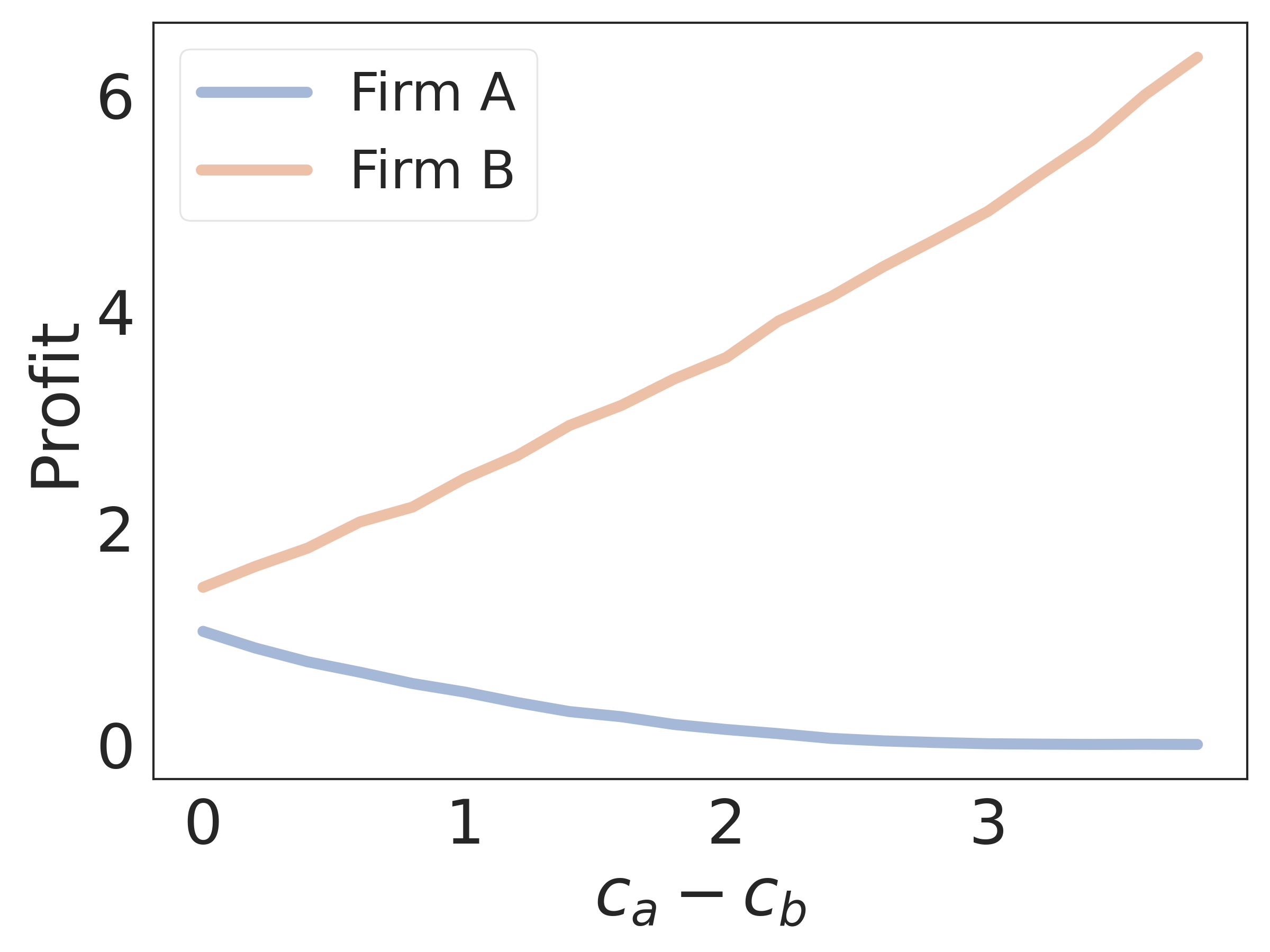}
	\caption{Profit}
	\label{fig:ml_ih_fm_c_profit}
	\end{subfigure}
	\begin{subfigure}{0.3\textwidth}
	\includegraphics[width=0.9\linewidth]{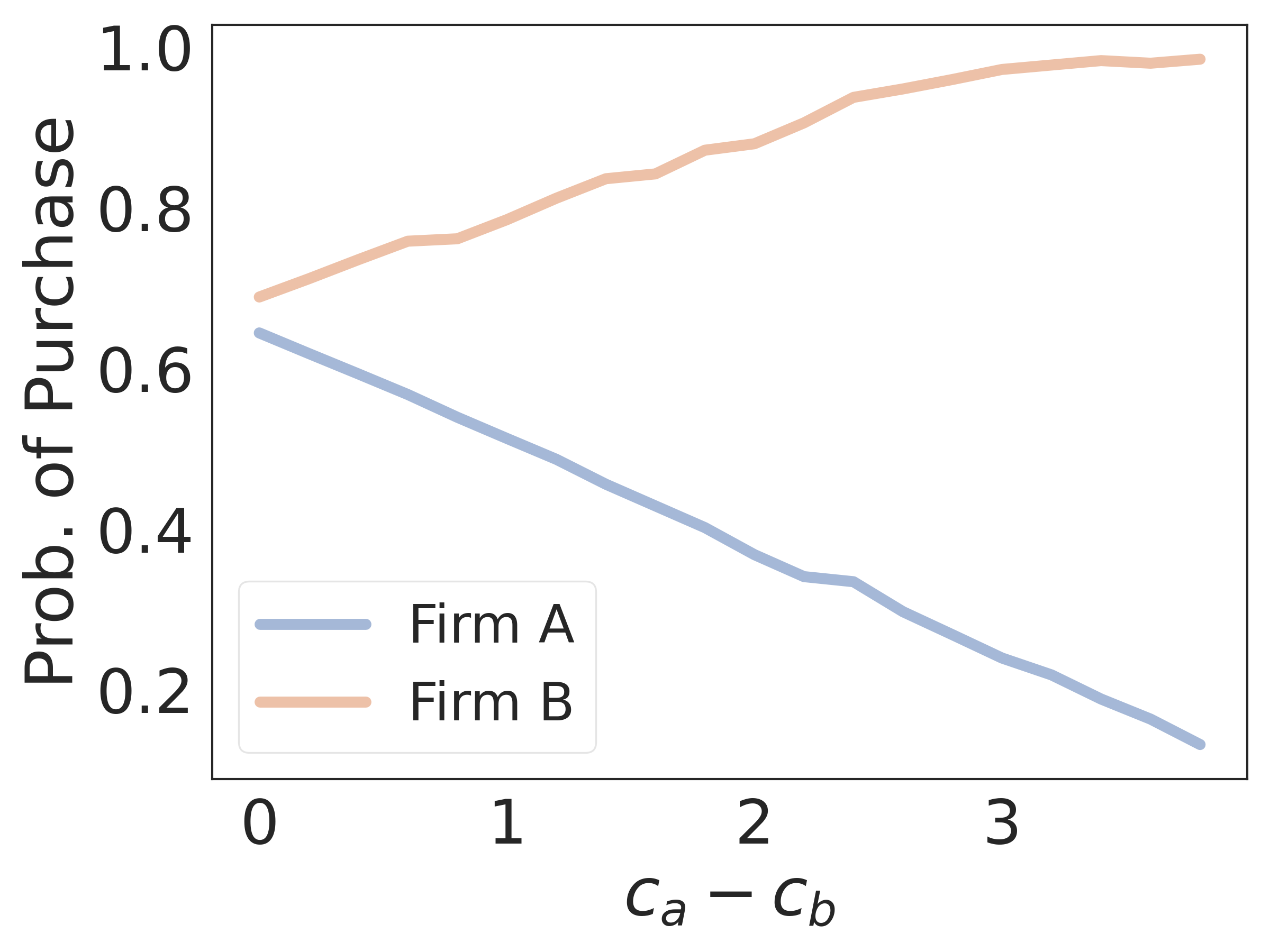}
	\caption{Probability of purchase}
	\label{fig:ml_ih_fm_c_probabilities}
	\end{subfigure}
\caption{Infinite horizon setting market outcomes using numerical simulations under multiplicative loyalty where the constraints can be binding: Optimal prices, market shares, profits of firms, and probability of purchase of customers of types $\alpha$ and $\beta$ as a function of $\ca-\cb$. Here, $\cb=.2, \ta=3, \tb=4, \df =.4$. In contrast to Figure~\ref{fig:ml_ih_fm_unc}, the equilibrium is computed for cost differences up to $4$ units.}
\label{fig:ml_ih_fm_c}
\end{figure}

When the prices are constrained, the situation changes quite a bit (see Figure~\ref{fig:ml_ih_fm_c}). For example, in Figure~\ref{fig:ml_ih_fm_c_pbb_pab}, we can observe that $\pbb$ converges to $\pab$ in a nonlinear way. Similarly, the rate of change of market share and profit as a function of cost asymmetry is also non-linear (see Figures~\ref{fig:ml_ih_fm_unc_mkt} and~\ref{fig:ml_ih_fm_unc_profit}).

For both the constrained and the unconstrained instances, we omit characterization of distinct regions (due to the non availability of closed-form expressions describing the boundaries as seen in Section~\ref{sec:ml-ss} Figure~\ref{fig:ml-ss-regions}). Nonetheless, the nonlinear trends of various market outcomes seen in Figure~\ref{fig:ml_ih_fm_c} provides convincing evidence of the non-trivial impact that cost asymmetry and the loyalty model can have.

\section{Additive Special Case of Linear Loyalty: Analysis and Insights}\label{sec:al}

In the additive special case (AL) of the linear loyalty model, the loyalty function is given by $\ga(\xi) = \xi + \oa$ (its inverse is given by $h_{\alpha}(y) = y-\oa$). Thus, given prices $\paa$ and $\pba$, the probability of a customer belonging to the set $\alpha$ purchasing from firm $A$ is $1 - F(\xi^{\alpha})$, where $\xia = \paa - \pba - \oa$. The parameter $\oa \geq 0$ can be interpreted as the bias in the loyalty level (which is driven by $\oa$ and additionally by the random variable $\xi$ as well). For instance, if $\mathbb{E}[\xi] = 0$, then $\oa$ represents the overall non-random loyalty or inclination of a customer from set $\alpha$ to purchase from firm $A$. From a different point of view, if $\xi$ is supported on the interval $[-B,B]$ for some positive scalar $B$, then $B+\oa$ can be interpreted as the maximum loyalty level exhibited by any customer. The parametric model $\ga(\xi) = \xi + \oa$ has been used in prior work such as ~\citep{somaini2013model,rhodes2014re,villas2015short} and ~\citep{cabral2016dynamic}, where the symbol $s$ (without market segment subscript) is used and is referred to as the sub-market agnostic \emph{switching cost}. We will comment on the differences between our approach and these prior works whenever relevant below, although note that our unified general treatment of the loyalty in this paper (with multiple parametric models of which the additive version is but one) and their impact on market outcomes in conjunction with non-zero product costs significantly extends these prior works. 

\subsection{Single Stage Setting}\label{sec:al-ss}

Just as before, in this setting, there is a one-shot competition between the firms. Specializing the demand functions to the AL gives us the following expressions (the demand function related to the firm $B$ is analogous):
\begin{gather}
D_{A}^{ss}(\paa, \pba) = \theta(1-F(\paa - \pba - \oa)), \text{ and}\\
D_{A}^{ws}(\pab, \pbb) = (1-\theta)F(\pbb - \pab - \ob).
\end{gather}

\begin{figure}[ht]
\centering
\includegraphics[width=0.65\textwidth]{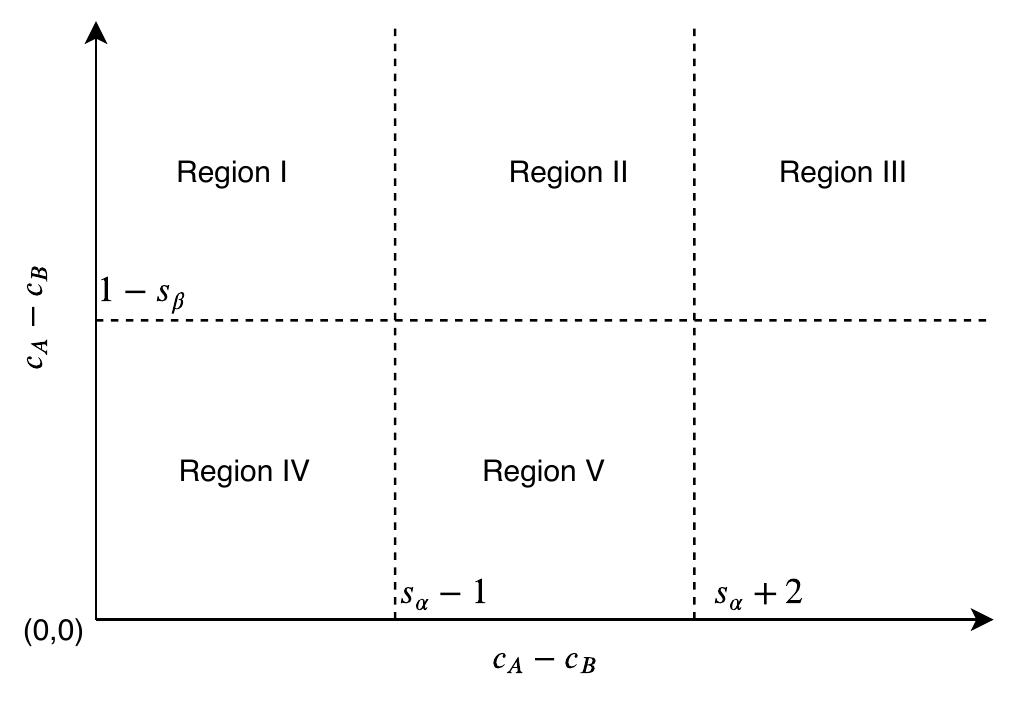}
\caption{Five regions that determine different equilibrium prices depending on the relationship between loyalty levels and product costs in the single stage AL setting.\label{fig:al-ss-regions}}
\end{figure}

Under the choices made for $F$, $\ga$ and $\gb$ above, our analysis reveals six distinct price discrimination regimes (one of which is infeasible) based on the interplay of maximum loyalty levels ($\oa$ and $\ob$) and the magnitude of product cost difference ($\ca-\cb$).  Equilibrium conditions are determined for each of the following sub-cases in Propositions~\ref{prop:al-ss-r1}-\ref{prop:al-ss-r5} below, which are mutually exclusive and exhaustive (see Figure~\ref{fig:al-ss-regions}): 
\begin{itemize}
\item Region I: $1-\ob \leq \camcb \leq \oa-1$ (see Proposition~\ref{prop:al-ss-r1}).
\item Region II: $\max(1-\ob,\oa-1) \leq \camcb \leq \oa+2$ (see Proposition~\ref{prop:al-ss-r2}).
\item Region III: $\max(1-\ob,\oa+2) \leq \camcb$ (see Proposition~\ref{prop:al-ss-r3}).
\item Region IV: $\camcb \leq \min(\oa-1,1-\ob)$ (see Proposition~\ref{prop:al-ss-r4}).
\item Region V: $\oa-1 \leq \camcb \leq \min(\oa+2,1-\ob)$ (see Proposition~\ref{prop:al-ss-r5}).
\item Region VI: Is an empty set, as there are no parameters $(\camcb,\oa,\ob)$ for which the region constraint $\oa+2 \leq \camcb \leq 1-\ob $ holds. We will drop this region from the subsequent analysis and discussion.
\end{itemize}

\begin{proposition}
\label{prop:al-ss-r1}
Under Region I, with Assumptions~\ref{assume:customer-type} and~\ref{assume:uniform-01}, the unique pure Nash equilibrium prices for the strong and weak sub-markets for firms $A$ and $B$ for the AL special case are as follows:
\begin{gather}
\paa = \cb+\oa,\;
\pab = \ca,\;
\pbb = \ca+\ob, \text{and }\;
\pba = \cb.
\end{gather}
\end{proposition}

\begin{proposition}
\label{prop:al-ss-r2}
Under Region II, with Assumptions~\ref{assume:customer-type} and~\ref{assume:uniform-01}, the unique pure Nash equilibrium prices for the strong and weak sub-markets for firms $A$ and $B$ for the AL special case are as follows:
\begin{gather}
\paa = \frac{1}{3}(2\ca+\cb+\oa+2),\;
\pab = \ca,\;\nonumber\\
\pbb = \ca+\ob, \text{and }\;
\pba = \frac{1}{3}(\ca+2\cb-\oa+1).
\end{gather}
\end{proposition}

\begin{proposition}
\label{prop:al-ss-r3}
Under Region III, with Assumptions~\ref{assume:customer-type} and~\ref{assume:uniform-01}, the unique pure Nash equilibrium prices for the strong and weak sub-markets for firms $A$ and $B$ for the AL special case are as follows:
\begin{gather}
\paa = \ca,\;
\pab = \ca,\;
\pbb = \ca+\ob, \text{and }\;
\pba = \ca-\oa-1.
\end{gather}
\end{proposition}

\begin{proposition}
\label{prop:al-ss-r4}
Under Region IV, with Assumptions~\ref{assume:customer-type} and~\ref{assume:uniform-01}, the unique pure Nash equilibrium prices for the strong and weak sub-markets for firms $A$ and $B$ for the AL special case are as follows:
\begin{gather}
\paa = \cb+\oa,\;
\pab = \frac{1}{3}(\cb+2\ca-\ob+1),\;\nonumber\\
\pbb = \frac{1}{3}(2\cb+\ca+\ob+2), \text{and }\;
\pba = \cb.
\end{gather}
\end{proposition}

\begin{proposition}
\label{prop:al-ss-r5}
Under Region V, with Assumptions~\ref{assume:customer-type} and~\ref{assume:uniform-01}, the unique pure Nash equilibrium prices for the strong and weak sub-markets for firms $A$ and $B$ for the AL special case are as follows:
\begin{gather}
\paa = \frac{1}{3}(2\ca+\cb+\oa+2),\;
\pab = \frac{1}{3}(\cb+2\ca-\ob+1),\;\nonumber\\
\pbb = \frac{1}{3}(2\cb+\ca+\ob+2), \text{and  }\;
\pba = \frac{1}{3}(\ca+2\cb-\oa+1).
\end{gather}
\end{proposition}

As can be inferred from Propositions~\ref{prop:al-ss-r1}-\ref{prop:al-ss-r5}, the general sum game between the two firms stays in equilibrium throughout. The possible combinations of product cost differences and loyalty model parameters lead to a variety of market outcomes, which is visually captured in Figure~\ref{fig:al-ss-regions}.

\begin{figure}
\centering
	\begin{subfigure}{0.3\textwidth}
	\includegraphics[width=0.9\linewidth]{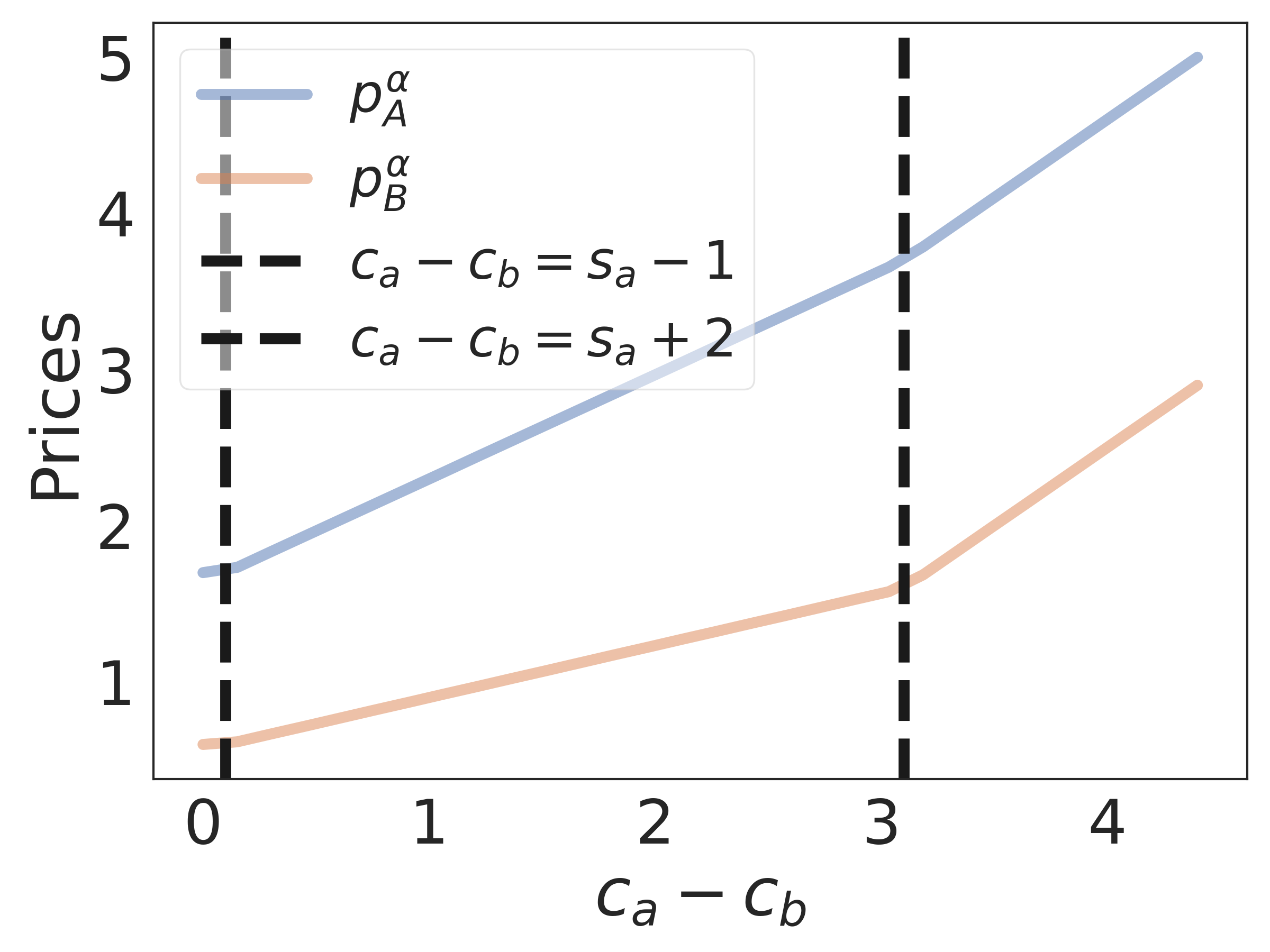} 
	\caption{Prices seen by firm $A$'s strong sub-market}
	\label{fig:al_single_stage_paa_pba}
	\end{subfigure}
	\begin{subfigure}{0.3\textwidth}
	\includegraphics[width=0.9\linewidth]{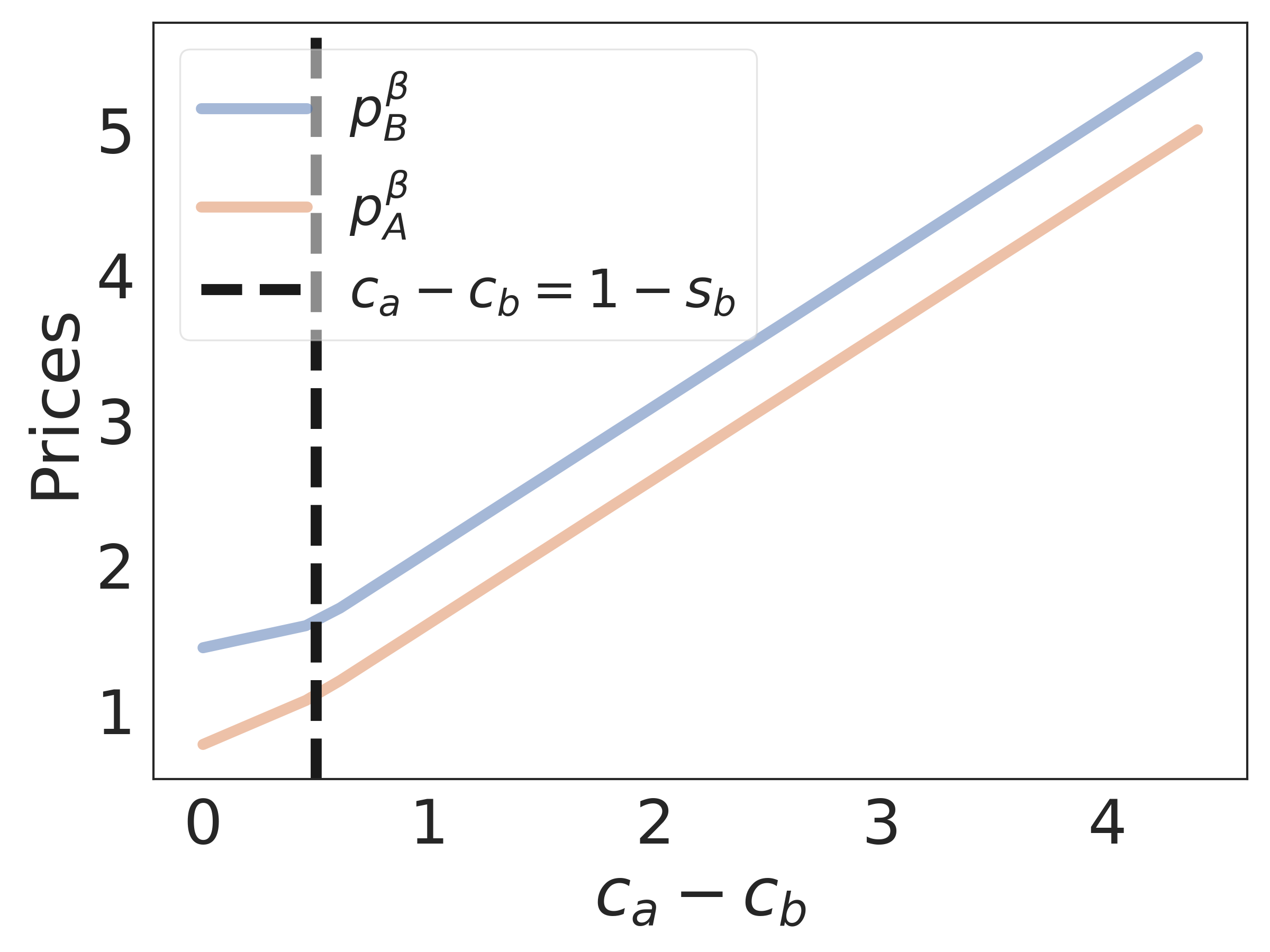}
	\caption{Prices seen by firm $A$'s weak sub-market}
	\label{fig:al_single_stage_pbb_pab}
	\end{subfigure}
	\begin{subfigure}{0.3\textwidth}
	\includegraphics[width=0.9\linewidth]{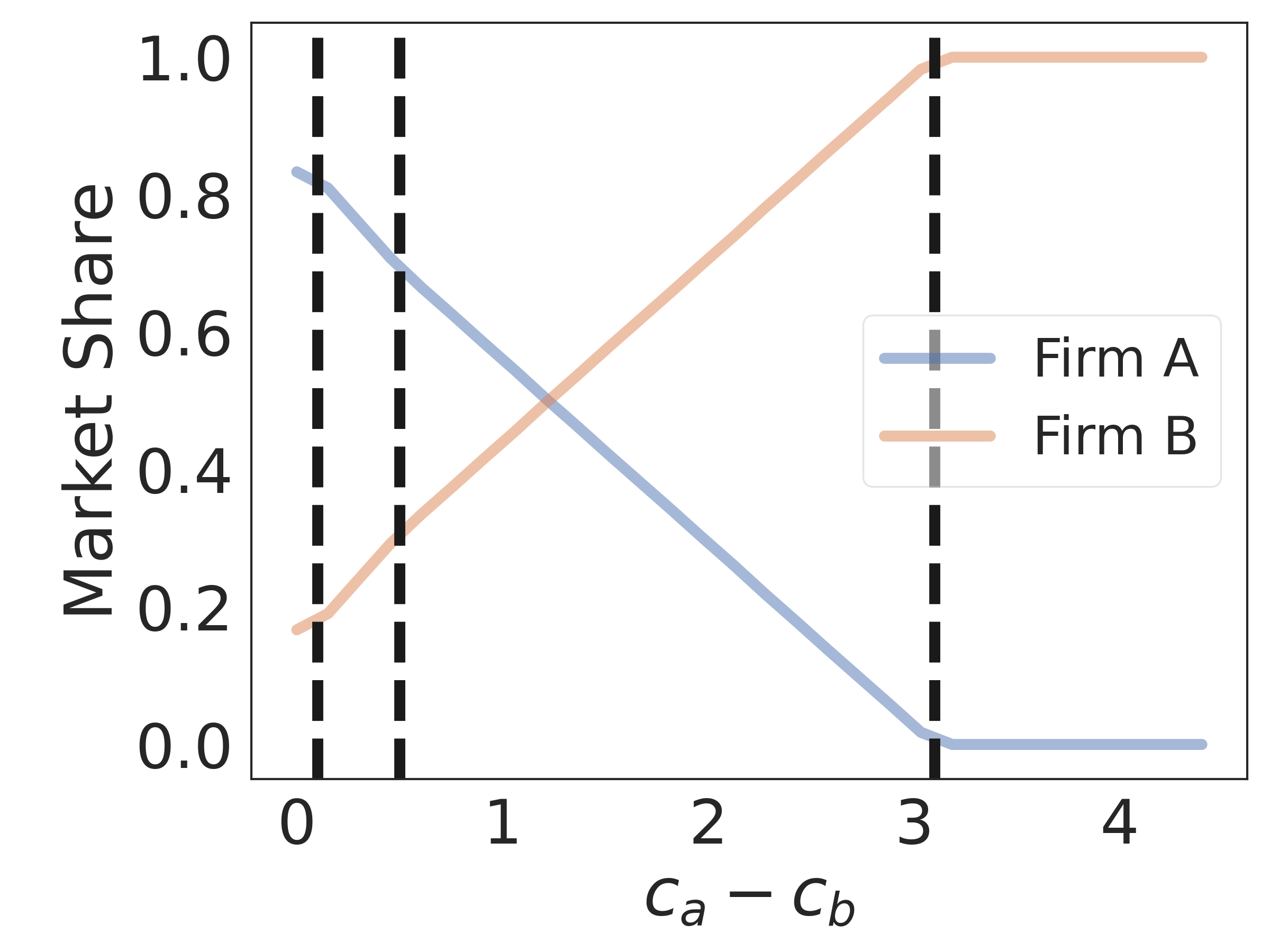} 
	\caption{Market share}
	\label{fig:al_single_stage_market_share}
	\end{subfigure}
	\begin{subfigure}{0.3\textwidth}
	\includegraphics[width=0.9\linewidth]{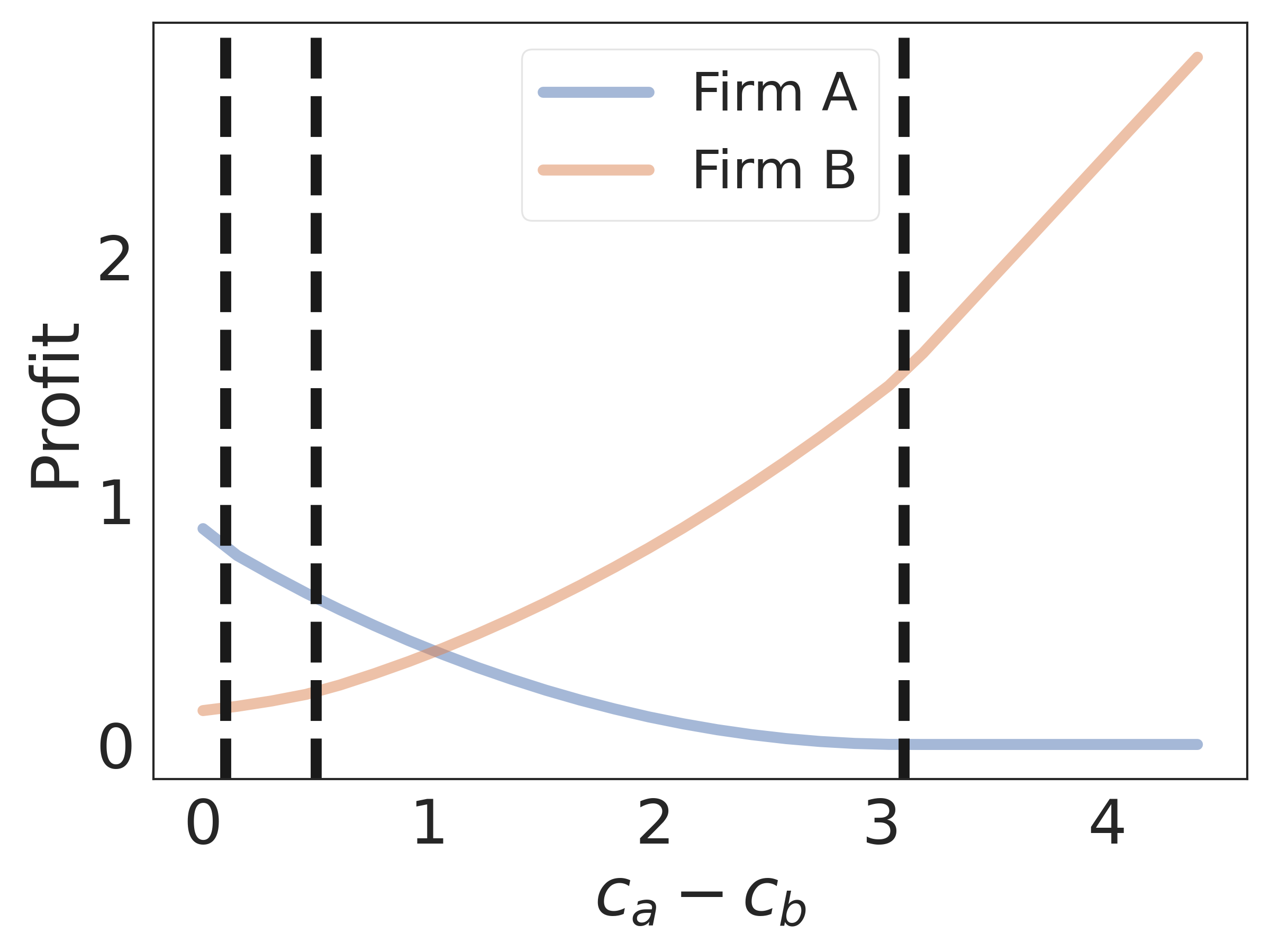}
	\caption{Profit}
	\label{fig:al_single_stage_profits}
	\end{subfigure}
	\begin{subfigure}{0.3\textwidth}
	\includegraphics[width=0.9\linewidth]{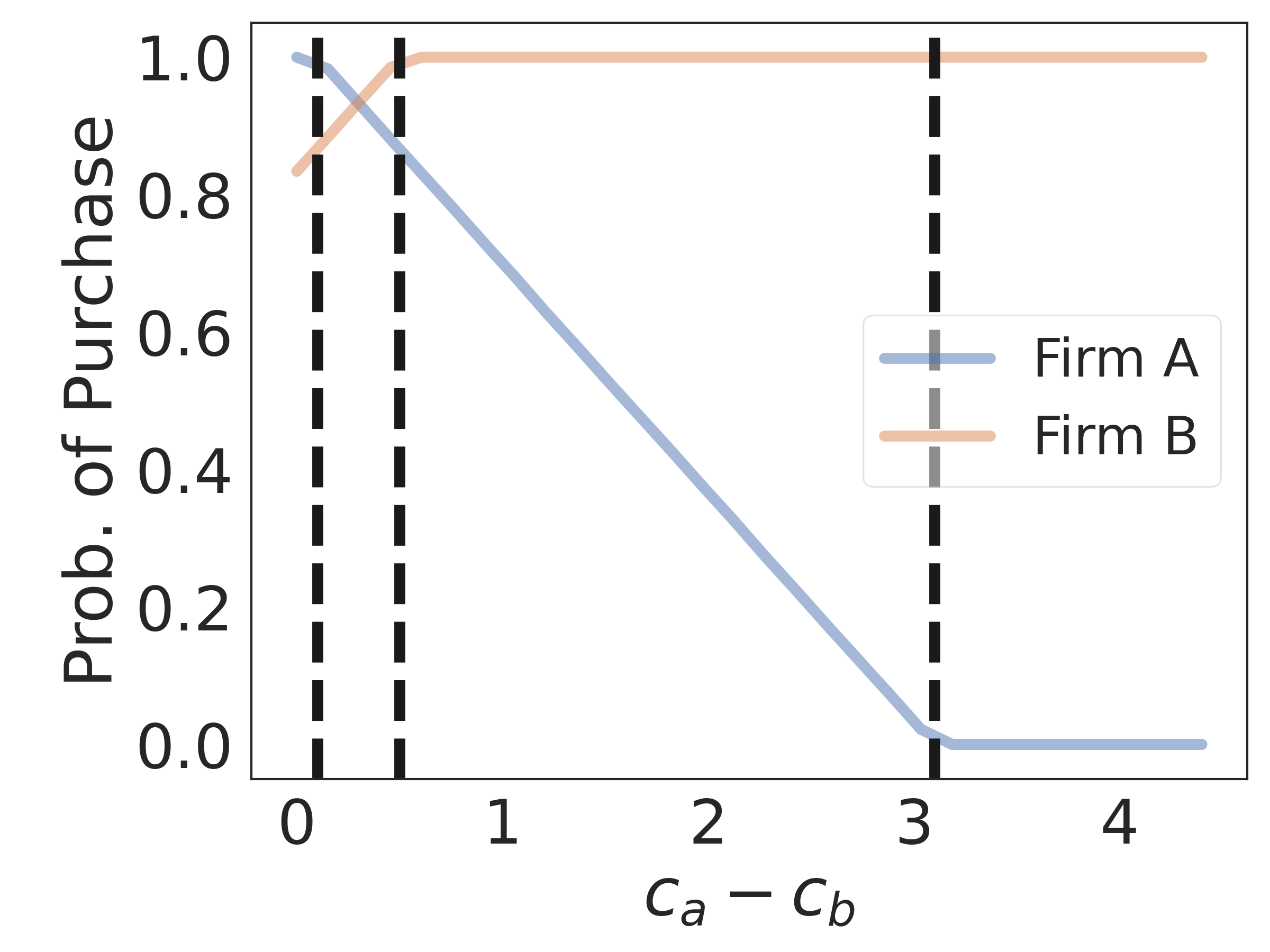}
	\caption{Probability of purchase}
	\label{fig:al_single_stage_probabilities}
	\end{subfigure}
\caption{Single stage market outcomes under additive loyalty special case: Optimal prices, market shares, profits of firms, and probability of purchase of customers of types $\alpha$ and $\beta$ as a function of $\ca-\cb$. Here, $\cb=0.6, \oa=1.1, \ob=.5$ and $\theta=0.8$.}
\label{fig:al-ss-market-outcomes}
\end{figure}

\subsubsection{Insights}

Similar to the single stage setting for the ML special case, nonlinear trends of market outcomes with respect to cost asymmetry are also observed in this setting. From Propositions~\ref{prop:al-ss-r1}-\ref{prop:al-ss-r5} and Figure~\ref{fig:al-ss-market-outcomes}, we can again see that a variety of market outcomes that were not previously considered in the literature can realize. In fact, the regions of parameter space where they happen are significant. The previously studied regime where there was no cost asymmetry would fall under Region IV (and to be more specific, is represented by the origin). In some of the regions, both firm $A$ and $B$ are able to sell to their weak sub-markets in addition to their strong sub-markets, carefully trading off profitability and market shares. At the same time, the high cost firm ($A$) is unable to sell to its loyal following in its strong sub-market while also not being able to make inroads into its weak sub-market in other regions. Thus, Regions I, II, III, and V together represent a class of competitive price discrimination market equilibrium cases that have hitherto gone unnoticed. For brevity, we omit detailed insights similar to the ones presented in Section~\ref{sec:ml-ss}.

\subsection{Infinite Horizon Setting}\label{sec:al-ih}

We start with the setting when both $\da=\db=0$, and then discuss the setting where $\da= \db (=\df \textrm{, a common discount value}) >0$.

\subsubsection{Myopic Firms}

As we have seen before, the equilibrium prices computed in the single stage setting remain valid here. Using Lemma~\ref{lemma:gl-ih-mm}, we can obtain the following expressions for the market shares (and profits can be derived analogously). With $\theta$ as the initial market share of firm $A$ at time $t=0$, we get the following results.

\begin{lemma} \label{lemma:al-ih-mm} With Assumptions~\ref{assume:customer-type} and~\ref{assume:uniform-01}, the market share of firm $A$ at any time index $t$, namely $\theta_t$, under the AL is given as:

\begin{itemize}
\item Region I: $\theta_t = \theta, \theta_{\infty} = \theta.$ 
\item Region II: $\theta_t = \theta\left(\frac{2-\ca +\cb +\oa}{3}\right)^t, \theta_{\infty} = 0.$ 
\item Region III: $\theta_t = 0, \theta_{\infty} = 0.$ 
\item Region IV:
\begin{align*}
\theta_t &= 1-(1-\theta)\left(\frac{\camcb +\ob +2 }{3}\right)^t.\\
\textrm{ Further, } \theta_{\infty} &= 1 \;\;\textrm{ if } \camcb +\ob +2 < 3, \textrm{ else }\;\; \theta.
\end{align*}
\item Region V:
\begin{align*}
\theta_t &= \theta(\oa+\ob-1)^t \\
& + \frac{\camcb +\ob -1}{\oa+\ob -2}\left(1+(\oa+\ob-1)^t\right).\\
\textrm{ Further, } \theta_{\infty} &= \frac{\camcb +\ob -1}{\oa+\ob -2}.
\end{align*}
\end{itemize}

The market share of firm $B$ at the end of time period $t$ is simply $1-\theta_t$ (similarly at steady state it is $1-\theta_{\infty}$). 
\end{lemma}

From the above lemma, it is clear that the steady state market shares of the high-cost firm (firm $A$) are $0$ in Regions I, II and III. In Region IV, the market share is $1$ when $\camcb +\ob +2 < 3$ and $\theta$ otherwise. Finally, in Region V, the eventual market share depends on the loyalty parameters ($\oa,\ob$) and the cost asymmetry. In particular, in Region V, which corresponds to $\oa-1 \leq \camcb \leq \min(1-\ob,\oa+2)$, the steady state market share of firm $A$ is proportional to how large the product cost difference is and the loyalty level of its weak sub-market ($\ob$). If either of them are large, then its market share will likely be large. Noticeably, its market share also depends inversely on the loyalty levels of all customers (i.e., on $\oa$ and $\ob$), unlike Region IV of the multiplicative loyalty setting (see Lemma~\ref{lemma:ml-ih-mm}).

\subsubsection{Forward Looking Firms}\label{sec:al-ih-fm}

Similar to the multiplicative setting in Section~\ref{sec:ml-ih-fm}, we characterize the equilibrium conditions when $F$ satisfies Assumption~\ref{assume:cdf} and when there are no restrictions on the prices (Assumption~\ref{assume:gl-ih-unconstrained}).

\begin{proposition} For the additive loyalty special case (AL), under Assumptions~\ref{assume:customer-type},~\ref{assume:gl-ih-unconstrained} and ~\ref{assume:cdf}, there exists a unique Markov equilibrium when $\da=\db=\df >0$, where firms price based on whether the customer bought their product in the immediate preceding time period. This equilibrium is characterized by the following fixed point equations for thresholds $\xia$ and $\xib$:
\begin{align*}
(\xia - &(\camcb-\oa))\left(\frac{1-\df}{\df} +\Fb+1\right) \\
&+  \frac{2\Fa-1}{\fa}\left(\frac{1-\df}{\df} +\Fa+\Fb\right) +  \frac{\Fa}{\fa} \\
	& = \frac{1-\Fb}{\fb} - \Fb\left(\xib + \camcb +\ob \right),
\end{align*}
and
\begin{align*}
(\xib - &(\cb-\ca-\ob))\left(\frac{1-\df}{\df} +\Fa+1\right)\\
&+  \frac{2\Fb-1}{\fb}\left(\frac{1-\df}{\df} +\Fa+\Fb\right) +  \frac{\Fb}{\fb} \\
	& = \frac{1-\Fa}{\fa} - \Fa\left(\xia + \cb - \ca +\oa \right).
\end{align*}

\label{prop:al-ih-fm}
\end{proposition}

Similar to the Proposition~\ref{prop:ml-ih-fm}, the above thresholds can be used in conjunction with Equations~\ref{eqn:paa}-\ref{eqn:pba} to obtain the optimal prices, profits and resulting market shares. Because the thresholds are implicitly defined, in the following, we numerically solve for their values and characterize the dependence of key market metrics on cost asymmetry and loyalty parameters.

\subsubsection{Insights}

Recall that a customer who bought firm $A$'s product in the immediate preceding time period belongs to their strong sub-market. Similarly, a customer who bough firm $B$'s product in the immediate preceding time period belongs to firm $A$'s weak sub-market. Also note that, similar to the analysis in Section~\ref{sec:ml-ih}, our analysis is at the level of an individual customer. In order to get market level metrics, we need to aggregate over the total number of customers (or take into account their proportion).

We see the following intuitive relationship between firm $A$'s pricing across its strong and weak markets when compared to firm $B$'s from Proposition~\ref{prop:gl-ih-eq}. In particular, Equations~\ref{eqn:paa}-\ref{eqn:pba} yield the following for the AL special case:
\begin{gather}
\paa-\pab - \frac{1}{\fa} = \pbb-\pba - \frac{1}{\fb}.
\end{gather}
This indicates that when $\xi$ is uniform, the difference in the optimal prices charged by firm $A$ is the same as the difference in optimal prices changed by firm $B$. Further, this equivalence is invariant to the costs $\ca,\cb$ as well as the loyalty parameters $\oa$ and $\ob$. Further, this holds even when the long term discounting done by each of the firm is different (i.e., when $\da \neq \db$). 

\begin{figure}
\centering
	\begin{subfigure}{0.3\textwidth}
	\includegraphics[width=0.9\linewidth]{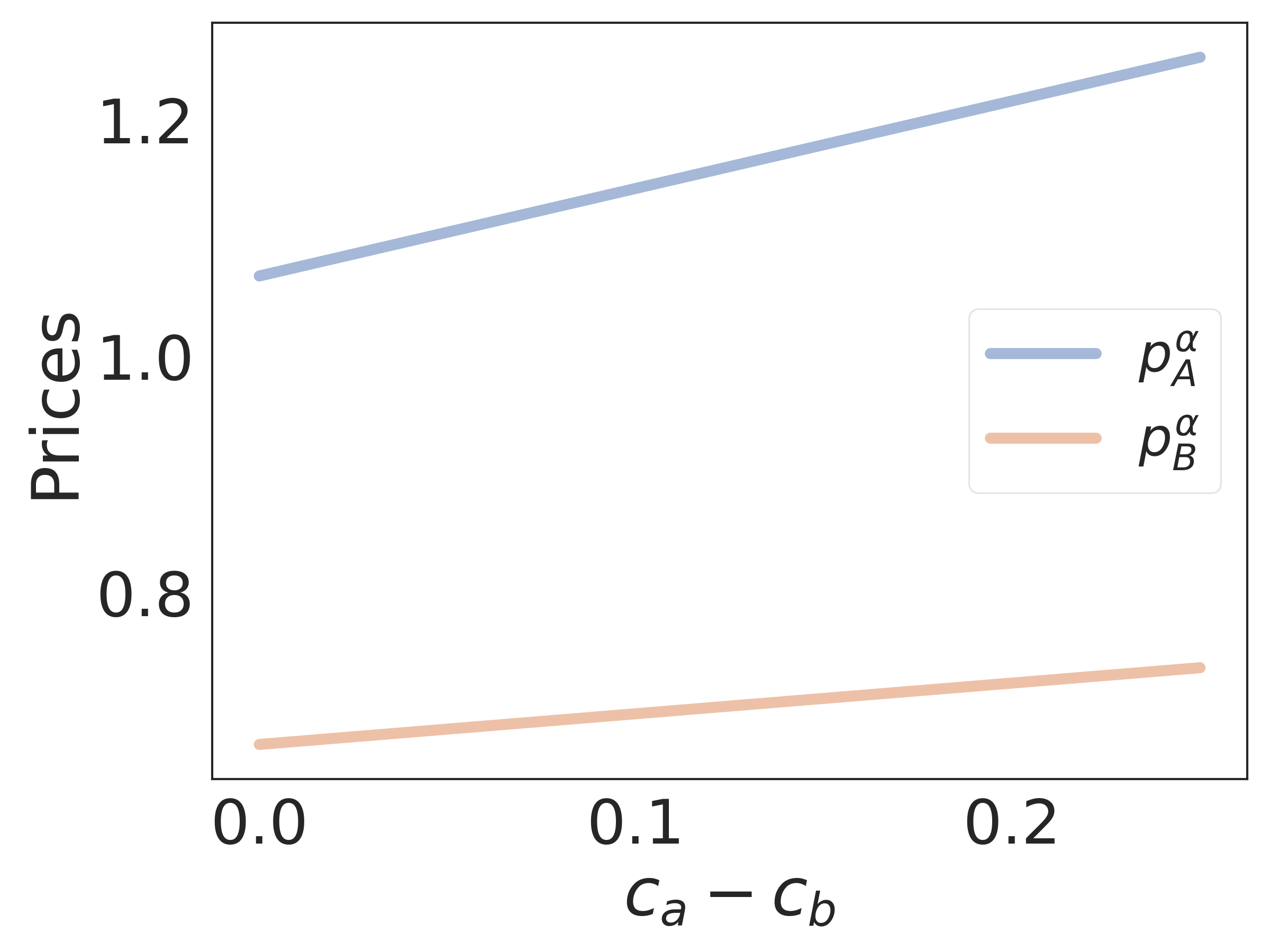} 
	\caption{Prices seen by firm $A$'s strong sub-market}
	\label{fig:al_ih_fm_unc_paa_pba}
	\end{subfigure}
	\begin{subfigure}{0.3\textwidth}
	\includegraphics[width=0.9\linewidth]{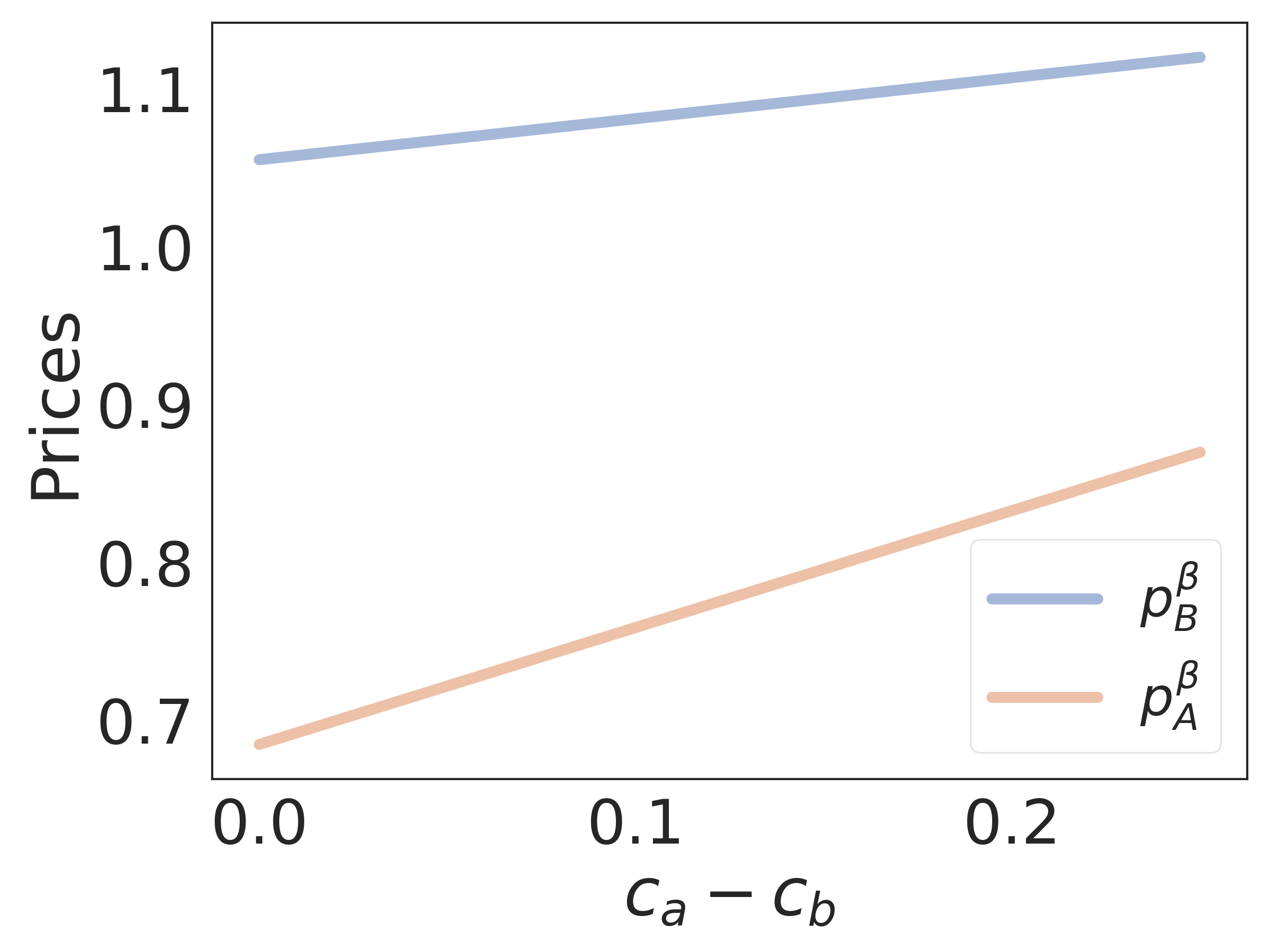}
	\caption{Prices seen by firm $A$'s weak sub-market}
	\label{fig:al_ih_fm_unc_pbb_pab}
	\end{subfigure}
	\begin{subfigure}{0.3\textwidth}
	\includegraphics[width=0.9\linewidth]{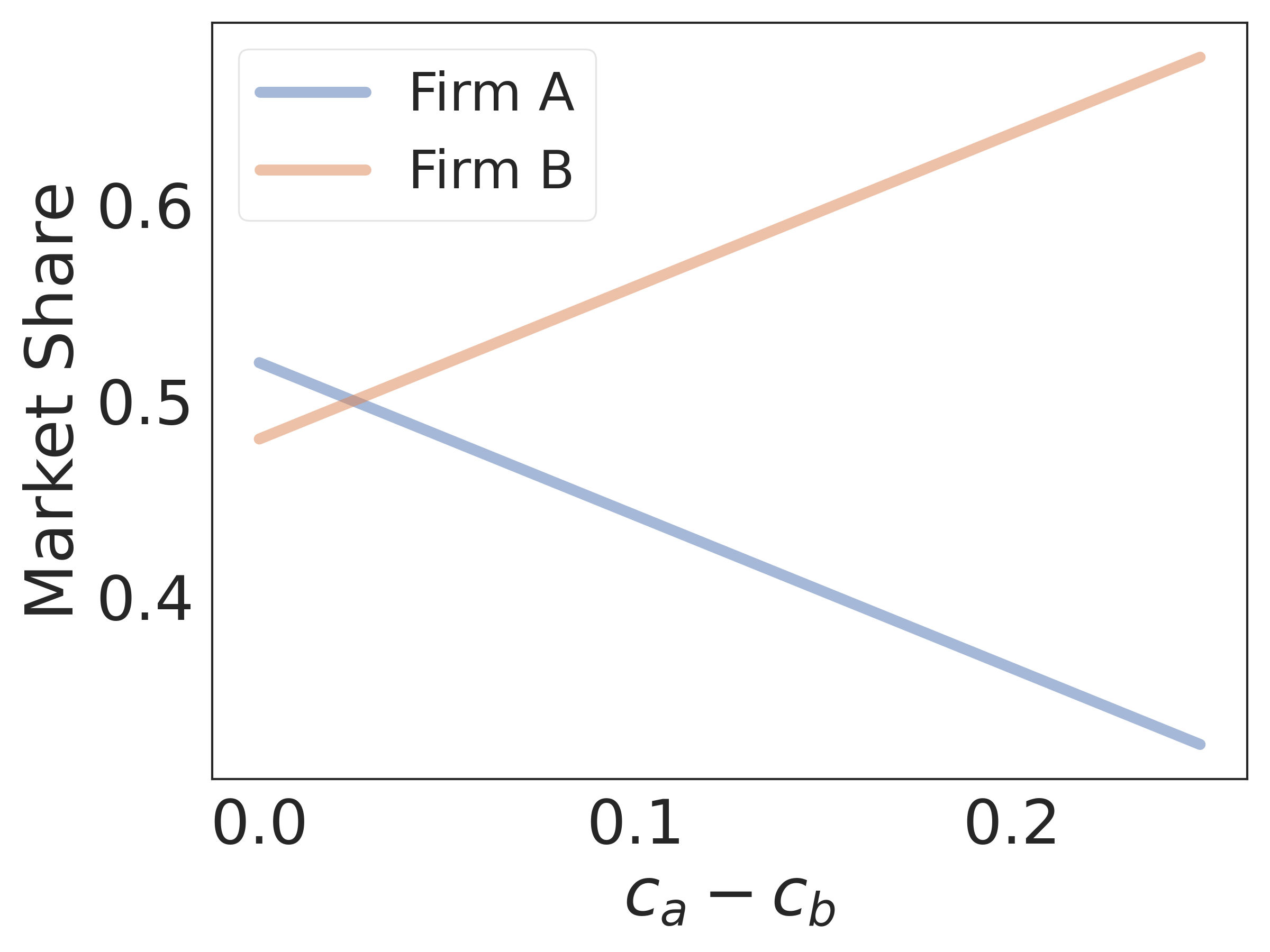} 
	\caption{Market share}
	\label{fig:al_ih_fm_unc_mkt}
	\end{subfigure}
	\begin{subfigure}{0.3\textwidth}
	\includegraphics[width=0.9\linewidth]{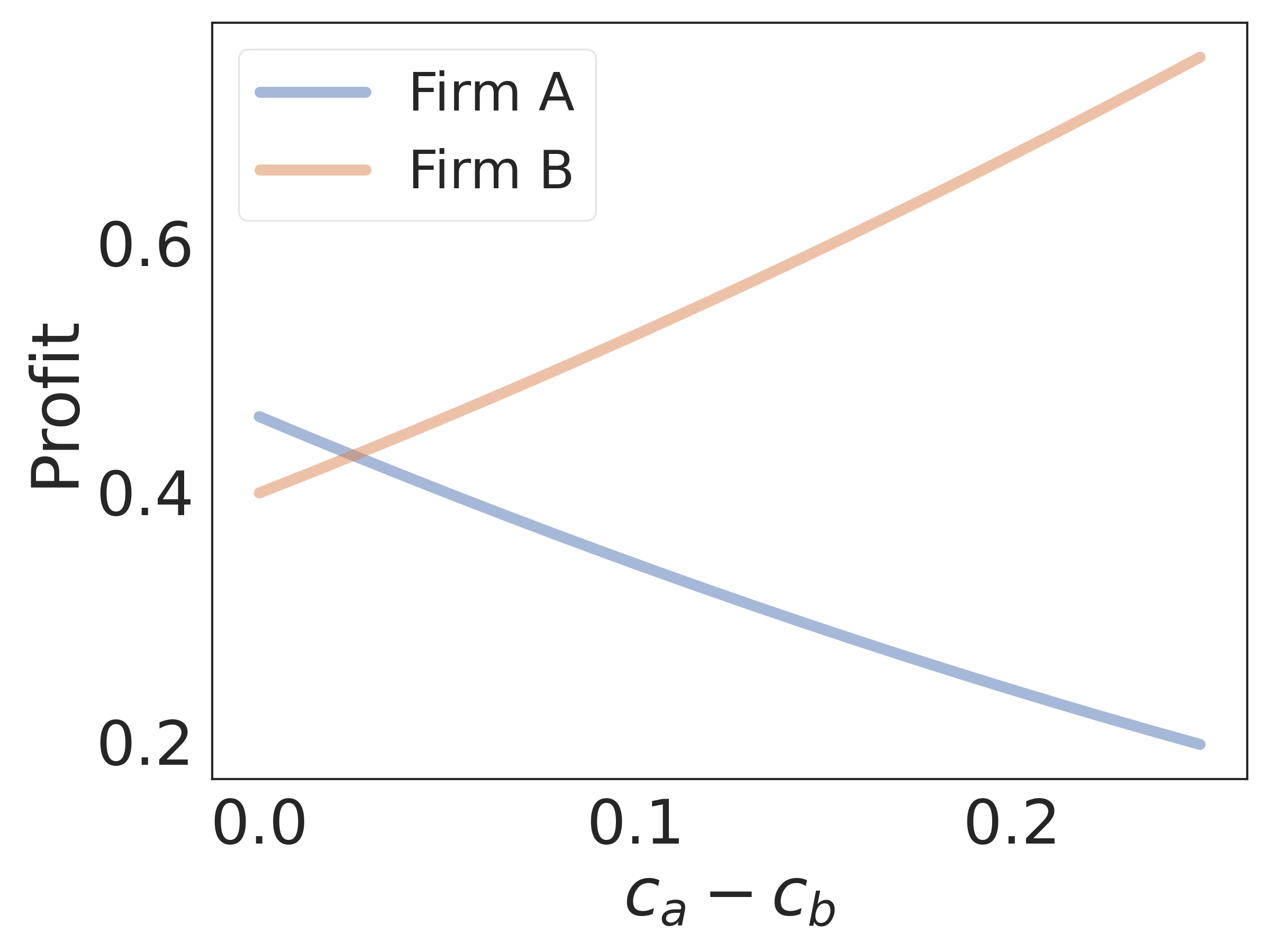}
	\caption{Profit}
	\label{fig:al_ih_fm_unc_profit}
	\end{subfigure}
	\begin{subfigure}{0.3\textwidth}
	\includegraphics[width=0.9\linewidth]{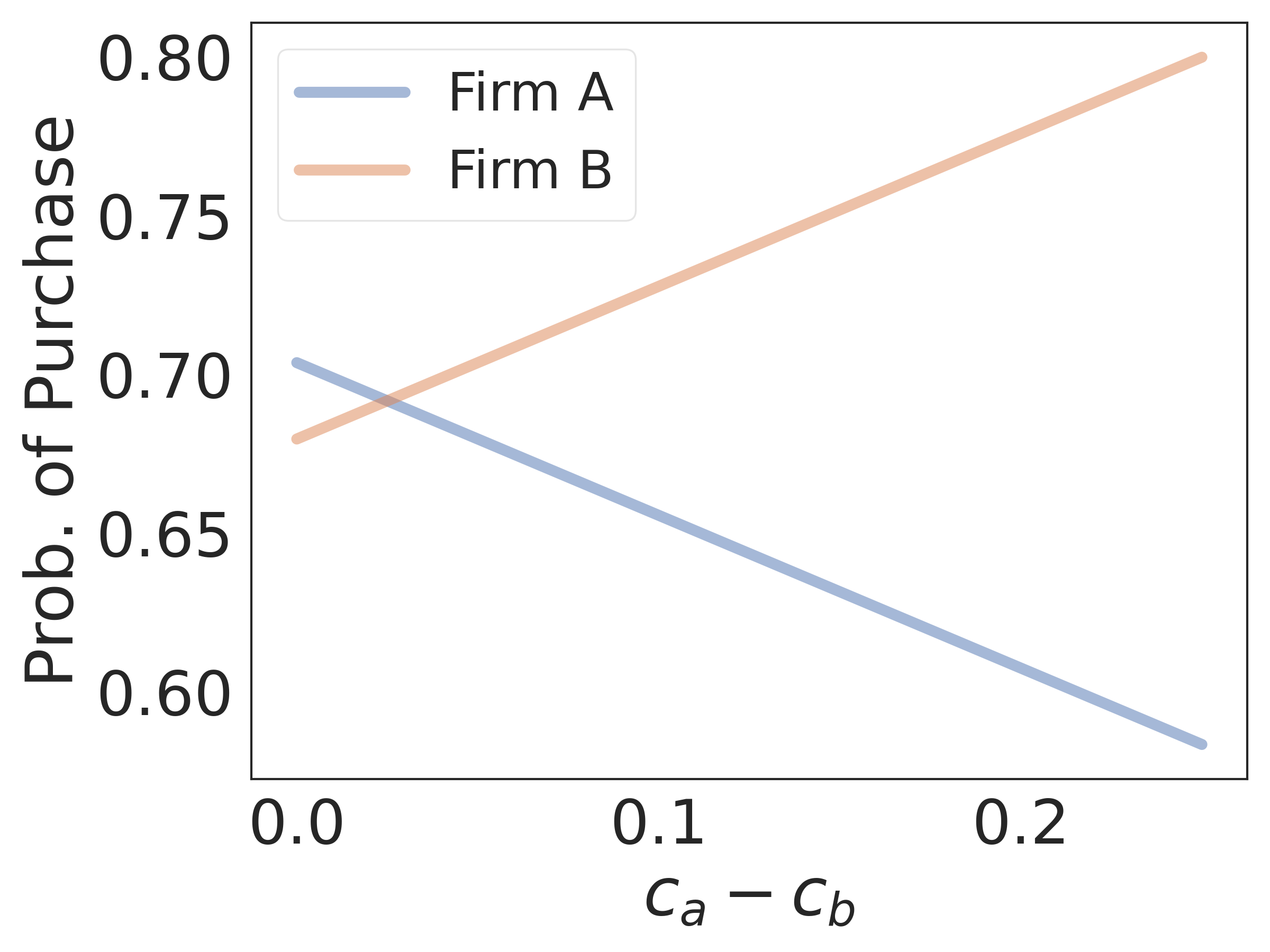}
	\caption{Probability of purchase}
	\label{fig:al_ih_fm_unc_probabilities}
	\end{subfigure}
\caption{Infinite horizon setting market outcomes under additive loyalty where the constraints are non-binding: Optimal prices, market shares, profits of firms, and probability of purchase of customers of types $\alpha$ and $\beta$ as a function of $\ca-\cb$. Here, $\cb=.6, \oa=.1, \ob=.05, \df=.6$.}
\label{fig:al_ih_fm_unc}
\end{figure}

Figure~\ref{fig:al_ih_fm_unc} shows how the prices, market shares and profits of firms evolve when $\da=\db=\df=0.6$ and prices are not constrained (i.e., Assumption~\ref{assume:gl-ih-unconstrained} holds). This is observed by choosing a suitable range of cost differences and appropriate loyalty parameters. From Figures~\ref{fig:al_ih_fm_unc_paa_pba} and ~\ref{fig:al_ih_fm_unc_pbb_pab}, we can infer that the prices vary linearly with different slopes as the cost asymmetry increases (similar to the multiplicative loyalty (ML) special case). In the depicted (narrow) regime of cost asymmetry, firm $A$ starts losing significant market share and profit, as its myopic loyal consumers increasingly prefer to purchase from its rival. 

\begin{figure}
\centering
	\begin{subfigure}{0.3\textwidth}
	\includegraphics[width=0.9\linewidth]{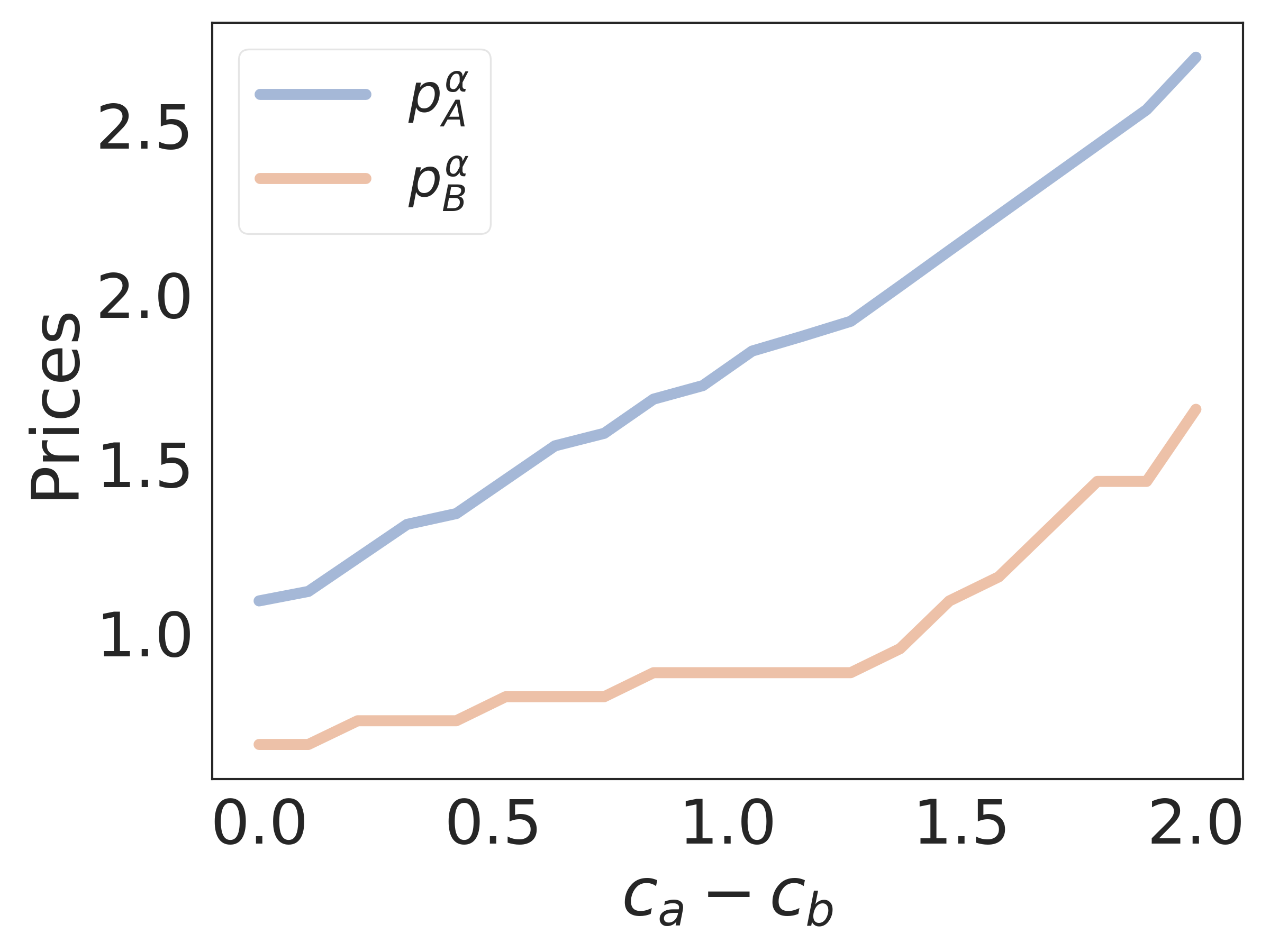} 
	\caption{Prices seen by firm $A$'s strong sub-market}
	\label{fig:al_ih_fm_c_paa_pba}
	\end{subfigure}
	\begin{subfigure}{0.3\textwidth}
	\includegraphics[width=0.9\linewidth]{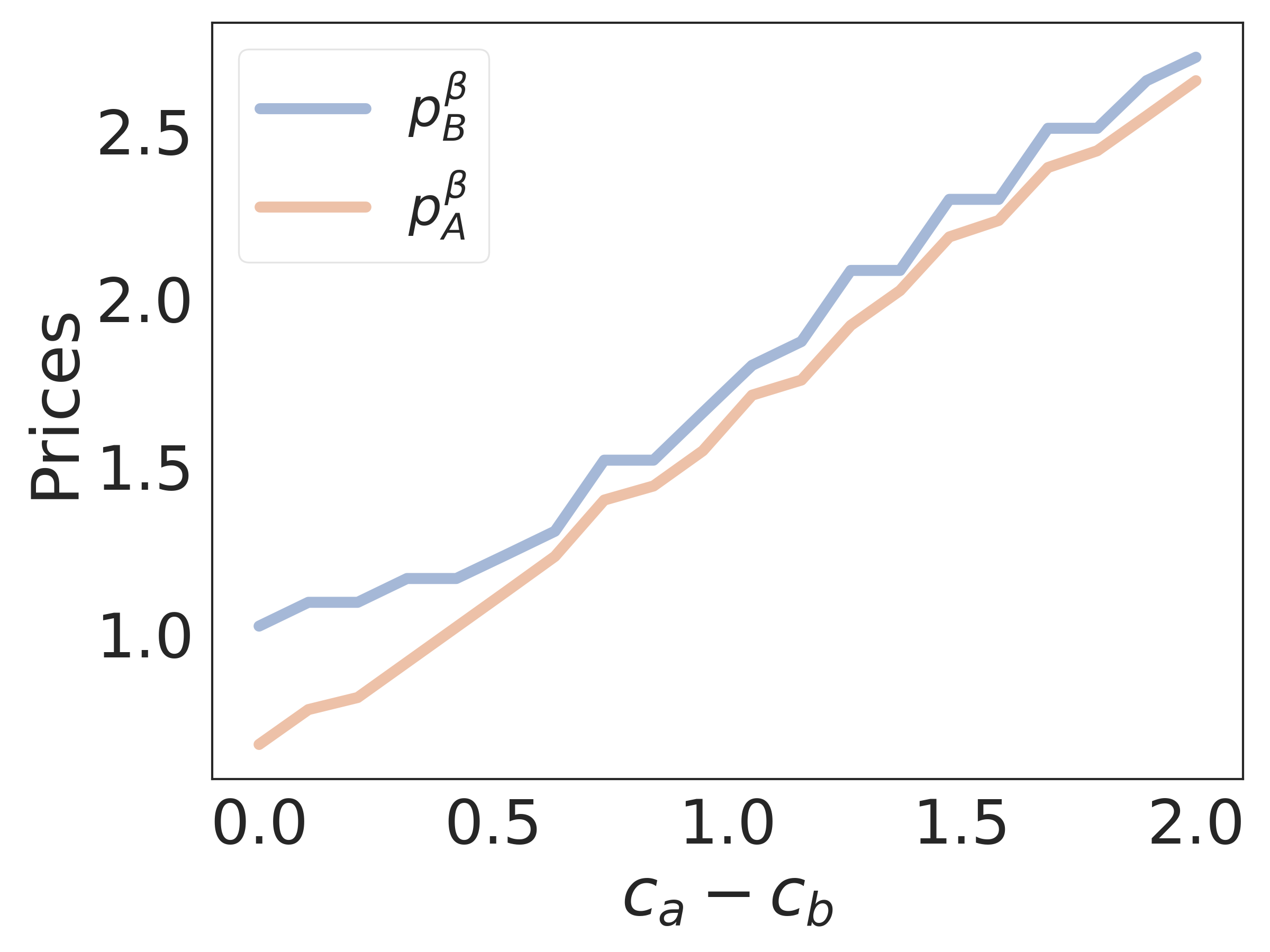}
	\caption{Prices seen by firm $A$'s weak sub-market}
	\label{fig:al_ih_fm_c_pbb_pab}
	\end{subfigure}
	\begin{subfigure}{0.3\textwidth}
	\includegraphics[width=0.9\linewidth]{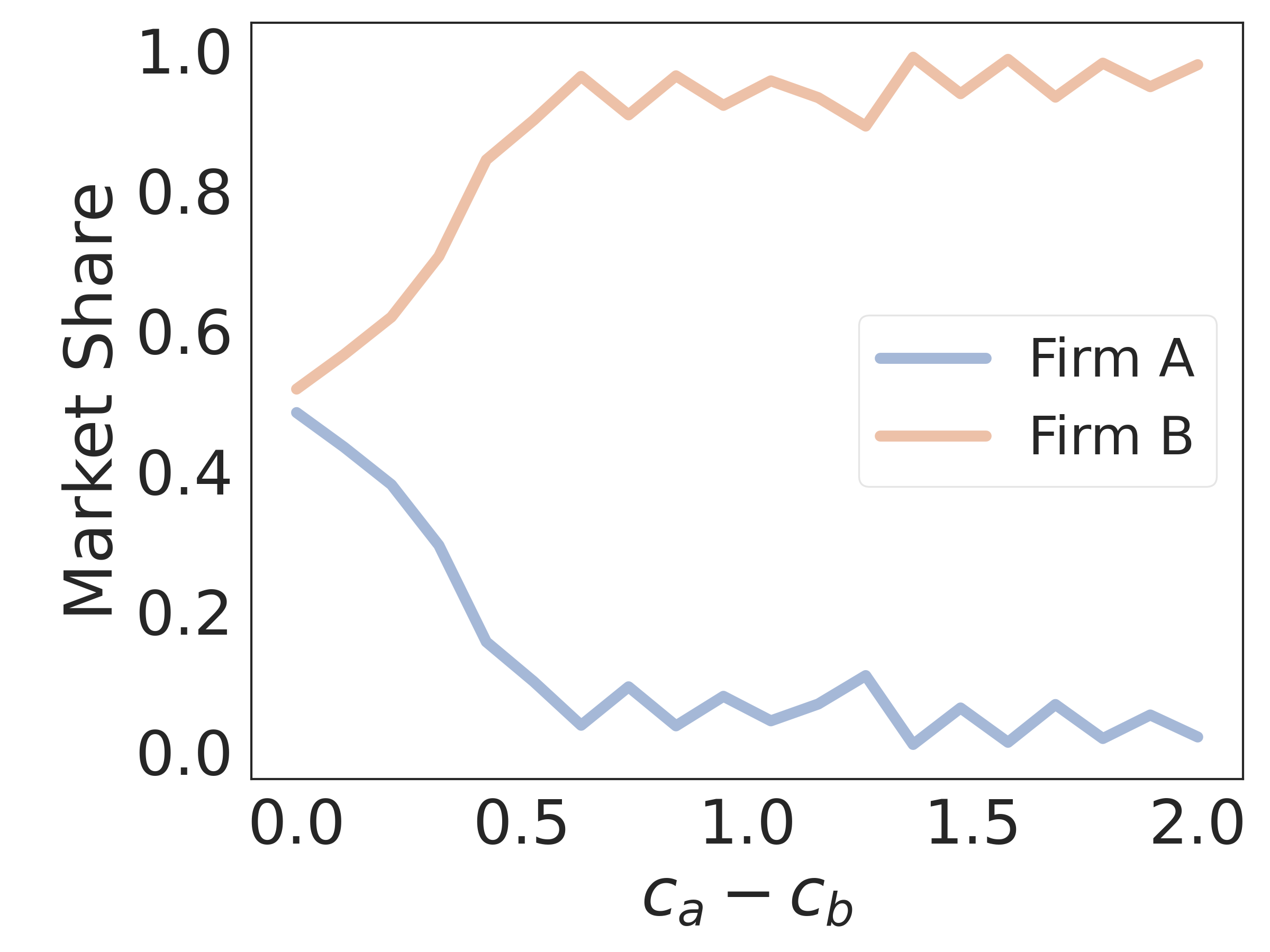} 
	\caption{Market share}
	\label{fig:al_ih_fm_c_mkt}
	\end{subfigure}
	\begin{subfigure}{0.3\textwidth}
	\includegraphics[width=0.9\linewidth]{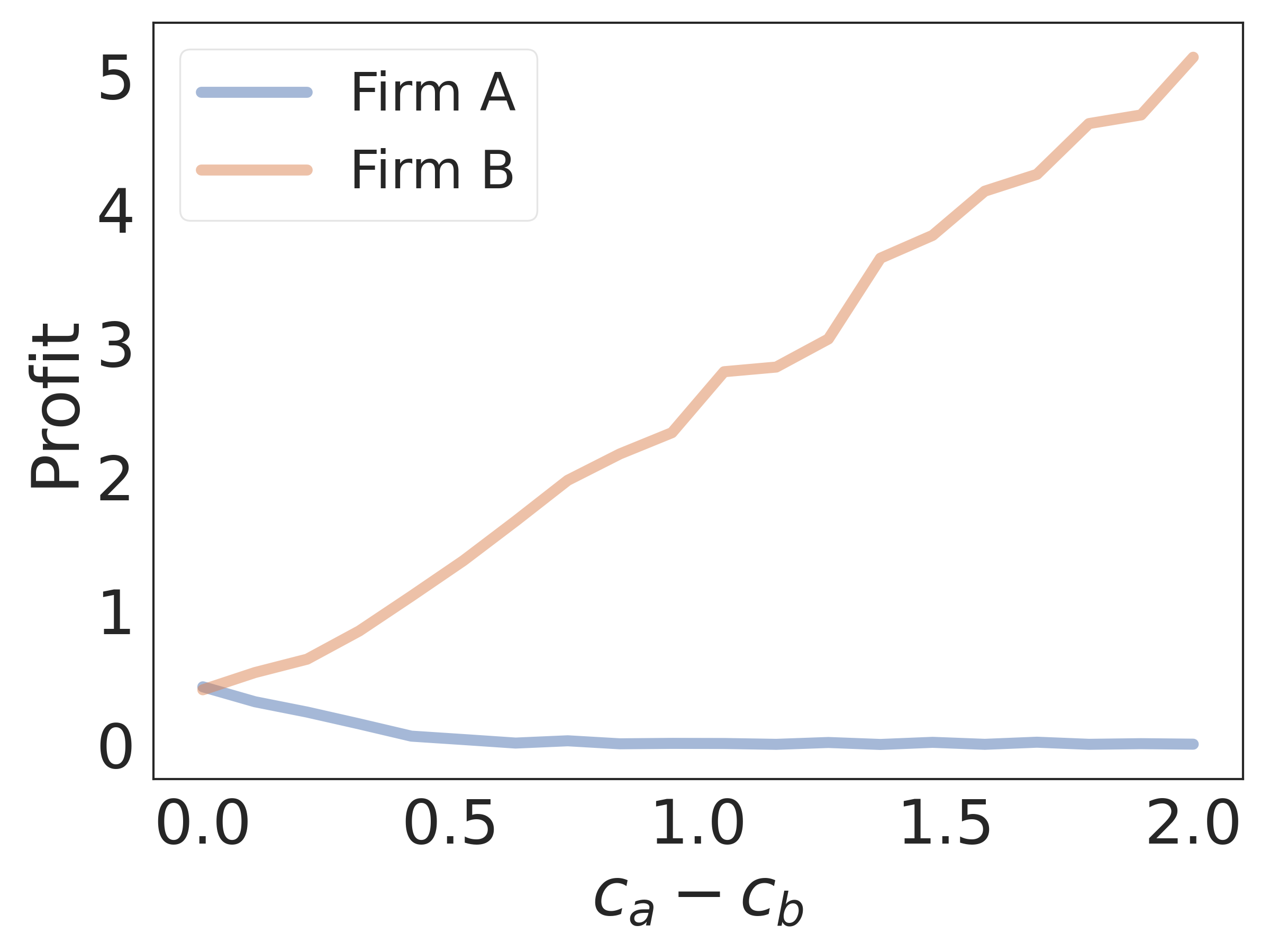}
	\caption{Profit}
	\label{fig:al_ih_fm_c_profit}
	\end{subfigure}
	\begin{subfigure}{0.3\textwidth}
	\includegraphics[width=0.9\linewidth]{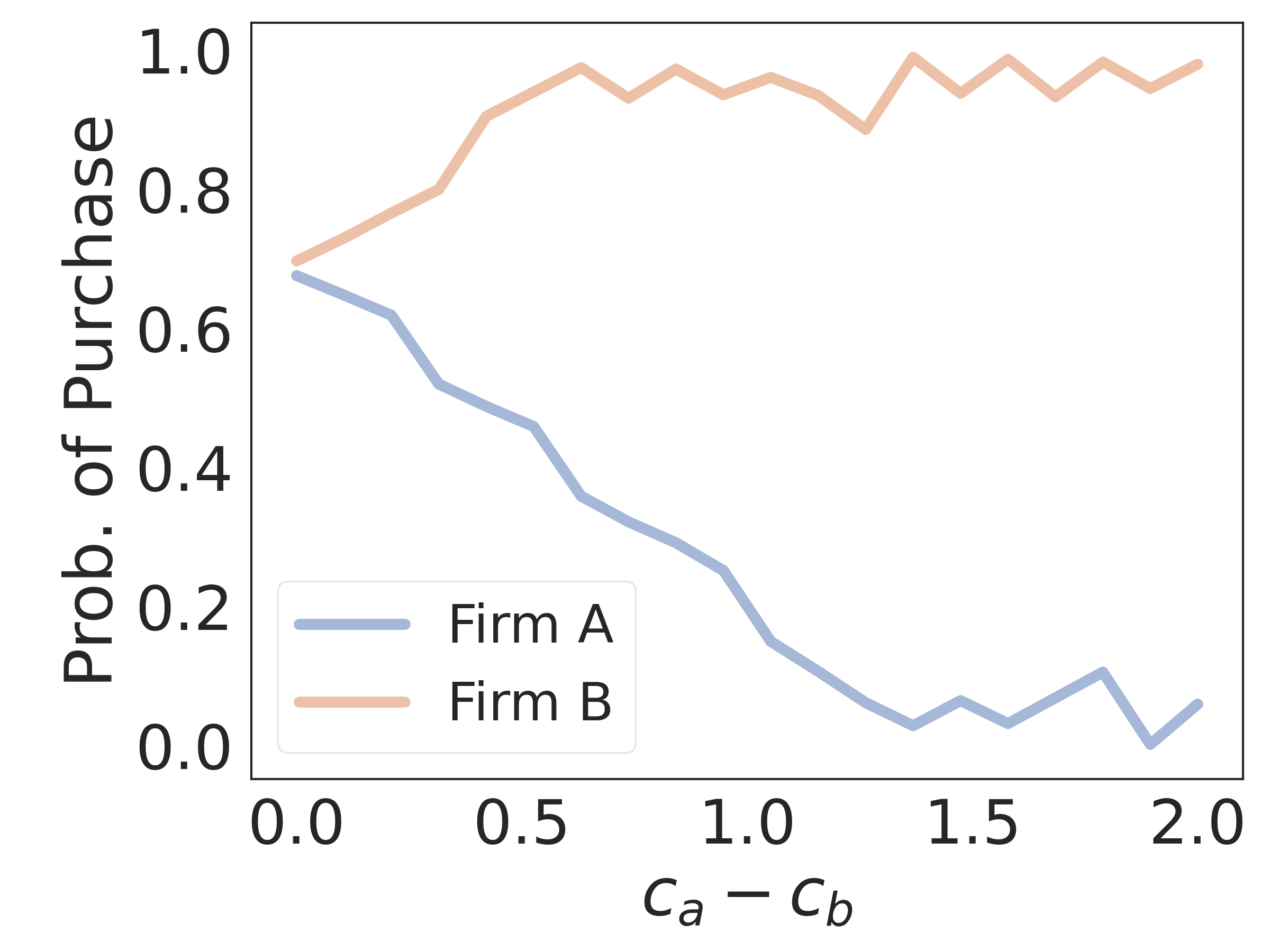}
	\caption{Probability of purchase}
	\label{fig:al_ih_fm_c_probabilities}
	\end{subfigure}
\caption{Infinite horizon setting market outcomes using numerical simulations under additive loyalty where the constraints can be binding: Optimal prices, market shares, profits of firms, and probability of purchase of customers of types $\alpha$ and $\beta$ as a function of $\ca-\cb$. Here, $\cb=.6, \oa=.1, \ob=.05, \df=.6$. In contrast to Figure~\ref{fig:al_ih_fm_unc}, the equilibrium is computed for cost differences up to $2$ units.}
\label{fig:al_ih_fm_c}
\end{figure}

When the prices can be constrained (i.e., Assumption~\ref{assume:gl-ih-unconstrained} does not hold), the market outcomes start evolving non-linearly with cost asymmetry (see Figure~\ref{fig:al_ih_fm_c}). For example, in Figure~\ref{fig:al_ih_fm_c_paa_pba}, we observe that $\paa$ has to initially increase faster than $\pba$ for firm $A$ to maximize its profit, and in Figure~\ref{fig:al_ih_fm_c_pbb_pab}, we observe that $\pbb$ converges to $\pab$ in a nonlinear way signifying that the lower cost firm $B$ just needs to match rival firm's prices to maximize profit and market share. Similarly, the rate of change of market share and profit as a function of cost asymmetry is also non-linear (Figures~\ref{fig:al_ih_fm_unc_mkt} and~\ref{fig:al_ih_fm_unc_profit}) with saturating trends for the former and runaway trends for the latter (i.e., with firm $B$ making much more profit). Similar to the multiplicative loyalty special case setting, numerical computation of equilibria driving the plots in Figure~\ref{fig:al_ih_fm_c} is performed using the dynamic stochastic game solver.

Although we once again omit the characterization of distinct regions (due to the non availability of closed-form expressions describing the boundaries as seen in Section~\ref{sec:al-ss} Figure~\ref{fig:al-ss-regions}), it is  evident that the interaction between cost asymmetry and additive loyalty has a significant impact on market outcomes, an aspect that was under-explored in prior literature.

\section{Proofs}\label{sec:proofs}

In this section, we aim to provide proofs for most claims made in the paper. For brevity, we omit ones that are very similar to other previously derived proofs.

\subsection{Proof of Proposition~\ref{prop:gl-ss-eq}}

\begin{proof} The result follows by noting that the profit optimization problem for each firm separates across variables. We are interested in the unconstrained variant of this optimization. Thus, firm $A$ can maximize with respect to $\paa$ and $\pab$ independent of each other. Taking the derivatives with respect to $\paa$ and $\pab$ and setting them to $0$ gives:
\begin{gather*}
-(\paa-\ca)\fa\hap + (1-\Fa) = 0, \textrm{ and}\\
-(\pab-\ca)\fb\hbp + \Fb = 0. 
\end{gather*}
A similar set of equations can also be obtained for firm $B$. Rearranging the terms yields the desired first order conditions (FOC).  
\end{proof}

\subsection{Proof of Lemma~\ref{lemma:gl-ih-mm}}

\begin{proof} The proof follows by a straightforward substitution of the definition of market shares for the single stage setting from Section~\ref{sec:gl-ss}. Given the initial market share $\theta$ and the optimal prices, the market share at $t=1$, i..e, $\theta_1$ is given as:
\begin{gather*}
\theta_1 = \theta(1-\Fa) + (1-\theta)\Fb,\\
\theta_2 = \theta_1(1-\Fa) + (1-\theta_1)\Fb,
\end{gather*}
and so on. As a result,
\begin{gather*}
\theta_t = \theta(1-\Fa-\Fb)^t + \Fb \sum_{j=0}^{t-1}(1-\Fa-\Fb)^j.
\end{gather*}
Finally using the fact that $\Fa+\Fb \in (0,2)$, we get the desired result. The result when $t \rightarrow \infty$ also follows naturally.
\end{proof}

\subsection{Proof of Proposition~\ref{prop:gl-ih-eq}}

\begin{proof} 
Because we are assuming that there are no price constraints, we can differentiate Equations~\ref{eqn:vaa}-\ref{eqn:vba} with respect to price variables to obtain the following implicit equations:
\begin{gather*}
{\scriptstyle
-(\paa-\ca)\fa\hap + (1-\Fa) -\da \fa\hap(\vaao-\vabo) = 0,}\\
{\scriptstyle-(\pab-\ca)\fb\hbp + \Fb -\da\fb\hbp (\vaao-\vabo) = 0,} \\
{\scriptstyle-(\pbb-\cb)\fb\hbp + (1-\Fb) -\db \fb\hbp(\vbbo-\vbao) = 0, \textrm{ and}}\\
{\scriptstyle-(\pba-\cb)\fa\hap + \Fa -\db\fa\hap (\vbbo-\vbao) = 0. }
\end{gather*}
Rearranging the terms gives us the desired implicit equations.
\end{proof}

\subsection{Proof of Proposition~\ref{prop:ll-ss-r1}}
\begin{proof} 
Our proof approach for Proposition~\ref{prop:ll-ss-r1} is similar to proof of Proposition~\ref{prop:ml-ss-r1}.
\end{proof}

\subsection{Proof of Proposition~\ref{prop:ll-ss-r2}}
\begin{proof}
In this proof, we will treat the two sub-markets separately. For each one, we will show that the candidate prices in the proposition indeed form a unique pure Nash equilibrium (PNE).

Our proof strategy is the same for both settings. To show that a pair of candidate prices constitute a PNE, we show that neither firm $A$ or $B$ would benefit by changing their prices unilaterally, assuming that the other firm doesn't change its price. Further, we also demonstrate that these prices are unique (by contradiction or otherwise).

\noindent\textbf{Establishing Unique PNE for the $\alpha$ market}

For the $\alpha$ sub-market, we have the candidate pair of PNE prices:

\begin{gather}
\paa = \frac{1}{3}(2\ca+\cb+\oa+2\ta),\;
\pba = \frac{1}{3}(\ca+2\cb-\oa+\ta).
\end{gather}

$\paa,\pba$ are a pair unique PNE prices because they are the only critical points of their objective profit functions and the second derivatives at these critical points are negative. In depth proof of these prices is provided in proof of proposition 5.

\noindent\textbf{Establishing Unique PNE for the $\beta$ market}

\noindent\textit{PNE prices}:
For the $\beta$ sub-market, we have the candidate pair of PNE prices:

\begin{gather}
\pab = \ca,\;
\pbb = \ca+\ob.
\end{gather}

At these prices, firm $A$ makes 0 dollars in profit:
\begin{align*}
(\pab - \ca)(1-\theta)&F(\xib)\\
 &=(\ca - \ca)(1-\theta)F\left(\frac{\ca +\ob - \ca - \ob}{\tb}\right),\\
 &=0
\end{align*}

And firm $B$ makes $\theta(\ca +\ob - \cb)\left(1-\frac{\ob}{\tb}\right)$ dollars in profit:
\begin{align*}
(\pbb - \cb)\theta&(1-F(\xib))\\
 &=(\ca +\ob - \cb)\theta\left(1-F\left(\frac{\ca +\ob - \ca - \ob}{\tb}\right)\right),\\
 &=\theta(\ca +\ob - \cb).
\end{align*}

Next, we show that neither firm can unilaterally deviate and be better off.\\ 
\noindent\textit{Deviation by Firm $A$}: Let fix $\pbb = \ca+\ob$. Firm $A$ cannot decrease its price $\pab$ to lower than its cost $\ca$. So it can only increase its price by $\pab = \ca + \epsilon$ where $\epsilon \in \mathbb{R}_{+}$. We see that the profit of firm $A$ in the $\beta$ sub-market is still 0 which means that A can't improve its profit by increasing its price:
\begin{align*}
(\pab - \ca)(1-\theta)&F(\xib)\\
 &=(\ca + \epsilon - \ca)(1-\theta)F\left(\frac{\ca +\ob - \ca - \epsilon - \ob}{\tb}\right),\\
 &=0.
\end{align*}
     
\noindent\textit{Deviation by Firm $B$}: Also, let fix $\pab = \ca$ and assume firm $B$ attempts to increase or decrease $\pbb$ to $\pbb = \ca + \ob + \epsilon$ or $\pbb = \ca + \ob - \epsilon$ respectively (again $\epsilon \in \mathbb{R}_{+}$). We show below that firm $B$ can't improve its profit.

By increasing $\pbb = \ca + \ob + \epsilon$, we have the following profit:
\begin{align*}
(\pbb - \cb)\theta&(1-F(\xib))\\
 &=(\ca + \ob + \epsilon - \cb)\theta\left(1-F\left(\frac{\ca + \ob + \epsilon - \ca - \ob}{\tb}\right)\right),\\
 &=\theta(\ca + \ob + \epsilon - \cb)\left(1-\frac{\epsilon}{\tb}\right).\\
\end{align*}

To show that the above profit is not a strict improvement over the current profit of $\theta(\camcb)$ is equivalent to show the following statements:
\begin{align*}
\theta(\ca + \ob + \epsilon - \cb)(\tb -\epsilon)& \leq \theta\tb(\ca +\ob - \cb), \\
\Leftrightarrow -\ca\epsilon - \ob\epsilon + \epsilon\tb - \epsilon^2 + \cb\epsilon &\leq 0\\
\Leftrightarrow \tb - \ob - \epsilon &\leq \camcb.
\end{align*}
The last inequality is true because $\max(\tb-\ob,\oa-\ta) \leq \camcb \leq \oa+2\ta$ (the definition of Region I).

By decreasing $\pbb = \ca + \ob - \epsilon$, we have the following profit:
\begin{align*}
(\pbb - \cb)\theta&(1-F(\xib))\\
 &=(\ca + \ob - \epsilon - \cb)\theta\left(1-F\left(\frac{\ca + \ob - \epsilon - \ca -\ob}{\tb}\right)\right),\\
 &=\theta(\ca + \ob - \epsilon - \cb)\leq\theta(\ca +\ob - \cb).
\end{align*}
Thus, $\pab = \ca$ and $\pbb = \ca+\ob$ are PNE prices.

\noindent\textit{Uniqueness of the above equilibrium}:

We now show that any other candidate price pairs in the $\beta$ sub-market cannot be a PNE by considering the following cases that are mutually exclusive and exhaustive, when combined with the above PNE:
\begin{itemize}
    \item Case 0: $\pbb = \ca+\ob, \pab > \ca$.
    \item Case 1: $\pbb > \ca+\ob, \pab \geq \ca$.
\end{itemize}
There are no other cases because $\pab < \ca$ is not feasible as firm $A$ cannot price below cost. Also, setting $\pbb < \ca+\ob$ is not a desirable option for firm $B$ because it will lower its profit since it already captures all the market share when $\pbb = \ca+\ob$.

\underline{Case 0}: Let $\pbb = y, \pab = x$ that satisfy the constraints $y = \ca+\ob$ and $x > \ca$. We show below that firm $B$ can unilaterally deviate and increase its profit. This invalidates the pair ($\pbb = y, \pab = x$) being a PNE.

\noindent\textit{Deviation by Firm $B$}: Let fix $\pab = \ca + \epsilon$ where $\epsilon \in \mathbb{R}_{+}$ and assume firm $B$ attempts to increase $\pbb$ from $\pbb = \ca + \ob$ to $\pbb = \ca + \ob + \eta$ where $\eta \in \mathbb{R}_{+}$ and $\eta \leq \epsilon$. We show below that firm $B$'s profit increases by increasing its price to any value between  $\ca + \ob$ and $\ca + \ob +\epsilon$. In fact, by matching $\eta = \epsilon$, firm $B$ maximizes its profit by the exact amount firm $A$ increases its price.

At $\pab = \ca + \epsilon$ and $\pbb = \ca+\ob$, firm $B$ makes $\theta(\ca - \cb)$ dollars in profit:
\begin{align*}
(\pbb - \cb)&\theta(1-F(\xib))\\
 &=(\ca + \ob - \cb)\theta\left(1-F\left(\frac{\ca +\ob - \ca - \epsilon - \ob}{\tb}\right)\right),\\
 &=\theta(\ca +\ob - \cb). 
\end{align*}

When firm $B$ increases its price, it increases its profit from $\theta(\ca + \ob - \cb)$ to $\theta(\ca + \ob + \eta - \cb)$:
\begin{align*}
(\pbb &- \cb)\theta(1-F(\xib))\\
 &=(\ca + \ob + \eta - \cb)\theta\left(1-F\left(\frac{\ca +\ob + \eta - \ca - \epsilon - \ob}{\tb}\right)\right),\\
 &=\theta(\ca + \ob + \eta - \cb) \geq \theta(\ca +\ob - \cb).
\end{align*}

Thus, the case of $\pbb = \ca + \ob, \pab > \ca$ cannot be a PNE.

\underline{Case 1}: Let $\pbb = y, \pab = x$ that satisfy the constraints $y > \ca + \ob$ and $x \geq \ca$ and that $y$ is a valid best response to $x$. This is without loss of generality because if $y$ is not the best response, firm $B$ can unilaterally deviate and improve. Further, there always exists such a valid best response because $x \geq \ca$. In particular, when $x=\ca$, $y = \ca + \ob$ is the best response (as shown at the beginning of this proof). And when $x > \ca$ by some $\epsilon$, $y = \ca + \ob + \epsilon$ is the best response (as shown in \emph{case 0} above).

Now, we show that firm $A$ can unilaterally deviate and increase its profit given the above pair of prices. This invalidates the given pair ($\pbb = y, \pab = x$) from being a PNE.

\noindent\textit{Deviation by Firm $A$}: Lets fix $y = \pbb  = \ca +\ob + \eta$ where $\eta \in \mathbb{R}_{+}$ and assume firm $A$ attempts to deviate $x = \pab = \ca + \eta$ to a new value.

At $\pab = \ca + \eta$ and $\pbb = \ca + \ob + \eta$, firm $A$ makes $0$ dollars in profit:
\begin{align*}
(\pab - \ca)&(1-\theta)F(\xib)\\
 &=(\ca + \eta - \ca)(1-\theta)F\left(\frac{\ca + \ob + \eta - \ca - \eta - \ob}{\tb}\right),\\
 & = 0.
\end{align*}

When $\pbb$ is fixed, Firm $A$ can find its best response $\ca + \epsilon=\ca + \frac{\eta}{2}$ by finding the critical point of the objective and checking that the second derivative is negative.

At the new value $\pab = \ca + \frac{\eta}{2}$, and $\pbb = \ca + \ob + \eta$, firm $A$ makes $(1-\theta)\frac{\eta^2}{4\tb}$ dollars in profit:
\begin{align*}
(\pab - \ca)&(1-\theta)F(\xib)\\
 &=(\ca + \frac{\eta}{2} - \ca)(1-\theta)F\left(\frac{\ca + \ob + \eta - \ca - \frac{\eta}{2} - \ob}{\tb}\right),\\
 &=\frac{\eta}{2}\left(\frac{\frac{\eta}{2}}{\tb}\right)(1-\theta) = \frac{\eta^2}{4\tb}(1-\theta) > 0.
\end{align*}

Thus, the case of $\pbb > \ca+\ob, \pab \geq \ca$ cannot lead to a PNE.

From the examined cases above, we can conclude that $\pab = \ca$ and $\pbb = \ca+\ob$ are the unique PNE prices.

\end{proof}

\subsection{Proof of Proposition~\ref{prop:ll-ss-r3}}
\begin{proof} 
Our proof approach for Proposition~\ref{prop:ll-ss-r3} is similar to proof of Proposition~\ref{prop:ml-ss-r1}.
\end{proof}

\subsection{Proof of Proposition~\ref{prop:ll-ss-r4}}
\begin{proof} 
Our proof approach for Proposition~\ref{prop:ll-ss-r4} is similar to proof of Proposition~\ref{prop:ml-ss-r1}.
\end{proof}

\subsection{Proof of Proposition~\ref{prop:ll-ss-r5}}
\begin{proof}

In this proof, we will show that the candidate prices in the proposition indeed form a unique PNE.

Our proof strategy is the same for both settings. We show that they are valid with their region definition and they are the only critical points of their objective profit functions and the second derivatives at these critical points are negative.

For the $\alpha$ market, we have the candidate pair of PNE prices:

\begin{gather}
\paa = \frac{1}{3}(2\ca+\cb+\oa+2\ta),\;
\pba = \frac{1}{3}(\ca+2\cb-\oa+\ta).
\end{gather}

Given Region III's definition $\oa-\ta \leq \camcb \leq \min(\oa+2\ta,\tb-\ob)$ or $\ca \leq \min\{\oa + 2\ta + \cb,\tb - \ob +\cb\}$, the candidate prices satisfy assumption 1 thus are valid, as shown below:

\begin{align*}
\paa &= \frac{1}{3}(2\ca+\cb+\oa+2\ta),\\
&= \frac{2}{3}\ca  +\frac{1}{3}\left(\cb +\oa +2\ta \right) \geq \frac{2}{3}\ca  +\frac{1}{3}\ca \geq \ca.
\end{align*}
\vspace{-1.5cm}
\begin{align*}
\pba &= \frac{1}{3}(\ca+2\cb-\oa+\ta),\\
&= \frac{2}{3}\cb  +\frac{1}{3}\left(\ca -\oa +\ta \right) \geq \frac{2}{3}\cb  +\frac{1}{3}\ca \geq \cb.
\end{align*}

A similar claim can be made for the $\beta$ market, which shows that the candidate prices satisfy assumption 1 thus are valid.

First, $\paa,\pba$ are a pair of unique extremums because they are the only critical points of their objective profit functions (see proof of Proposition 1).

We then show below that $\paa,\pba$ are maximums by obtaining negative values of the second derivatives of their objective functions. 

\begin{align*}
&\frac{d}{d\paa} (- (\paa-\ca)\fa\hap + (1-\Fa))\\
&= \frac{d}{d\paa}(-(\paa-\ca)\frac{1}{\ta} + 1 - \frac{\paa-\pba-\oa}{\ta}),\\
&= -2\frac{1}{\ta} < 0.
\end{align*}
Similarly,
\begin{align*}
&\frac{d}{d\pba} (-(\pab-\cb)\fa\hap + \Fa)\\
&= \frac{d}{d\pba}(-(\pab-\cb)\frac{1}{\ta} + \frac{\paa-\pba-\oa}{\ta}),\\
&= -2\frac{1}{\ta} < 0
\end{align*}

Thus, $\paa$ and $\pba$ are a pair of unique PNE prices.

A similar claim can be made for the $\beta$ market, which shows that $\pbb$ and $\pab$ as defined above are a pair of unique PNE prices.
\end{proof}

\subsection{Proof of Proposition~\ref{prop:ll-ss-r6}}
\begin{proof} 
Our proof approach for Proposition~\ref{prop:ll-ss-r5} is similar to proof of Proposition~\ref{prop:ml-ss-r1}.
\end{proof}

\subsection{Proof of Lemma~\ref{lemma:ll-ih-mm}}
\begin{proof} 
The five market share expressions are obtained by using Lemma~\ref{lemma:gl-ih-mm} with Propositions~\ref{prop:ll-ss-r1}-\ref{prop:ll-ss-r5}.
\end{proof}

\subsection{Proof of Proposition~\ref{prop:ll-ih-fm}}

\begin{proof} Our proof approach is similar to the proofs of Proposition~\ref{prop:ml-ih-fm} and Proposition~\ref{prop:al-ih-fm}, the only differences are the definition of $\gamma,\xia,\xib$. Let $\gamma = \ca - \cb -\oa$ and we know that $\frac{\paa - \pba - \oa}{\ta} = \ha = \xia$ and $\frac{\pbb - \pab - \ob}{\tb} = \hb = \xib$. We then can again show that we have a Markov equilibrium by proving $\xia$ is unique for any fixed $\xib$ and vice versa; after that, we assert $\vaao,\vabo,\vbbo$ and $\vbao$ are also unique, implying that the corresponding  prices (shown in Equations~\ref{eqn:paa}-\ref{eqn:pba}) constitute a unique Markov equilibrium.

For Linear Loyalty Model case, we get:
\begin{align*}
\Big(\xia -& \left.\frac{\camcb-\oa}{\ta}\right)\left(\frac{1-\df}{\df} +\Fb+1\right)\\
& + \frac{2\Fa-1}{\fa}\left(\frac{1-\df}{\df} +\Fa+\Fb\right) +  \frac{\Fa}{\fa} \\
& = \frac{(1-\Fb)\tb}{\fb\ta} - \Fb\left(\frac{\tb}{\ta}\xib +\frac{\camcb}{\ta}+\frac{\ob}{\ta}\right),
\end{align*}

We then can again see that for any fixed $\xib$ implies that there is a unique solution for $\xia$ and vice versa. These two results imply that there are unique solutions for $\vaao,\vabo,\vbbo$ and $\vbao$, and further imply the unique Markov prices $(\paa,\pab,\pbb,\pba)$ given in Equations~\ref{eqn:paa}-\ref{eqn:pba}.

\end{proof}

\subsection{Proof of Proposition~\ref{prop:ml-ss-r1}}
\begin{proof}
In this proof, we will treat the two sub-markets separately. For each one, we will show that the candidate prices in the proposition indeed form a unique pure Nash equilibrium (PNE).

Our proof strategy is the same for both settings. To show that a pair of candidate prices constitute a PNE, we show that neither firm $A$ or $B$ would benefit by changing their prices unilaterally, assuming that the other firm doesn't change its price. Further, we also demonstrate that these prices are unique (by contradiction or otherwise).

\noindent\textbf{Establishing Unique PNE for the $\alpha$ market}

For the $\alpha$ sub-market, we have the candidate pair of PNE prices:

\begin{gather}
\paa = \frac{1}{3}\left(2\ca  +\cb +2\ta \right),\;
\pba = \frac{1}{3}\left(2\cb  +\ca +\ta \right).
\end{gather}

$\paa,\pba$ are a pair unique PNE prices because they are the only critical points of their objective profit functions and the second derivatives at these critical points are negative. In depth proof of these prices is provided in proof of proposition 5.

\noindent\textbf{Establishing Unique PNE for the $\beta$ market}

\noindent\textit{PNE prices}:
For the $\beta$ sub-market, we have the candidate pair of PNE prices:

\begin{gather}
\pab = \ca,\;
\pbb = \ca.
\end{gather}

At these prices, firm $A$ makes 0 dollars in profit:
\begin{align*}
(\pab - \ca)(1-\theta)&F(\xib)\\
 &=(\ca - \ca)(1-\theta)F\left(\frac{\ca - \ca}{\tb}\right),\\
 &=0
\end{align*}

And firm $B$ makes $\theta(\ca - \cb)$ dollars in profit:
\begin{align*}
(\pbb - \cb)\theta&(1-F(\xib))\\
 &=(\ca - \cb)\theta\left(1-F\left(\frac{\ca - \ca}{\tb}\right)\right),\\
 &=\theta(\ca - \cb).
\end{align*}

Next, we show that neither firm can unilaterally deviate and be better off.\\ 
\noindent\textit{Deviation by Firm $A$}: Let fix $\pbb = \ca$. Firm $A$ cannot decrease its price $\pab$ to lower than its cost $\ca$. So it can only increase its price by $\pab = \ca + \epsilon$ where $\epsilon \in \mathbb{R}_{+}$. We see that the profit of firm $A$ in the $\beta$ sub-market is still 0 which means that A can't improve its profit by increasing its price:
\begin{align*}
(\pab - \ca)(1-\theta)&F(\xib)\\
 &=(\ca + \epsilon - \ca)(1-\theta)F\left(\frac{\ca - \ca - \epsilon}{\tb}\right),\\
 &=0.
\end{align*}

\noindent\textit{Deviation by Firm $B$}: Also, let fix $\pab = \ca$ and assume firm $B$ attempts to increase or decrease $\pbb$ to $\pbb = \ca + \epsilon$ or $\pbb = \ca - \epsilon$ respectively (again $\epsilon \in \mathbb{R}_{+}$). We show below that firm $B$ can't improve its profit.

By increasing $\pbb = \ca + \epsilon$, we have the following profit:
\begin{align*}
(\pbb - \cb)\theta&(1-F(\xib))\\
 &=(\ca + \epsilon - \cb)\theta\left(1-F\left(\frac{\ca + \epsilon - \ca}{\tb}\right)\right),\\
 &=\theta(\ca + \epsilon - \cb)\left(1-\frac{\epsilon}{\tb}\right).\\
\end{align*}

To show that the above profit is not a strict improvement over the current profit of $\theta(\camcb)$ is equivalent to show the following statements:
\begin{align*}
\theta(\ca + \epsilon - \cb)(\tb-\epsilon)& \leq \theta(\ca\tb - \cb\tb), \\
\Leftrightarrow -\ca\epsilon + \epsilon\tb - \epsilon^2 + \cb\epsilon &\leq 0, \\
\Leftrightarrow \tb - \epsilon &\leq \camcb.
\end{align*}
The last inequality is true because $\tb \leq \camcb \leq 2\ta$ (the definition of Region I).

By decreasing $\pbb = \ca - \epsilon$, we have the following profit:
\begin{align*}
(\pbb - \cb)\theta&(1-F(\xib))\\
 &=(\ca - \epsilon - \cb)\theta\left(1-F\left(\frac{\ca - \epsilon - \ca}{\tb}\right)\right),\\
 &=\theta(\ca - \epsilon - \cb)\leq\theta(\camcb).
\end{align*}
Thus, $\pab = \ca$ and $\pbb = \ca$ are PNE prices.

\noindent\textit{Uniqueness of the above equilibrium}:

We now show that any other candidate price pairs in the $\beta$ sub-market cannot be a PNE by considering the following cases that are mutually exclusive and exhaustive, when combined with the above PNE:
\begin{itemize}
    \item Case 0: $\pbb = \ca, \pab > \ca$.
    \item Case 1: $\pbb > \ca, \pab \geq \ca$.
\end{itemize}
There are no other cases because $\pab < \ca$ is not feasible as firm $A$ cannot price below cost. Also, setting $\pbb < \ca$ is not a desirable option for firm $B$ because it will lower its profit since it captures all the market share when $\pbb = \ca$.

\underline{Case 0}: Let $\pbb = y, \pab = x$ that satisfy the constraints $y = \ca$ and $x > \ca$. We show below that firm $B$ can unilaterally deviate and increase its profit. This invalidates the pair ($\pbb = y, \pab = x$) being a PNE.

\noindent\textit{Deviation by Firm $B$}: Let fix $\pab = \ca + \epsilon$ where $\epsilon \in \mathbb{R}_{+}$ and assume firm $B$ attempts to increase $\pbb$ from $\pbb = \ca$ to $\pbb = \ca + \eta$ where $\eta \in \mathbb{R}_{+}$ and $\eta \leq \epsilon$. We show below that firm $B$'s profit increases by increasing its price to any value between  $\ca$ and $\ca+\epsilon$. In fact, by matching $\pbb$ to $\pab$ ($\eta = \epsilon$), firm $B$ maximizes its profit by the exact amount firm $A$ increases its price.

At $\pab = \ca + \epsilon$ and $\pbb = \ca$, firm $B$ makes $\theta(\ca - \cb)$ dollars in profit:
\begin{align*}
(\pbb - \cb)\theta(1-F(\xib))\\
 &=(\ca - \cb)\theta\left(1-F\left(\frac{\ca - \ca - \epsilon}{\tb}\right)\right),\\
 &=\theta(\ca - \cb). 
\end{align*}

When firm $B$ increases its price, it increases its profit from $\theta(\ca - \cb)$ to $\theta(\ca + \eta - \cb)$:
\begin{align*}
(\pbb - \cb)\theta(1-F(\xib))\\
 &=(\ca + \eta - \cb)\theta\left(1-F\left(\frac{\ca + \eta - \ca - \epsilon}{\tb}\right)\right),\\
 &=\theta(\ca + \eta - \cb) \geq \theta(\ca - \cb).
\end{align*}

Thus, the case of $\pbb = \ca, \pab > \ca$ cannot be a PNE.

\underline{Case 1}: Let $\pbb = y, \pab = x$ that satisfy the constraints $y > \ca$ and $x \geq \ca$ and that $y$ is a valid best response to $x$. This is without loss of generality because if $y$ is not the best response, firm $B$ can unilaterally deviate and improve. Further, there always exists such a valid best response because $x \geq \ca$. In particular, when $x=\ca$, $y = \ca$ is the best response (as shown at the beginning of this proof). And when $x > \ca$, $y = x $ is the best response (as shown in \emph{case 0} above).

Now, we show that firm $A$ can unilaterally deviate and increase its profit given the above pair of prices. This invalidates the given pair ($\pbb = y, \pab = x$) from being a PNE.

\noindent\textit{Deviation by Firm $A$}: Lets fix $y = \pbb  = \ca + \eta$ where $\eta \in \mathbb{R}_{+}$ and assume firm $A$ attempts to deviate $x = \pab = \ca + \eta$ to a new value.

At $\pab = \ca + \eta$ and $\pbb = \ca + \eta$, firm $A$ makes $0$ dollars in profit:
\begin{align*}
(\pab - \ca)(1-\theta)F(\xib)\\
 &=(\ca + \eta - \ca)(1-\theta)F\left(\frac{\ca + \eta - \ca - \eta}{\tb}\right),\\
 & = 0.
\end{align*}

When $\pbb$ is fixed, Firm $A$ can find its best response $\ca + \epsilon=\ca + \frac{\eta}{2}$ by finding the critical point of the objective and checking that the second derivative is negative.

At the new value $\pab = \ca + \frac{\eta}{2}$, and $\pbb = \ca + \eta$, firm $A$ makes $(1-\theta)\frac{\eta^2}{4\tb}$ dollars in profit:
\begin{align*}
(\pab - \ca)(1-\theta)F(\xib)\\
 &=(\ca + \frac{\eta}{2} - \ca)(1-\theta)F\left(\frac{\ca + \eta - \ca - \frac{\eta}{2}}{\tb}\right),\\
 &=\frac{\eta}{2}\left(\frac{\frac{\eta}{2}}{\tb}\right)(1-\theta) = \frac{\eta^2}{4\tb}(1-\theta) > 0.
\end{align*}

Thus, the case of $\pbb > \ca, \pab \geq \ca$ cannot lead to a PNE.

From the examined cases above, we can conclude that $\pab = \ca$ and $\pbb = \ca$ are the unique PNE prices.

\end{proof}

\subsection{Proof of Proposition~\ref{prop:ml-ss-r2}}
\begin{proof} 
Our proof approach for Proposition~\ref{prop:ml-ss-r2} is similar to proof of Proposition~\ref{prop:ml-ss-r1}.
\end{proof}

\subsection{Proof of Proposition~\ref{prop:ml-ss-r3}}
\begin{proof}
In this proof, we will show that the candidate prices in the proposition indeed form a unique PNE.

Our proof strategy is the same for both settings. We show that they are valid with their region definition and they are the only critical points of their objective profit functions and the second derivatives at these critical points are negative.

For the $\alpha$ market, we have the candidate pair of PNE prices:

\begin{gather}
\paa = \frac{1}{3}\left(2\ca  +\cb +2\ta \right),\;
\pba = \frac{1}{3}\left(2\cb  +\ca +\ta \right).
\end{gather}

Given Region III's definition, i.e., $\camcb \leq \min\{2\ta,\tb\}$ or $\ca \leq \min\{2\ta + \cb,\tb+\cb\}$, the candidate prices satisfy assumption 1 thus are valid.

\begin{align*}
\paa &= \frac{1}{3}\left(2\ca  +\cb +2\ta \right),\\
&= \frac{2}{3}\ca  +\frac{1}{3}\left(\cb +2\ta \right) \geq \frac{2}{3}\ca  +\frac{1}{3}\ca \geq \ca.
\end{align*}

\begin{align*}
\pba &= \frac{1}{3}\left(2\cb  +\ca +\ta \right),\\
&= \frac{2}{3}\cb  +\frac{1}{3}\left(\ca + \ta \right) \geq \frac{2}{3}\cb  +\frac{1}{3}\ca \geq \cb.
\end{align*}

A similar claim can be made for the $\beta$ market, which shows that the candidate prices satisfy assumption 1 thus are valid.

$\paa,\pba$ are a pair of unique extremums because they are the only critical points of their objective profit functions (see proof of Proposition 1).

We then show below that $\paa,\pba$ are maximums by obtaining negative values of the second derivatives of their objective functions. 

\begin{align*}
&\frac{d}{d\paa} (- (\paa-\ca)\fa\hap + (1-\Fa))\\
&= \frac{d}{d\paa}(-(\paa-\ca)\frac{1}{\ta} + 1 - \ha),\\
&= -2\frac{1}{\ta} < 0.
\end{align*}
Similarly,
\begin{align*}
&\frac{d}{d\pba} (-(\pab-\cb)\fa\hap + \Fa)\\
&= \frac{d}{d\pba}(-(\pab-\cb)\frac{1}{\ta} + \ha),\\
&= -2\frac{1}{\ta} < 0
\end{align*}

Thus, $\paa$ and $\pba$ are a pair of unique PNE prices.

A similar claim can be made for the $\beta$ market, which shows that  $\pbb$ and $\pab$ as defined above are a pair of unique PNE prices.

\end{proof}

\subsection{Proof of Proposition~\ref{prop:ml-ss-r4}}
\begin{proof} 
Our proof approach for Proposition~\ref{prop:ml-ss-r4} is similar to proof of Proposition~\ref{prop:ml-ss-r1}.
\end{proof}

\subsection{Proof of Lemma~\ref{lemma:ml-ih-mm}}
\begin{proof} 
The four market share expressions are obtained by using Lemma~\ref{lemma:gl-ih-mm} with Propositions~\ref{prop:ml-ss-r1}-\ref{prop:ml-ss-r4}.
\end{proof}

\subsection{Proof of Proposition~\ref{prop:ml-ih-fm}}

\begin{proof}
To show that we have a Markov equilibrium, we first show that $\xia$ is unique for any fixed $\xib$ and vice versa. And then we assert that $\vaao,\vabo,\vbbo$ and $\vbao$ are also unique, implying that the corresponding  prices (shown in Equations~\ref{eqn:paa}-\ref{eqn:pba}) constitute a unique Markov equilibrium.

Let $\gamma = (\ca - \cb)/\ta$. Noting that $\da=\db=\df$, $\ha = (\paa-\pba)/\ta$ and $\hap = 1/\ta$, we can simplify the expression for $\xia$ in Equation~\ref{eqn:xia} as follows:
\small
\begin{align*}
&\xia\\
\;&
= h_{\alpha}\left(\ca -\cb + \frac{1-2\Fa}{\fa\hap} - \da(\va)+\db(\vb)\right)\\
\;&= \frac{1}{\ta}\left(\ca -\cb + \frac{1-2\Fa}{\fa}\ta - \da(\va)+\db(\vb)\right)\\
\;& = \gamma +  \frac{1-2\Fa}{\fa} -\frac{\df}{\ta}(\va-(\vb)).
\end{align*}
\normalsize
Substituting candidate prices back into the value function expressions in Equations~\ref{eqn:vaa}-\ref{eqn:vba}, we can get the following equations for the difference of optimal values (i.e., the difference between the optimal value from the strong sub-market and the weak sub-market), one for each firm:
\small
\begin{gather}
\vaao-\vabo \nonumber\\
= \frac{1}{1-\da +\da(\Fa+\Fb)}\left( (1-\Fa)(\paa-\ca) - \Fb(\pab-\ca)\right), \label{eqn:va}\\
\vbbo -\vbao \nonumber\\
= \frac{1}{1-\db +\db(\Fa+\Fb)}\left( (1-\Fb)(\pbb-\cb) - \Fa(\pba-\cb)\right). \label{eqn:vb}
\end{gather}
\normalsize
Further, the optimal values can also be obtained ad a function of prices by solving the following system of equations:
\begin{align*}
(1-\da(1-\Fa))\vaao - \da\Fa\vabo &= (1-\Fa)(\paa-\ca),\\
-\da\Fb\vaao + (1-\da(1-\Fb))\vabo &= \Fb(\pab - \ca),\\
(1-\db(1-\Fb))\vbbo - \db\Fb\vbao &= (1-\Fb)(\pbb-\cb), \textrm{ and}\\
-\db\Fa\vbbo + (1-\db(1-\Fa))\vbao &= \Fa(\pba - \cb).
\end{align*}

From Equations~\ref{eqn:va} and ~\ref{eqn:vb}, we know that:
\begin{align*}
&\vaao-\vabo \\
&\quad\ = \frac{1}{1-\df +\df(\Fa+\Fb)}\\
&\quad\quad\quad\times\left( (1-\Fa)(\paa-\ca) - \Fb(\pab-\ca)\right), \text{ and}\\
&\vbbo -\vbao\\
&\quad = \frac{1}{1-\df +\df(\Fa+\Fb)}\\
&\quad\quad\quad\times\left( (1-\Fb)(\pbb-\cb) - \Fa(\pba-\cb)\right).
\end{align*}
Substituting these in the above expression for $\xia$, we get:
\begin{align}
\xia = &\gamma + \frac{1-2\Fa}{\fa} \nonumber\\
	& -\frac{\df}{\ta(1-\df +\df(\Fa+\Fb))} \nonumber\\
	&\quad\quad \Bigg(\Big( (1-\Fa)(\paa-\ca)
	 - \Fb(\pab-\ca)\Big) \nonumber\\
	&\quad\quad\quad -\Big( (1-\Fb)(\pbb-\cb) - \Fa(\pba-\cb)\Big)\Bigg).\label{eqn:prop:ml-ih-fm} 
\end{align}

In the above expression, we intend to replace prices with $\xia$ and $\xib$ and then segregate all terms involving $\xia$ to the left. This allows us to inspect if the left hand side is monotonic for every value of $\xib$. As we show below, this is the case. We start with replacing terms involving prices with terms involving $\xia$ and $\xib$. Let
{\small
\begin{align*}
T &= \Big( (1-\Fa)(\paa-\ca)- \Fb(\pab-\ca)\Big)\\
 &\quad\quad -\Big( (1-\Fb)(\pbb-\cb) - \Fa(\pba-\cb)\Big),\\
 &= \paa -\ca -\Fa\paa + \Fa\ca -\Fb\pab +\Fb\ca \\
 &\quad\quad -(\pbb-\cb -\Fb\pbb + \Fb\cb -\Fa\pba + \Fa\cb)\\
 &=  \paa -\ca -\Fa\paa + \Fa\ca -\Fb\pab +\Fb\ca \\
 &\quad\quad -\pbb + \cb +\Fb\pbb -\Fb\cb +\Fa\pba - \Fa\cb\\
 &= \paa -\pbb -\ta\gamma -\Fa(\paa-\pba) +\Fa\ta\gamma\\
 &\quad\quad\quad\quad +\Fb(\pbb-\pab) +\Fb\ta\gamma,
\end{align*}
}%
where in the last step we substituted the definition of $\gamma$ and grouped a few terms together. Further, from the definition of $\xia$ and $\xib$, we have $\paa - \pba = \ta\xia$ and $\pbb-\pab = \tb\xib$. Thus,
\begin{align*}
T = \paa -\pbb -\ta\gamma -\Fa\ta\xia +\Fa\ta\gamma +\Fb\tb\xib +\Fb\ta\gamma.
\end{align*} 

From Equations~\ref{eqn:pbb} and~\ref{eqn:pba} of optimal prices, we also know that:
\begin{gather*}
\pbb = \frac{1-\Fb}{\fb\hbp} + \pba - \frac{\Fa}{\fa\hap} \\
 =  \frac{1-\Fb}{\fb}\tb + \pba - \frac{\Fa}{\fa}\ta.
\end{gather*}
Again using the identity $\pba = \paa - \ta\xia$ we get,
\begin{align*}
\pbb &=  \frac{1-\Fb}{\fb}\tb + \paa - \ta\xia - \frac{\Fa}{\fa}\ta,\\
\Rightarrow \paa - \pbb &=  -\frac{1-\Fb}{\fb}\tb + \ta\xia + \frac{\Fa}{\fa}\ta.\\
\end{align*}
Thus, the term $T$ can be updated as:
\begin{align*}
T &= -\frac{1-\Fb}{\fb}\tb + \ta\xia + \frac{\Fa}{\fa}\ta \\
&\quad\quad -\ta\gamma -\Fa\ta\xia +\Fa\ta\gamma +\Fb\tb\xib +\Fb\ta\gamma,\\
&= -\frac{1-\Fb}{\fb}\tb + \ta(\xia-\gamma) \\
&\quad\quad + \frac{\Fa}{\fa}\ta -\Fa\ta(\xia -\gamma) +\Fb\tb\xib +\Fb\ta\gamma.
\end{align*} 
Rearranging terms in Equation~\ref{eqn:prop:ml-ih-fm} and using the definition of $T$ above, we get:
\begin{align*}
(\xia - \gamma)&\left(\frac{1-\df}{\df} +\Fa+\Fb\right) \\
&\quad\quad +  \frac{2\Fa-1}{\fa}\left(\frac{1-\df}{\df} +\Fa+\Fb\right) = -\frac{1}{\ta}T.
\end{align*}
Bringing the terms involving $\xia$ in the expression $T$ to the left hand side, we get:
\begin{align*}
(\xia - \gamma)&\left(\frac{1-\df}{\df} +\Fb+1\right) \\
	& +  \frac{2\Fa-1}{\fa}\left(\frac{1-\df}{\df} +\Fa+\Fb\right) +  \frac{\Fa}{\fa} \\
	& = -\frac{1}{\ta}\left(-\frac{1-\Fb}{\fb}\tb +\Fb\tb\xib +\Fb\ta\gamma\right),\\
	& = \frac{(1-\Fb)\tb}{\fb\ta} - \Fb\left(\frac{\tb}{\ta}\xib +\gamma\right).
\end{align*}

Under Assumption~\ref{assume:cdf}, we know that $\frac{2\Fa-1}{\fa}$ and $\frac{\Fa}{\fa}$ are increasing functions of $\xia$. Thus, each term on the left hand side of the above expression is monotonically increasing with $\xia$. Equating the left hand side to a constant on the right hand side (for any fixed $\xib$) implies that there is a unique solution for $\xia$. An analogous claim can be made for $\xib$ as well. These two results imply that there are unique solutions for $\vaao,\vabo,\vbbo$ and $\vbao$, and further imply the unique Markov prices $(\paa,\pab,\pbb,\pba)$ given in Equations~\ref{eqn:paa}-\ref{eqn:pba}.

\end{proof}

\subsection{Proof of Proposition~\ref{prop:al-ss-r1}}
\begin{proof} 
Our proof approach for Proposition~\ref{prop:al-ss-r1} is similar to proof of Proposition~\ref{prop:ml-ss-r1}.
\end{proof}

\subsection{Proof of Proposition~\ref{prop:al-ss-r2}}
\begin{proof} 
Our proof approach for Proposition~\ref{prop:al-ss-r2} is similar to proof of Proposition~\ref{prop:ml-ss-r1}.
\end{proof}

\subsection{Proof of Proposition~\ref{prop:al-ss-r3}}
\begin{proof} 
Our proof approach for Proposition~\ref{prop:al-ss-r3} is similar to proof of Proposition~\ref{prop:ml-ss-r1}.
\end{proof}

\subsection{Proof of Proposition~\ref{prop:al-ss-r4}}
\begin{proof} 
Our proof approach for Proposition~\ref{prop:al-ss-r4} is similar to proof of Proposition~\ref{prop:ml-ss-r1}.
\end{proof}

\subsection{Proof of Proposition~\ref{prop:al-ss-r5}}
\begin{proof} 
Our proof approach for Proposition~\ref{prop:al-ss-r5} is similar to proof of Proposition~\ref{prop:ml-ss-r3}.
\end{proof}

\subsection{Proof of Lemma~\ref{lemma:al-ih-mm}}
\begin{proof} 
The five market share expressions are obtained by using Lemma~\ref{lemma:gl-ih-mm} with Propositions~\ref{prop:al-ss-r1}-\ref{prop:al-ss-r5}.
\end{proof}

\subsection{Proof of Proposition~\ref{prop:al-ih-fm}}

\begin{proof} Similar to the proof of Proposition~\ref{prop:ml-ih-fm} above, to show that we have a Markov equilibrium, we first show that $\xia$ is unique for any fixed $\xib$ and vice versa. And then we assert that $\vaao,\vabo,\vbbo$ and $\vbao$ are also unique, implying that the corresponding  prices (shown in Equations~\ref{eqn:paa}-\ref{eqn:pba}) constitute a unique Markov equilibrium.

Let $\gamma = \ca - \cb -\oa$. Noting that $\da=\db=\df$, $\ha = \paa-\pba -\oa$ and $\hap = 1$, we can simplify the expression for $\xia$ in Equation~\ref{eqn:xia} as follows:
\begin{gather*}
\xia  = \gamma +  \frac{1-2\Fa}{\fa} -\df(\va-(\vb)).
\end{gather*}
From Equations~\ref{eqn:va} and ~\ref{eqn:vb}, we know that:
\begin{align*}
&\vaao-\vabo \\
&\quad = \frac{1}{1-\df +\df(\Fa+\Fb)}\\
&\quad\quad\quad\times \left( (1-\Fa)(\paa-\ca) - \Fb(\pab-\ca)\right), \text{ and}\\
&\vbbo -\vbao \\
&\quad = \frac{1}{1-\df +\df(\Fa+\Fb)}\\
&\quad\quad\quad\times\left( (1-\Fb)(\pbb-\cb) - \Fa(\pba-\cb)\right).
\end{align*}
Substituting these in the above expression for $\xia$, we get:
\begin{align}
\xia& = \gamma \nonumber\\
&+  \frac{1-2\Fa}{\fa} \nonumber\\
&-\frac{\df}{1-\df +\df(\Fa+\Fb)}\Big(\Big( (1-\Fa)(\paa-\ca)\nonumber\\
&- \Fb(\pab-\ca)\Big)-\Big( (1-\Fb)(\pbb-\cb) - \Fa(\pba-\cb)\Big)\Big).\label{eqn:prop:al-ih-fm} 
\end{align}

In the above expression, we intend to replace prices with $\xia$ and $\xib$ and then segregate all terms involving $\xia$ to the left. This allows us to inspect if the left hand side is monotonic for every value of $\xib$. As we show below, this is the case. We start with replacing terms involving prices with terms involving $\xia$ and $\xib$. Let
{\small
\begin{align*}
T &= \Big( (1-\Fa)(\paa-\ca)- \Fb(\pab-\ca)\Big)\\
  &-\Big( (1-\Fb)(\pbb-\cb) - \Fa(\pba-\cb)\Big),\\
  &= \paa -\pbb -\gamma-\oa -\Fa(\paa-\pba) \\
  &+\Fa\gamma +\Fa\oa +\Fb(\pbb-\pab) +\Fb\gamma +\Fb\oa,
\end{align*}
}%
where we substituted the definition of $\gamma$ and grouped a few terms together. Further, from the definition of $\xia$ and $\xib$, we have $\paa - \pba - \oa= \xia$ and $\pbb-\pab -\ob = \xib$. Thus,
\begin{align*}
T &= \paa -\pbb - \gamma -\oa -\Fa\xia +\Fa\gamma \\
&\quad\quad\quad +\Fb\xib +\Fb(\gamma + \oa +\ob).
\end{align*} 

From Equations~\ref{eqn:pbb} and~\ref{eqn:pba} of optimal prices, we also know that:
\begin{gather*}
\pbb = \frac{1-\Fb}{\fb} + \pba - \frac{\Fa}{\fa}.
\end{gather*}
Again using the identity $\pba = \paa - \oa - \xia$ we get,
\begin{align*}
\Rightarrow \paa - \pbb -\oa &=  \xia -\frac{1-\Fb}{\fb} + \frac{\Fa}{\fa}.\\
\end{align*}
Thus, the term $T$ can be updated as:
\begin{align*}
T &= \xia -\frac{1-\Fb}{\fb} + \frac{\Fa}{\fa}- \gamma \\
  &-\Fa\xia +\Fa\gamma +\Fb\xib +\Fb(\gamma + \oa +\ob).
\end{align*} 
Rearranging terms in Equation~\ref{eqn:prop:al-ih-fm} and using the definition of $T$ above, we get:
\begin{align*}
(\xia - \gamma)&\left(\frac{1-\df}{\df} +\Fa+\Fb\right) +  \\
& \frac{2\Fa-1}{\fa}\left(\frac{1-\df}{\df} +\Fa+\Fb\right) = -T.
\end{align*}
Bringing the terms involving $\xia$ in the expression $T$ to the left hand side, we get:
\begin{align*}
(\xia - \gamma)&\left(\frac{1-\df}{\df} +\Fb+1\right) \\
	&+  \frac{2\Fa-1}{\fa}\left(\frac{1-\df}{\df} +\Fa+\Fb\right) +  \frac{\Fa}{\fa} \\
	& = \frac{(1-\Fb)}{\fb} - \Fb\left(\xib +\gamma + \oa +\ob \right).
\end{align*}

Under Assumption~\ref{assume:cdf}, we know that $\frac{2\Fa-1}{\fa}$ and $\frac{\Fa}{\fa}$ are increasing functions of $\xia$. Thus, each term on the left hand side of the above expression is monotonically increasing with $\xia$. Equating the left hand side to a constant on the right hand side (for any fixed $\xib$) implies that there is a unique solution for $\xia$. An analogous claim can be made for $\xib$ as well. These two results imply that there are unique solutions for $\vaao,\vabo,\vbbo$ and $\vbao$, and further imply the unique Markov prices $(\paa,\pab,\pbb,\pba)$ given in Equations~\ref{eqn:paa}-\ref{eqn:pba}.

\end{proof}

\end{document}